\documentclass[longauth]{aa} 

\usepackage{color}
\newcommand{\gtsim}{\protect\raisebox{-0.5ex}{$\:\stackrel{\textstyle >}
        {\sim}\:$}}
\newcommand{\ltsim}{\protect\raisebox{-0.5ex}{$\:\stackrel{\textstyle <}
        {\sim}\:$}}
\newcommand{\teff}{$T_{\rm eff}$}
\newcommand{\logg}{$\log g$}

\newcommand{\csi}{$\xi$}
\newcommand{\feh}{[Fe/H]}

\newcommand{\vsini}{$v\sin i$}

\usepackage[flushleft]{threeparttable}

\usepackage[colorinlistoftodos,prependcaption]{todonotes}
\presetkeys%
    {todonotes}%
    {inline,backgroundcolor=yellow}{}
    
\usepackage{lipsum}

\usepackage{graphicx}
\usepackage{hyperref}
\usepackage{txfonts} 
\usepackage{lscape}   
\usepackage{natbib}

\usepackage{orcidlink}

\usepackage{comment}

\usepackage{multirow}

\usepackage{soul}

\bibpunct{(}{)}{;}{a}{}{,} 
\makeatother

\usepackage[normalem]{ulem}

\begin{document}

\title{The GAPS Programme at the TNG}
\subtitle{LXX. TOI-5734b: A hot sub-Neptune orbiting a relatively young K dwarf with an Earth-like density\thanks{Based on observations made with the Italian Telescopio Nazionale {\it Galileo}  operated on the island of La Palma by the INAF - {\it Fundaci\'on Galileo Galilei} at the {\it Roque de los Muchachos} Observatory of the {\it Instituto de Astrof\'isica de Canarias} (IAC) in the framework of the Large Programme Global Architecture of Planetary Systems (GAPS; PI: Micela, ID: A37TAC\_31) and the Large Programme \textit{Ariel} Mass Survey (ArMS) programme (PI: Benatti, ID: A48TAC$\_$48).}}
   
\author{
S.~Filomeno\inst{\ref{oaroma}, \ref{unitor}, \ref{sapienza}} \orcidlink{0009-0000-5623-5237} \and
T.~Trifonov\inst{\ref{mpia}, \ref{unisofia}, \ref{uniheidelberg}} \and
M.~Damasso\inst{\ref{oatorino}} \and
M.~Baratella\inst{\ref{esochile}}\orcidlink{0000-0002-1027-5003} \and
S.~Benatti\inst{\ref{oapalermo}} \and
K.~Biazzo\inst{\ref{oaroma}} \orcidlink{0000-0002-1023-6821} \and
K.A.~Collins\inst{\ref{harvard}}\orcidlink{0000-0001-6588-9574}\and
R.~Cosentino\inst{\ref{tng}} \and
S.~Desidera\inst{\ref{oapadova}} \and
C.~Di Maio\inst{\ref{oapalermo}}\and
D.~Locci\inst{\ref{oapalermo}}\and
A.~Maggio\inst{\ref{oapalermo}}\and
L.~Mancini\inst{\ref{unitor}, \ref{oatorino}, \ref{mpia}} \and
S.~Messina\inst{\ref{oacatania}}\and
L.~Naponiello\inst{\ref{oatorino}}\and
D.~Nardiello\inst{\ref{unipadova}, \ref{oapadova}} \and
K.~G.~Stassun\inst{\ref{nashville_aa}} \orcidlink{0000-0002-3481-9052} \and
F.~Amadori\inst{\ref{oatorino}}\and
S.~Antoniucci\inst{\ref{oaroma}} \and
F.~Biassoni\inst{\ref{uniinsubria},\ref{oabrera}}\and 
A.S.~Bonomo\inst{\ref{oatorino}}\and 
L.~Cabona\inst{\ref{oapadova}}\and
C.A.~Clark\inst{\ref{nasa-ipac}}\orcidlink{0000-0002-2361-5812}\and
M.~Gonzalez\inst{\ref{tng}}\and
A.~F.~Lanza\inst{\ref{oacatania}}\and
D.W.~Latham\inst{\ref{harvard}}\orcidlink{0000-0001-9911-7388}\and
V.~Lorenzi\inst{\ref{tng}}\and
L.~Malavolta\inst{\ref{unipadova}, \ref{oapadova}}\and
G.~Mantovan\inst{\ref{unipadova},\ref{oapadova}}\and
M.~Pedani\inst{\ref{tng}} \and
G.~Scandariato\inst{\ref{oacatania}}\and
A.~Shporer\inst{\ref{mit}} \orcidlink{0000-0002-1836-3120} \and
A.~Sozzetti\inst{\ref{oatorino}}\and 
T.~Zingales\inst{\ref{unipadova},\ref{oapadova}} \and
D.~Ciardi\inst{\ref{nasa-ipac}}\orcidlink{0000-0002-5741-3047}\and
M.~Everett\inst{\ref{noirlab}}\and
M.B.~Lund\inst{\ref{nasa-ipac}}\orcidlink{0000-0003-2527-1598}\and
H.~Osborn\inst{\ref{csh_unibe}}\and
I.~A.\,Strakhov\inst{\ref{sai_moscow}}\orcidlink{0000-0003-0647-6133}\and
J.~Villase{\~ n}or\inst{\ref{mit}}\and
D.~Watanabe\inst{\ref{Fredericksburg}}\and 
A.~Youngblood\inst{\ref{nasa_goddard}}\and
R.~Zambelli\inst{\ref{sal_ita}}
}

\offprints{simone.filomeno@inaf.it}
\mail{simone.filomeno@inaf.it}
\institute{
INAF-Osservatorio Astronomico di Roma, Via Frascati 33, I-00040 Monte Porzio Catone (RM), Italy \label{oaroma}
\and Dipartimento di Fisica, Universit\`a di Roma Tor Vergata, Via della Ricerca Scientifica 1, I-00133 Roma, Italy \label{unitor}
\and Dipartimento di Fisica, Sapienza Università di Roma, Piazzale Aldo Moro 5, I-00185 Roma, Italy \label{sapienza}
\and Max Planck Institute for Astronomy, K\"{o}nigstuhl 17, D-69117, Heidelberg, Germany \label{mpia}
\and Department of Astronomy, Sofia University "St Kliment Ohridski", 5 James Bourchier Blvd, BG-1164 Sofia, Bulgaria \label{unisofia}
\and Landessternwarte, Zentrum f\"ur Astronomie der Universt\"at Heidelberg, K\"onigstuhl 12, 69117 Heidelberg, Germany \label{uniheidelberg}
\and INAF - Osservatorio Astrofisico di Torino, Via Osservatorio 20, I-10025 Pino Torinese, Italy, \label{oatorino}
\and ESO - European Southern Observatory, Alonso de Córdova 3107, Casilla 19, Santiago 19001, Chile \label{esochile}
\and INAF - Osservatorio Astronomico di Palermo, Piazza del Parlamento 1, I-90134 Palermo, Italy \label{oapalermo}
\and Centre for Astrophysics \textbar \ Harvard \& Smithsonian, 60 Garden Street, Cambridge, MA 02138, USA \label{harvard}
\and Fundaci{\'o}n Galileo Galilei - INAF, Rambla Jos{\'e} Ana Fernandez P{\'e}rez 7, E-38712 Bre$\tilde{\rm n}$a Baja (TF), Spain \label{tng}
\and INAF - Osservatorio Astronomico di Padova, Vicolo dell'Osservatorio 5, I-35122 Padova, Italy \label{oapadova}
\and INAF - Osservatorio Astrofisico di Catania, Via S. Sofia 78, I-95123 Catania, Italy \label{oacatania}
\and Dipartimento di Fisica e Astronomia, Università degli Studi di Padova, Vicolo dell’Osservatorio 3, I-35122 Padova, Italy \label{unipadova}
\and Department of Physics and Astronomy, Vanderbilt University, Nashville, TN 37235, USA \label{nashville_aa}
\and DISAT, Università degli Studi dell’Insubria, via Valleggio 11, 22100 Como, Italy \label{uniinsubria}
\and INAF – Osservatorio Astronomico di Brera, Via E. Bianchi 46, 23807 Merate (LC), Italy \label{oabrera}
\and NASA Exoplanet Science Institute-Caltech/IPAC, Pasadena, CA 91125, USA \label{nasa-ipac}
\and Department of Physics and Kavli Institute for Astrophysics and Space Research, Massachusetts Institute of Technology, Cambridge, MA 02139, USA \label{mit}
\and National Optical-Infrared Astronomy Research Laboratory, 950 N. Cherry Avenue, Tucson, AZ 85719, USA \label{noirlab}
\and Center for Space and Habitability, University of Bern, Bern, Switzerland \label{csh_unibe}
\and Sternberg Astronomical Institute, Lomonosov Moscow State University, 119992 Universitetskii prospekt 13, Moscow, Russia \label{sai_moscow}
\and Planetary Discoveries in Fredericksburg, VA 22405, USA \label{Fredericksburg}
\and NASA Goddard Space Flight Centre, 8800 Greenbelt Rd, Greenbelt, MD 20771, USA \label{nasa_goddard}
\and Società Astronomica Lunae, Castelnuovo Magra, Italy \label{sal_ita}
}

\date{Received .../ accepted ...}

\abstract
{Increasing interest in young exoplanets is leading to a growing effort to understand the formation and evolutionary processes responsible for their different architectures. One interesting target is TOI-5734, a relatively young K3--K4 dwarf star ($500_{-150}^{+300}$\,Myr) showing a transiting candidate in photometric observations followed up with high-resolution spectroscopic data.
}
{Using both Transiting Exoplanet Survey Satellite (TESS) photometry and High Accuracy Radial velocity Planet Searcher for the Northern hemisphere (HARPS-N) radial-velocity (RV) data, we aim to validate the presence of the companion TOI-5734b, measure its planetary mass, size, and its orbital parameters after having characterised its host star. We then aim to study its possible planetary composition and atmospheric evolution.
}
{
By simultaneously modelling photometry and high-cadence RVs, we measured the radius, mass, and density of TOI-5734b precisely, which were needed in order to reconstruct the evolution of its atmosphere under the high-energy irradiation of its young host. In particular, we employed Gaussian processes (GPs) with a flexible kernel to discriminate between the stellar activity of the young host and planetary signals.
}
{We confirmed the planetary nature of TOI-5734b and measured its orbital period ($P_{\rm b}\sim6.18$\,d), radius ($R_{\rm b} = 2.10^{+0.12}_{-0.12}$\,$R_\oplus$), and mass ($M_{\rm b}=9.1^{+2.6}_{-2.6}$\,$M_\oplus$). By measuring its density ($\rho_{\rm b}=0.98_{-0.30}^{+0.36}$\,$\rho_\oplus$) and running atmospheric evolution modelling, we infer that TOI-5734b is close to having a rocky composition and an almost completely depleted primary envelope. We also modelled stellar activity, which points to a rotational period of the star of $P_{\rm rot} = 11.09_{-0.08}^{+0.07}$\,d; this is compatible with its young age. Our results point toward the possibility of considering the target for atmospheric studies with present and future ground- and space-based facilities.
}
{}

           
\titlerunning{The GAPS Programme at the TNG}
\authorrunning{S. Filomeno et al.}
\maketitle
\section{Introduction}
\label{sec:intro}

Thanks to the large number of exoplanets detected so far, we have entered the era of exoplanet demographics and started a systematic investigation of the large variety of physical properties and orbital architectures exhibited by extrasolar planets.
In particular, deep interest in their formation processes is rising; this requires solid observational verifications to confirm theoretical models. For this the description of the initial planetesimal and pebble accretion mechanisms is necessary (e.g. \citealt{tychoniec_2020_pebble_accr}), contributing to the formation of terrestrial and giant planets cores and the improvement of core-accretion (\citealt{ida_lin_2004_core_accretion_model, matsuo_2007_coreaccret_vs_diskinst}) and disk-instability (\citealt{boss_1997_grav_inst, currie_2022_disk_instab}) scenarios, which should be adopted to model the ensuing evolution and accretion of a primary atmosphere.
To unravel the mechanisms of planetary formation and evolution and their timescales, the study of infant planets, likely still at their original formation sites, is crucial. This is achieved by analysing and characterising planets in young systems (e.g. \citealt{carleo2020}). 

Since the Transiting Exoplanet Survey Satellite (TESS; \citealt{ricker_2014}) began scanning the solar neighbourhood (\citealt{benatti_2019_intro, newton_2019_intro}), the number of young planets discovered has increased. This led to the observational focus on high-resolution spectroscopic follow-ups of transiting systems detected by TESS, allowing for the constraint of planetary masses. However, it is still difficult to extract reliable masses for particularly active young stars with strong activity signals in their radial-velocity (RV) time series.
Spots, faculae, and other flux variations due to magnetic activity on the stellar surface can completely hide small planetary signals (e.g. \citealt{benatti_2021_intro, carleo2021, Damasso_2023_yo40}). Consequently, analytical techniques such as Gaussian processes (GPs; \citealt{Rasmussen2006GP}) must be used to model and remove the stellar influence so that an increasing number of young exoplanets with well-constrained radii and masses can be revealed in combination with a tailored observing strategy (e.g. \citealt{nardiello2022, barragan2022_intro, Mallorquin2023_toi1726}). Furthermore, the choice of the GP kernel is fundamental to face the system-specific data reduction. The most common kernels are the quasi-periodic and double stochastically driven, damped simple harmonic oscillator (dSHO) kernel (\citealt{GP_foreman-mackey_2017_celerite1}; \citealt{celerite2_GP}); however, in some cases it could be useful to apply a kernel taking into account the expected wavelength dependence of stellar activity to estimate RV signals below the activity RV noise level (see, e.g. \citealt{cale_2021_wav_rv_analysis}).

The study of small and close-in planets around solar-type stars has led to one of the most relevant observational results: the discovery of a bimodal distribution of planetary radii with a separation region, the `radius valley'. It is placed at $\sim$1.7--2.0\,$M_{\rm \oplus}$ (e.g. \citealt{Fulton17, vaneylen_2018_radiusvalley}) dividing two populations of planets: super-Earths ($\ltsim 1.8\,M_{\rm \oplus}$) and sub-Neptunes ($\gtsim 1.8\,M_{\rm \oplus}$). The exact location of the valley may vary depending on the relation of planetary radius with planetary orbital period and stellar mass (\citealt{vaneylen_2018_radiusvalley, fulton_petigura_2018, petigura_2022}).
Super-Earths are high-density planets with an expected rocky composition and a null or scarce atmospheric envelope. At the same time, sub-Neptunes are intermediate- and low-density planets whose composition is mostly explained as a core made of icy and rocky material with a shallow H-He atmospheric envelope (`gas dwarfs'). However, recent findings report greater diversity in the composition of sub-Neptunes, which can be modelled as `water worlds' or `steam worlds', which have a rocky core surrounded by liquid or icy water shells and an envelope of water vapour mixed with $H_{\rm 2}$ and high mean molecular weight volatiles (\citealt{luque_palle_2022, Piaulet_Ghorayeb_2024, burn_waterworlds_2024, bonomo2025}).

The Global Architecture of Planetary System (GAPS) programme at the Telescopio Nazionale \textit{Galileo} (TNG) was created more than ten years ago (\citealt{covino_2013_gapsI}) to provide a comprehensive characterisation of the architectural properties of planetary systems by exploiting RV follow-up observations and a precise and a homogeneous analysis of the stellar atmospheres (e.g. \citealt{baratellaetal2020a, Biazzoetal2022, filomenoetal2024}) and deriving planetary orbital and physical parameters (e.g. \citealt{nardiello2022, Naponiello2024, nardiello2025_toi1430}).
In the framework of GAPS, two observing programmes focused on planets around young stars, relying on high-cadence RV monitoring. The young objects (YOs) sub-programme (\citealt{carleo2020}) was aimed at observing young and intermediate-aged stars (from a few megayears to a few gigayears) to detect young exoplanets down to super-Earth masses, including validation and mass measurement of transiting candidates (see, e.g. \citealt{nardiello2025_toi1430, Mantovan+2024}). After the conclusion of the GAPS-YO sub-programme, the \textit{Ariel} Masses Survey (ArMS) large programme (PI: Benatti) started its operations, with the aim of measuring the masses of planets with radii spanning the `radius valley' suitable for atmospheric characterisation with the Atmospheric Remote-Sensing Infrared Exoplanet Large-survey (\textit{Ariel}; \citealt{Tinettietal2018_ariel}) Space Mission. In the ArMS sample, several young transiting planets were also included to predict their atmospheric evolution.

In this paper, we present the detection and characterisation of the transiting young planetary system TOI-5734 (TIC 9989136), selected both for the GAPS-YO and the ArMS programmes. The study was performed using a combination of three \textit{TESS} sectors and spectroscopic data from the High Accuracy Radial velocity Planet Searcher for the Northern hemisphere (HARPS-N, \citealt{cosentino2014harps}) spectrograph installed at the TNG. 

The paper is organised as follows: Sect.\,\ref{sec:obs} presents the photometric and
spectroscopic observations.
Sect.\,\ref{sec:stellar_analysis} reports the determination of the stellar fundamental parameters. Sect.\,\ref{sec:transit_validation} presents the planet photometric validation. In Sect.\,\ref{sec:planet_syst_analy}, we describe the methods for the planetary system analysis and the joint fit results with the retrieved planetary system parameters. We present our discussion in Sect.\,\ref{sec:discussion}, and in Sect.\,\ref{sec:conclusions} we summarise our results.

\section{Observations and data reduction}
\label{sec:obs}

\subsection{TESS data}\label{sec:TESS_data}
 
The \textit{TESS} mission observed TOI-5734 (\textit{2MASS} J07250142+3713516) with a cadence of 1800\,s, 600\,s, and 200\,s, respectively, for Sectors 20, 47 (starting on February 14, 2020 and February 11, 2022), and 60 (starting on February 7, 2023; programme ID: G05121).
The light curves were downloaded from the Mikulski Archive for Space Telescopes\footnote{\url{https://exo.mast.stsci.edu}} (MAST), extracted and corrected adopting the {\sc PATHOS} pipeline (see \citealt{nardiello+_2019_pathos, Nardiello+_2020_pathos, nardiello+_2021_pathos, nardiello_2020_pathos}). We analysed the light curve extracted with a three-pixel aperture photometry, which is optimal for our target with a \textit{TESS} magnitude of $T \sim 8.5$\,mag (see Table\,\ref{tab:stellar_prop}). Moreover, we excluded all the data points in the light curves with the data quality parameter {\tt DQUALITY>0} (see the \textit{TESS} Science Data Products Description Document for details). Further, we cleaned the light curves, removing the remaining outliers distant by more than 4$\sigma$ from the mean flux and the data points showing a bimodal trend at the beginning of the light curves.

\begin{figure}
    \centering
    \includegraphics[width=0.8\linewidth]{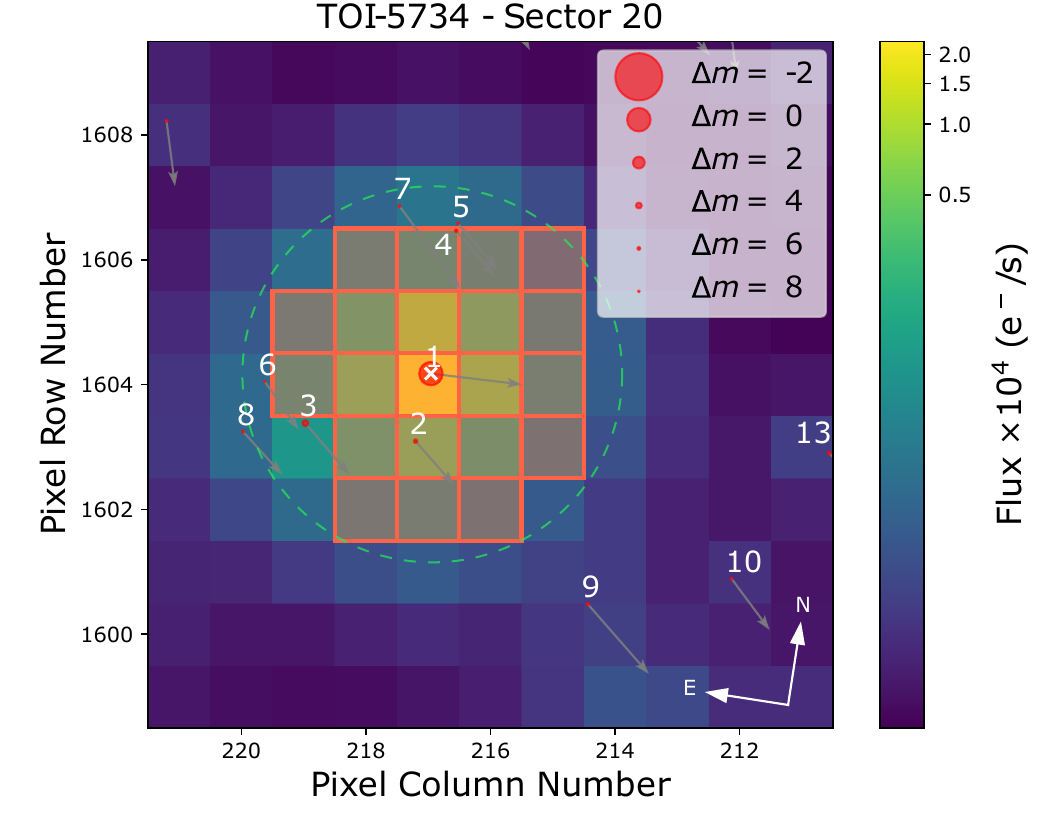}
    \caption{\textit{TESS} target pixel file of Sector 20 of TOI-5734, which is marked with `1'. The other sources, extracted from the \textit{Gaia}\,DR3 catalogue, have numbered circles of sizes proportional to the $G$-mag difference with our target. The colour bar shows the electron counts for each pixel. The orange squares represent the pixels used to construct the aperture photometry by the \textit{TESS} pipeline. Grey arrows indicate the direction of the proper motions for all the plotted sources. The dashed green circle represents the reference three-pixel aperture to select the targets relevant for the dilution factor estimate.}
    \label{fig:tpfplots}
\end{figure}

\begin{figure}
    \centering
    \includegraphics[width=0.9\linewidth]{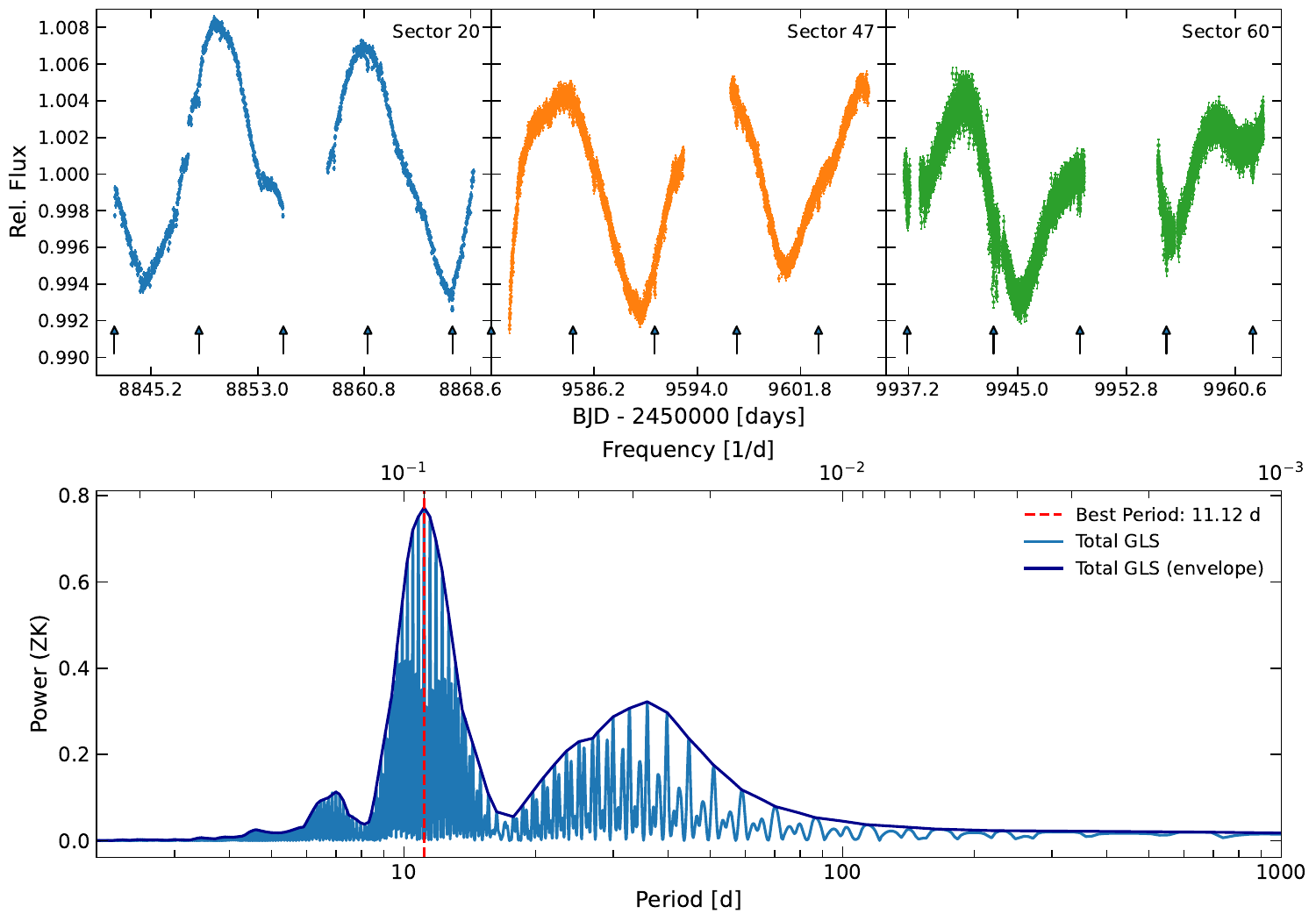}
        \caption{PATHOS light curves of TOI-5734 obtained using \textit{TESS} data in Sectors 20, 40, and 60, with arrows pointing to the mid-transit times of the planet. The plot below shows the overall GLS and its envelope. The dashed red line represents the signal due to the stellar rotation period peaking at 11.12\,d.}
    \label{fig:light_curves_with_gls}
\end{figure}

We verified the absence of nearby sources from Data Release 3 (DR3) of the \textit{Gaia} catalogue (\citealt{gaiacollaborationetal2016, Gaia_dr3_2023_j}), possibly contaminating the flux from TOI-5734 in the target pixel files using the code {\tt tpfplotter} (\citealt{tpfplotter_aller_2020}). We set a specific maximum magnitude contrast of $\Delta m$ = 8 for our targets and plotted the proper motion directions of all targets. As shown in Fig.\,\ref{fig:tpfplots} for Sector 20, the code highlights the aperture mask that the Science Processing Operations Centre (SPOC) pipeline (\citealt{jenkins_2016_spoc}) employs to measure the simple aperture photometry (SAP) flux. For TOI-5734, many targets are shown in the $10 \times 10$\,px field. The stars with the smallest contrast with TOI-5734 are identified as \textit{Gaia} DR3 898844472171624576 ($G= 14.6$\,mag) and \textit{Gaia} DR3 898844575250838400 ($G=13.2$\,mag), respectively numbered 2 and 3 in Fig.\,\ref{fig:tpfplots}.
We investigated the possibility that some targets could influence the signal from TOI-5734 by estimating the dilution factor. Since we used a three-pixel photometric aperture for our study, we chose to consider all the targets (7) within three pixels of the target star (i.e. inside the dashed green circle in Fig.\,\ref{fig:tpfplots}) for each sector, assuming that the faintest targets do not adequately influence the dilution. As reported in Eq. 6 of \cite{espinoza_2019}, the dilution factor is $D = 1/(1+\sum_n F_n/F_T)$,
where $n$ is the number of contaminating sources, $\sum_n F_n$ represents the flux of the contaminating sources in the photometric aperture, and $F_T$ is the out-of-transit flux of the target star in a given passband. As flux values, we used the \textit{Gaia} DR3 integrated RP mean flux in $e^-/s$. For each sector, the dilution factor resulted in $D$ = 0.969, and this value was applied to all light curves. The final light curves used in our analysis are shown in Fig.\,\ref{fig:light_curves_with_gls}.


\subsection{HARPS-N data}\label{sec:harpsn_data}

The target was observed with HARPS-N at TNG 97 times between 7 October, 2022 and 17 May, 2024, with a median exposure time ($t_{\rm exp}$) of 900\,s; so, an average [min, max] signal-to-noise ratio ($S/N$) at 550\,nm of 68 [21, 103] was achieved. We reduced the data with the new offline version of HARPS-N data-reduction software (DRS v3.2.0; Cosentino, priv. comm.).
The spectroscopic data of TOI-5734 were extracted using a K2 mask template and a cross-correlation-function (CCF) width of 30\,km/s. We obtained RV data with an average uncertainty of 1.6\,m/s and a dispersion of around 13.3\,m/s. To improve the quality of the analysis, we preferred not to consider the data points with an $S/N$ lower than one-third the maximum $S/N$ in the sample. Furthermore, the DRS provided the values of the bisector inverse slope (BIS) velocity span, line contrast, and line full width at half maximum (FWHM), which we used as indicators of stellar activity. The chromospheric activity index, $\log R^\prime_{\rm HK}$, was derived after extracting the S-index\footnote{The S-index was measured from the \ion{Ca}{II}\,H\&K emission using the code from Alejandro Suarez Mascareno and Christophe Lovis.}, which was calibrated on the Mount Wilson scale (\citealt{baliunas_1995_mtwilson}) with the use of the procedure described in \cite{lovis_2011_sindex_to_rhk}.

In addition, we exploited the software SpEctrum Radial Velocity AnaLyser ({\sc SERVAL}; \citealt{seval_rvcalc_zechmeister_2018}) to compare the extracted RVs and errors with the DRS ones. The SERVAL RVs have an average uncertainty of 1.3\,m/s and a dispersion of around 14.5\,m/s, which are compatible within 1\,$\sigma$ with the DRS data; therefore, we decided to use the latter values. The RVs are shown in Fig.\,\ref{fig:rv_rhk_bis_timmeseries} and presented in Table\,\ref{tab:toi_5734_rv_act_time_series} together with the BIS span and the $\log R^\prime_{\rm HK}$. Since the target was observed twice for some nights, we adopted the RV data with a 0.5-day binning width in the analysis, which is a value sufficiently smaller than the orbital period of the planet (see Sect.\,\ref{sec:transit_validation}).

\section{Stellar fundamental parameters}
\label{sec:stellar_analysis}

\subsection{Atmospheric parameters and iron abundance}\label{subsec:atm_parameters}

For the determination of the stellar atmospheric parameters (effective temperature \teff, surface gravity \logg, microturbulence velocity \csi) and iron abundance \feh\,of TOI-5734, we considered the co-added spectrum obtained through a combination of all spectra collected for the target. The final $S/N$ of the co-added spectrum was $\sim$390 at $\sim$6000\,\AA.

Given the relatively high level of chromospheric activity of our target, following the same procedure as in \cite{baratellaetal2020b} and \cite{filomenoetal2024} we used a combination of iron (Fe) and titanium (Ti) lines to derive the stellar parameters with the equivalent-width ($EW$) method.
We chose to use Ti lines because of the influence stellar activity might have on Fe lines; i.e. increasing their $EW$s if they form in the upper layers of the photosphere. This could cause an overestimation of the microturbulence velocity (\csi) and, consequently, an underestimation of the elemental abundances. Instead, Ti lines, which on average form deeper in the photosphere, led to more precise values of \csi (see also \citealt{baratellaetal2020a}).

We measured the $EW$s of both Fe and Ti lines in the neutral and ionised stages using the {\sc ares} code \citep{sousa_2015_ares_v2} and discarded those lines with uncertainties larger than 10$\%$. For lines with $EW$ > 120 m\AA, we double checked the measurements with IRAF\footnote{NOIRLab IRAF is distributed by the Community Science and Data Center at NSF NOIRLab, which is managed by the Association of Universities for Research in Astronomy (AURA) under a cooperative agreement with the U.S. National Science Foundation.} by fitting Voigt profiles, which better reproduce the line profiles. The line list comes from \cite{baratellaetal2020b}. We adopted the 1D LTE Kurucz model atmospheres linearly interpolated from the \cite{castelli_kurukz_2003} grid with solar-scaled chemical composition and new opacities (ODFNEW). To derive final stellar parameters and iron abundances, we used the pyMOOGi code by \cite{adamow_2017_pymoogi}, which is a Python wrapper of the {\sc MOOG} code (\citealt{sneden1973}, version 2019). The final results of this spectroscopic analysis are reported in Table\,\ref{tab:stellar_prop}.

Our final spectroscopic temperature agrees, within the uncertainties, with that obtained from photometry, with \textit{2MASS} (\citealt{2mass}) and \textit{Gaia} DR3 \citep{Gaia_dr3_2023_j} magnitudes, and considering the reddening derived through the extinction maps by \cite{lallement_2018_e_bv} is negligible ($E(B-V)$=$0.000^{+0.014}_{-0.000}$). Indeed, by using the {\sc colte} program developed by \cite{2021casagrande}, the photometric temperature ranges between $4604\pm80$\,K in $(G_{\rm RP}-J)$ and $4706\pm44$\,K in $(G_{\rm BP}-H)$ and has a weighted mean value of \teff\,(phot)\,$=4670\pm43$\,K, as also reported in Table\,\ref{tab:stellar_prop}. From the mass value reported in the TICv8.2 Catalogue \citep{TICv8.2_paegert_2022} and the photometric \teff, we obtained a \logg\,of $4.64\pm0.10$\,dex using the classical equation based on the luminosity of the star, in agreement with our spectroscopic value. With these values and from the relation published by \cite{dutra_ferreira_2016_microturb_rel}, the expected \csi\,is $0.66\pm0.10$\,km/s, again in agreement with the spectroscopic estimation (see Table\,\ref{tab:stellar_prop}). The derived stellar \teff\,is compatible with a K3--K4 dwarf star (see \citealt[v.2022]{pecaut_mamaj_2013_st_type_table}).

\subsection{Lithium abundance}\label{subsec:lithium}

We measured the $EW$ of the lithium line at 6707.8\,\AA\,with IRAF from the co-added spectrum, which resulted in $EW_{\rm Li}$ = $3.3 \pm 0.3$ m\AA. From this $EW$ and the adopted atmospheric parameters, we estimated the lithium abundance by applying the corrections for non-local thermodynamic equilibrium (NLTE) effects by following the prescriptions of \cite{Lind_2009_LiI_nlte}. We obtained an upper limit of 0.04 dex. Despite the almost depleted lithium value, this upper limit seems to support an age following the distribution of young clusters such as Hyades ($\sim$\,600\,Myr). Our age estimate is discussed in Sect.\,\ref{subsec:age_star}.

\subsection{Rotation period}
\label{subsec:stellar_rot_period}
 
We estimated the stellar rotation period, $P_{\rm rot}$, both from photometric and spectroscopic data, making use of the generalised Lomb--Scargle (GLS; \citealt{zechmeister_2009_gls}) periodogram\footnote{\url{https://github.com/mzechmeister/GLS}}. The results are shown in the bottom panel of Fig.\,\ref{fig:light_curves_with_gls} and in Fig.\,\ref{fig:gls_rv_activity} for the joint \textit{TESS} light curve and spectroscopic data, respectively. The former shows a peak at around 11.12\,d, while the latter shows the RV and BIS span periodograms both having a mean peak at $P_{\rm rot} = 11.09\pm0.02$\,d (respectively, FAP $\sim 10^{-10}$ and FAP $\sim 10^{-8}$), which is close to the values extracted from the photometric data. This result is consistent with the significant correlation measured between the RVs and BIS span (Spearman correlation coefficient $\rho_{\rm spear}= -0.69$), suggesting the influence of the stellar activity on the RVs (see also Sect.\,\ref{sec:planet_syst_analy}).

\subsection{Chromospheric and coronal activity}\label{subsec:chrom_act}

\begin{figure}[h]
    \centering
    \includegraphics[width=0.9\linewidth]{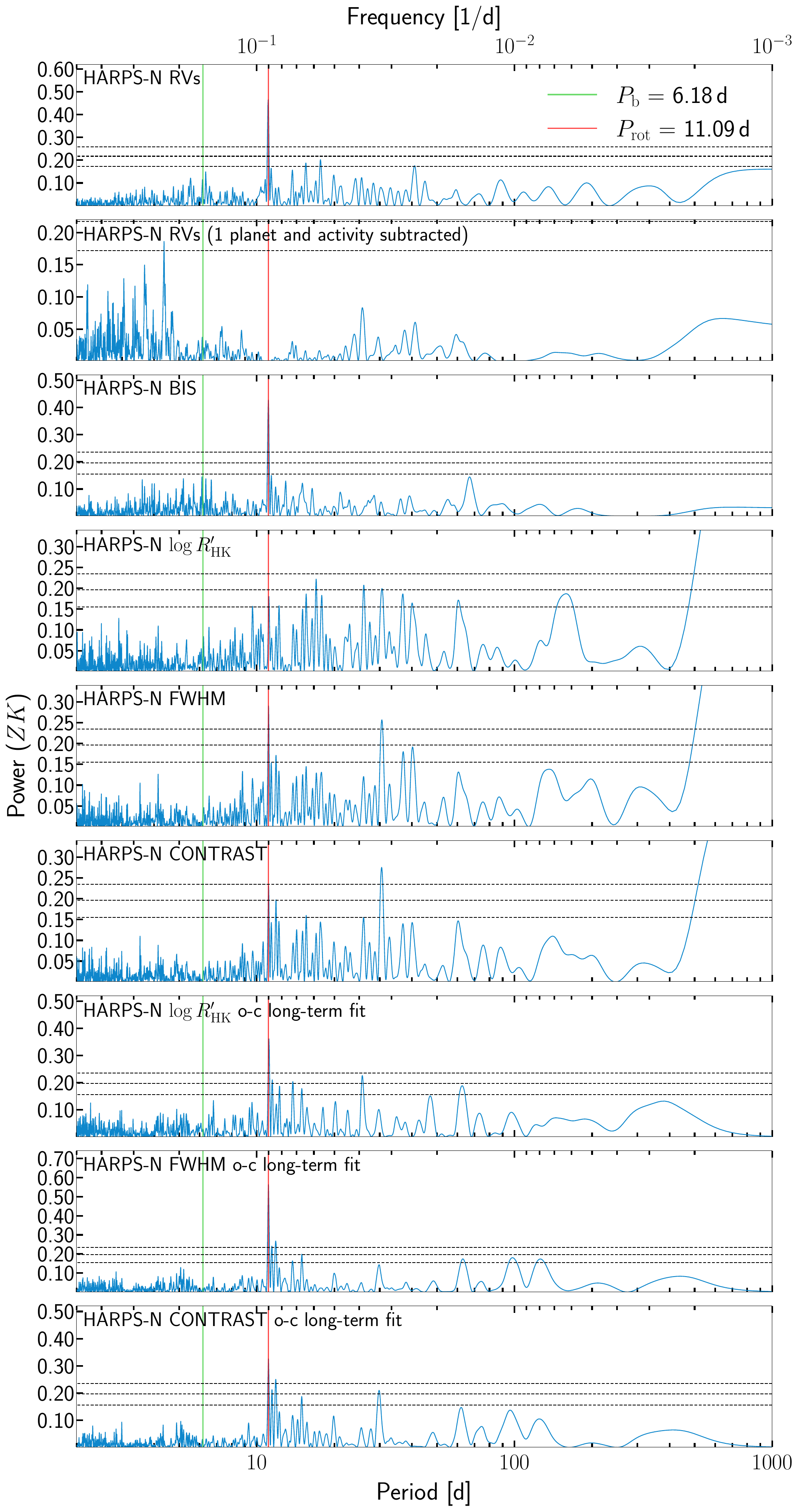}
    \caption{GLS power spectrum for TOI-5734 based on HARPS-N RVs, stellar activity indicators (BIS span, $\log R^\prime_{\rm HK}$, FWHM, contrast), and their residuals after removing the long-term trend. The horizontal lines in the GLS periodograms indicate the FAP levels at 10$\%$, 1$\%$, and 0.1$\%$ power. The vertical green and red lines, respectively, highlight the positions of the planetary orbital period and stellar rotation period.}
    \label{fig:gls_rv_activity}
\end{figure}

We measured the activity indices linked to \ion{Ca}{II}\,H\&K emission lines, which are the S-index and the $\log R^\prime_{\rm HK}$, and they show mean values of 0.72$\pm$0.12\,dex and $-$4.47$\pm$0.09\,dex, respectively. Despite the weak correlation between RVs and the $\log R^\prime_{\rm HK}$ (Spearman correlation coefficient $\rho_{\rm spear} = 0.27$), the GLS of $\log R^\prime_{\rm HK}$ in Fig.\,\ref{fig:gls_rv_activity} shows strong power at long periods, and an increasing trend in the $\log R^\prime_{\rm HK}$ is evident from its time series reported in the bottom panel of Fig.\,\ref{fig:rv_rhk_bis_timmeseries}. Here, a possible turnover is observed, and the minimum and maximum values of the time series are $\log R^\prime_{\rm HK}=-4.52$ and $-4.43$, respectively. These considerations were taken into account when calculating the uncertainties of these \ion{Ca}{II} activity indices. The star could have a long-term activity cycle, whose periodicity cannot be well defined because of the short time coverage of our dataset. After removing the long-term trend, a significant periodicity at 11.17 d, ascribable to the rotation period of the star, appears on the residuals of the $\log R^\prime_{\rm HK}$ time series.

The Chandra source catalogue v2.0 (source ID 2CXO J072501.4+371352; \citealt{Evans+2010}) includes TOI-5734, which is revealed at 5.4\,$\sigma$ in a shallow observation. 
More recently, it was also detected in the eROSITA All-Sky Survey (source ID 1eRASS J072501.0+371400; \citealt{eRosita2024}), with a count rate of 0.143 $\pm$ 0.045\,cts/s.
Adopting an isothermal model for an optically thin plasma, the flux in the standard ROSAT band ([0.1,\,2.4]\,keV) can be estimated in the [0.9,\,2.5]$\times 10^{-13}$\,erg/s/cm$^2$ range, depending on the assumed plasma temperature ([2,\,10]\,MK) and metallicity ([0.2,\,1] in solar units).
These values correspond to an X-ray luminosity of $L_{\rm X} = [2,\,3] \times 10^{28}$\,erg/s at the distance of the star, and to $\log L_{\rm X}/L_{\rm bol}$ in the [$-4.53$, $-4.36$] range from the stellar luminosity derived in Sect.\,\ref{subsec:star_mass_radius}.
In the following, we assume $L_{\rm X} = (2.0 \pm 0.5) \times 10^{28}$\,erg/s.
This value provides the best agreement with the independent predictions based on the measured rotational period \citep{Pizzolato+2003} and on the \ion{Ca}{II} $R'_{\rm HK}$ \citep{mamajek_age_est_2008} within a factor of 1.4. 

\subsection{Kinematic and multiplicity}\label{subsec:kinematics}

From the results obtained in the previous sections, TOI-5734 emerges as a relatively young star. We checked membership to nearby young moving groups exploiting the online tool BANYAN $\Sigma$ \footnote{\url{https://www.exoplanetes.umontreal.ca/banyan/banyansigma.php}} \citep{gagneetal2018}. The star results in a field object with a null membership probability to known groups.
The kinematic parameters $U$, $V$, and $W$ were derived following the prescriptions of \citet{johnson1987} and are listed in Table\,\ref{tab:stellar_prop}. TOI-5734 is close to the edge of the kinematic space typical of young stars \citep{montes2001}, with properties similar to the Ursa Major moving group. This is fully consistent with the age determinations of other methods. 

We also searched for co-moving objects in \textit{Gaia} DR3. None were found within 600\,arcsec (about 20000\,au), which is a plausible maximum limit for bound companions. 
Also, the standard diagnostic used to signify departure from a good single-star model in the \textit{Gaia} DR3 astrometric solution, the renormalised unit weight error (RUWE), is reported to be 0.930 for TOI-5734 (RUWE > 1.4 being the threshold for likely presence of orbital motion in the data), and the RV difference between \textit{Gaia} DR3 and HARPS-N is not significant. Hence, TOI-5734 appears to be a single star.

\subsection{Age}\label{subsec:age_star}

As expected for an unevolved K dwarf, isochrone fitting provides little constraint on stellar age (3.7$\pm$3.5\,Gyr) using the PARAM \footnote{\url{http://stev.oapd.inaf.it/cgi-bin/param_1.3}} \citep{dasilva_2006} web interface with the parameters described in Sect.\,\ref{subsec:star_mass_radius} and without priors on stellar age.
However, as mentioned in previous sections, all indirect age indicators agree quite well and support an age close to and likely slightly younger than that of the Hyades.

The gyrochronology calibration from \cite{mamajek_age_est_2008} yields an age of 524$\pm$25\,Myr. 
Using the \cite{mamajek_age_est_2008} activity--age relation and the activity values measured in Sect.\,\ref{subsec:chrom_act}, we derived a chromospheric age of 493$\pm$300\,Myr.
The X-ray emission of the star is slightly above the median value of Hyades members of similar colour but within the distribution of available measurements for cluster members, with the nominal age from \cite{mamajek_age_est_2008} of $360_{-70}^{+110}$\,Myr. 

Another age estimate was performed following the methodology described in \cite{jeffries_2023_eagles} and using the Empirical AGes from lithium Equivalent widthS ({\sc eagles}) code. In particular, the latter uses the lithium $EW$ at $\lambda$6707\,\AA\,and the stellar effective temperature. For TOI-5734, we measured the lithium $EW$ of 3.3$\pm$0.3 m\AA. Considering that the {\sc eagles} code is likely to derive lower limits in ages for $EW_{\rm Li}$ lower than 50\,m\AA, we estimated a lower age limit of $\sim$310\,Myr, which is in agreement with our other estimates.
Finally, while inconclusive in providing a well-defined age value, the kinematic parameters are in agreement with the age estimates above.
Considering the possible long-term variability of the chromospheric and coronal activity, and systematic uncertainties in the rotational evolution \citep[e.g.][]{curtis2020}, we adopted $500_{-150}^{+300}$\,Myr as a conservative estimate of the stellar age.

\subsection{Mass and radius}\label{subsec:star_mass_radius}

The stellar mass and radius were determined as in previous papers of the GAPS-YO series \citep{carleo2021,nardiello2025_toi1430}.
For the stellar mass, we exploited the PARAM web interface, adopting the mean between spectroscopic and photometric effective temperature, the [Fe/H] from our spectroscopic analysis, and the $V$ magnitude and parallax $\pi$ from Table\,\ref{tab:stellar_prop}, and constraining the age to the range allowed by the indirect methods \citep[300-800\,Myr, see][]{desidera2015}.
In this way, we found $M_{\rm \star} = 0.724 \pm 0.009 ~M_{\rm \odot}$, where the quoted error is the one obtained by the PARAM procedure, and it does not include possible systematics uncertainties in the input stellar models. 

The stellar radius was derived from the Stefan-Boltzmann law using the above parameters and bolometric corrections by \cite{pecaut_mamaj_2013_st_type_table}.
It is $R_{\rm \star} = 0.639 \pm 0.034 ~R_{\rm \odot}$. This determination is basically identical to the value obtained through PARAM ($R_{\rm \star} = 0.637 \pm 0.012 ~R_{\rm \odot}$). Both stellar mass and radius are compatible with an independent determination performed using the broadband spectral energy distribution (SED) of the star and \textit{Gaia} DR3 parallax (see App.\,\ref{sed_analysis}). The corresponding stellar luminosity is $L_{\rm \star} = 0.181 \pm 0.009 ~L_{\rm \odot}$.

\subsection{Projected rotational velocity and stellar inclination}\label{subsec:vsini_incl}

We used the spectral-synthesis method to estimate the stellar projected rotational velocity (see \citealt{Biazzoetal2022} for details). With the pyMOOGi code by \cite{adamow_2017_pymoogi} and the stellar parameters \teff, \logg, \csi, and [Fe/H] derived in Sect.\,\ref{subsec:atm_parameters}, we measured \vsini$_{\star}=2.9\pm0.7$\,km/s, taking into account the spectral regions around 5400, 6200, and 6700\,\AA. 
We also computed the projected rotational velocity by applying the empirical relations found in \cite{Rainer_2023} using the FWHM of the CCFs of the analysed spectra as input. Considering Equation 7 for the K5 mask, we obtained \vsini$_{\star}=3.2\pm0.5$\,km/s, which is comparable with the value from the spectral synthesis.
Using the previous estimates of $P_{\rm rot}$ and $R_{\star}$, we determined the stellar equatorial velocity $v_{\rm eq}=2\pi R_\star / P_{\rm rot} = 2.9 \pm 0.2$\,km/s. This value is compatible with our measurements of \vsini$_{\star}$; thus, the stellar inclination is consistent with $\simeq$ 90\,deg.
\setlength{\tabcolsep}{5.5pt}
\begin{table}[h]
\tiny
\caption{Stellar properties of TOI-5734 (TIC 9989136).}
\label{tab:stellar_prop}
\begin{tabular}{lrr}
\hline
Parameter & TOI-5734 & Reference \\ \hline
$\alpha$ $(\mathrm{J} 2000$; deg) & 111.25581 & \textit{Gaia} DR3 \\
$\delta$ $(\mathrm{J} 2000$; deg) & 37.23112 & \textit{Gaia} DR3 \\
$\mu_\alpha$ $(\mathrm{mas} / \mathrm{yr})$ & $-$27.71$\pm$0.02 & \textit{Gaia} DR3 \\
$\mu_\delta$ $(\mathrm{mas} / \mathrm{yr})$ & 14.51$\pm$0.01 & \textit{Gaia} DR3 \\
RV (km/s) & $-$9.18$\pm$0.18 & \textit{Gaia} DR3 \\
$\pi$ (mas) & 30.70$\pm$0.02 & \textit{Gaia} DR3 \\
$U$ (km/s) & 7.0$\pm$0.2 & This paper (Sect.\,\ref{subsec:kinematics}) \\
$V$ (km/s) & 3.6$\pm$0.1 & This paper (Sect.\,\ref{subsec:kinematics}) \\
$W$ (km/s) & $-$6.7$\pm$0.1 & This paper (Sect.\,\ref{subsec:kinematics}) \\
$B-V$ (mag) & 0.995$\pm$0.041 & APASS DR9 \\
$G$\,(mag) & 9.1974$\pm$0.0028 & \textit{Gaia} DR3 \\
$G_{\mathrm{BP}}-G_{\mathrm{RP}}$\,(mag) & 1.2775$\pm$0.0021 & \textit{Gaia} DR3 \\
T (mag) & 8.545$\pm$0.006 & TIC v8.2 \\
$J$\,(mag) & 7.613$\pm$0.019 & \textit{2MASS} \\
$H$\,(mag) & 7.114$\pm$0.017 & \textit{2MASS} \\
$K$\,(mag) & 6.987$\pm$0.016 & \textit{2MASS} \\
Spectral Type & K3-K4\,V & This paper (Sect.\,\ref{subsec:star_mass_radius}) \\
$T_{\text {eff }}(\mathrm{K})$ & 4750$\pm$100 & This paper (spec., adopted; Sect.\,\ref{subsec:atm_parameters}) \\
$T_{\text {eff }}(\mathrm{K})$ & 4670$\pm$43 & This paper (phot., Sect.\,\ref{subsec:atm_parameters}) \\
$\log g$ (cm/s$^2$) & 4.66$\pm$0.06 & This paper (spec., adopted; Sect.\,\ref{subsec:atm_parameters}) \\
$\log g$ (cm/s$^2$)& 4.64$\pm$0.10 & This paper (phot., Sect.\,\ref{subsec:atm_parameters}) \\
\csi\,(km/s) & 0.6$\pm$0.2 & This paper (spec., adopted; Sect.\,\ref{subsec:atm_parameters}) \\
\csi\,(km/s) & 0.66$\pm$0.10 & This paper (phot., Sect.\,\ref{subsec:atm_parameters}) \\
$[\mathrm{Fe} / \mathrm{H}]$ (dex) & $-$0.13$\pm$0.03 & This paper (Sect.\,\ref{subsec:atm_parameters}) \\
$E(B-V)$ & $0.000^{+0.014}_{-0.000}$ & \cite{lallement_2018_e_bv} \\
$S_{\mathrm{MW}}$ & 0.72$\pm$0.12 & This paper (Sect.\,\ref{subsec:chrom_act}) \\
$\log R_{\mathrm{HK}}^{\prime}$ & $-$4.47$\pm$0.09 & This paper (Sect.\,\ref{subsec:chrom_act}) \\
$L_{\mathrm{X}}$ (erg/s)& 2.0$\pm$0.5 $\times 10^{28}$ & This paper (Sect.\,\ref{subsec:chrom_act}) \\
$\log L_{\mathrm{X}}/L_{\mathrm{bol}} $ & $-4.54^{+0.09}_{-0.12}$ & This paper (Sect.\,\ref{subsec:chrom_act}) \\
$v \sin i_{\star}$ (km/s) & 2.9$\pm$0.7 & This paper (synt, Sect.\,\ref{subsec:vsini_incl}) \\
$v \sin i_{\star}$ (km/s) & 3.24$\pm$0.46 & Eq.\,7 in \cite{Rainer_2023} \\
$EW_{\rm Li}$ (m\AA) & 3.3$\pm$0.3 & This paper (Sect.\,\ref{subsec:lithium}) \\
$A_{\rm Li}$ (dex) & <0.04 & This paper (Sect.\,\ref{subsec:lithium}) \\
$M_{\star}$ ($M_{\odot}$) & 0.724$\pm$0.009 & This paper (adopted; Sect.\,\ref{subsec:star_mass_radius}) \\
$M_{\star}$ ($M_{\odot}$) & 0.725$\pm$0.044 & This paper (Sect.\,\ref{sed_analysis}) \\
$R_{\star}$ ($R_{\odot}$) & 0.639$\pm$0.034 & This paper (adopted; Sect.\,\ref{subsec:star_mass_radius}) \\
$R_{\star}$ ($R_{\odot}$) & 0.678$\pm$0.029 & This paper (Sect.\,\ref{sed_analysis}) \\
$L_{\rm \star}$ ($L_{\odot}$) & 0.181$\pm$0.009 & This paper (Sect.\,\ref{subsec:star_mass_radius}) \\
Age (Myr) & $500_{-150}^{+300}$ & This paper (Sect.\,\ref{subsec:age_star}) \\
$i_{\rm \star}$ (deg) & $\sim$90 & This paper (Sect.\,\ref{subsec:vsini_incl}) \\ 
$P_{\rm rot}$ (d; GP$_{\rm dSHO}$) & $11.09^{+0.07}_{-0.07}$ & This paper (Sect.\,\ref{sec:planet_syst_analy}) \\ \hline
\end{tabular}
\end{table}

\section{Transit validation}
\label{sec:transit_validation}

\begin{figure}
    \centering
    \includegraphics[width=0.9\linewidth]{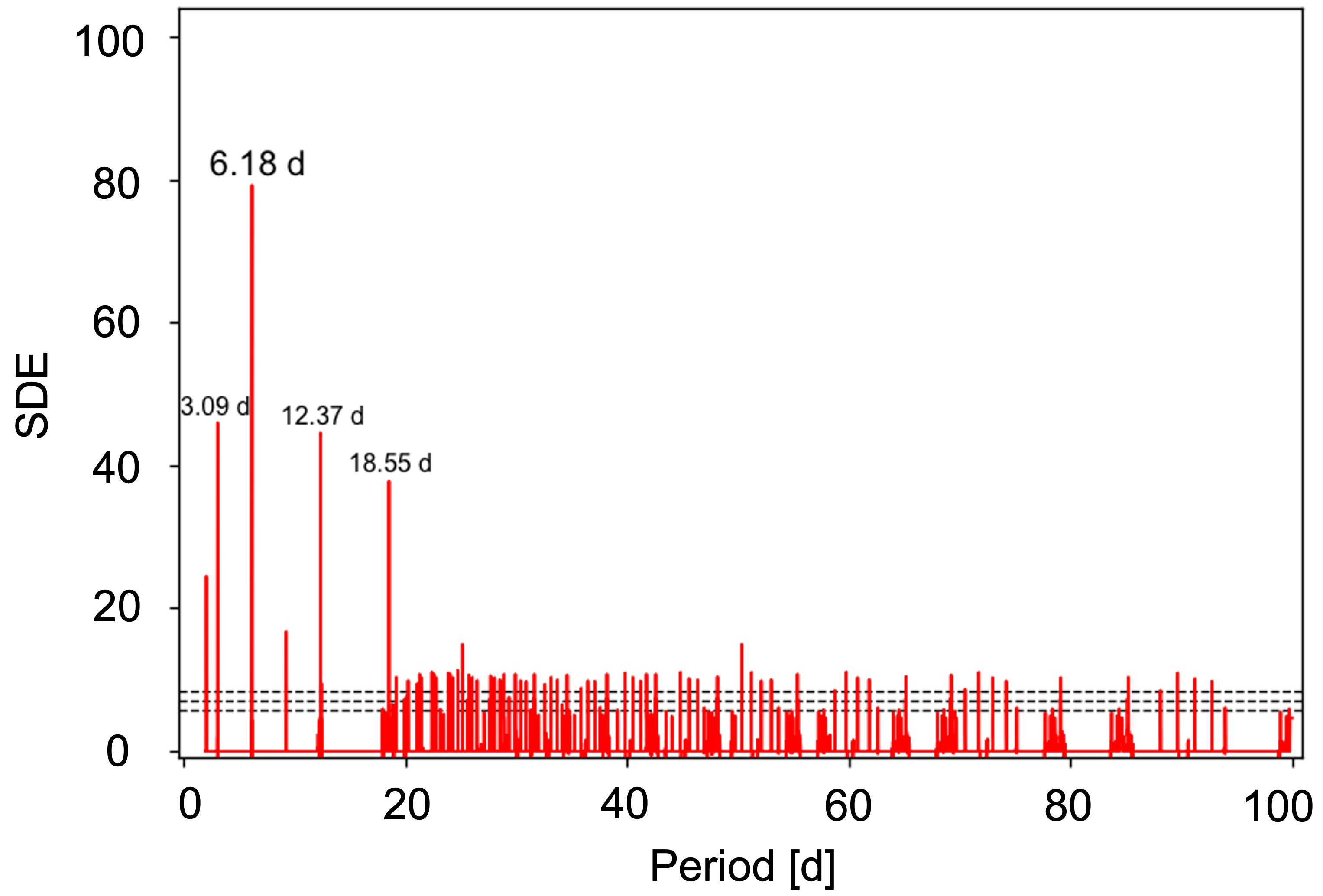}
    \caption{TLS power spectrum of light-curve data of TOI-5734. The transiting signal is confirmed by the peak at $P_{\rm b}$=6.18414\,d with an SDE value of 79, which is strongly above the thresholds (horizontal dashed lines) corresponding to the TLS false positive rates of 10$\%$, 1$\%$, and 0.1$\%$.}
    \label{fig:tls_transits}
\end{figure}

The \textit{TESS} mission released an alert about a planetary candidate around TOI-5734 with an orbital period around $P_{\rm rot} \sim 6.18$\,d measured by the SPOC pipeline \citep{jenkins_2016_spoc} and archived on the MAST on August 18, 2022.
We studied the transit event by first modelling and removing the stellar variability trend from the reduced light curves. We applied the {\sc wotan} detrending algorithm (\citealt{wotan_hippke_2019}) using, on each light curve, an \textit{rspline} interpolation with a window-length interval of 12 hours. We investigated the presence of transiting candidates using the {\sc Transit Least Squares} package (TLS; \citealt{tls_hippke_heller_2019}) to extract the periodogram of the joint detrended light curves, looking for transit periods from 1 to 100\,d. In Fig.\,\ref{fig:tls_transits} we show the TLS power spectrum with the main peak indicating an orbital period of $P_{\rm b} = 6.18414\pm0.00072$\,d with a signal-detection efficiency (SDE) of $\sim 79$, which strongly confirms the presence of a transiting candidate. The other relevant peaks represent only low-order harmonics of the main periodicity. Then, we followed the procedure described by \cite{Nardiello+_2020_pathos} for the in- and out-of-transit centroid analysis to check whether the transit is on our target or associated with a neighbour source. We calculated the differences between the mean centroids by averaging the in- and out-of-transit centroids of each sector. We found that the centroid and the target (at (0,0) coordinates) have compatible positions, within the errors (see Fig.\,\ref{fig:inout_centroid}), confirming that the transit signal originates from TOI-5734 itself.

To further validate the planet, we used \texttt{TRICERATOPS} (\citealt{Giacalone2021}), a Bayesian tool designed for \textit{TESS} candidates. 
It evaluates the probability of the nature of the transit signal being planetary by combining information on the transit signal, the target star, and nearby stars from \textit{Gaia} DR2, considering possible false-positive scenarios. The analysis also checks for instrumental artefacts and background contamination. \texttt{TRICERATOPS} can be applied to \textit{TESS} data alone or combined with high-resolution contrast imaging for improved constraints.
According to \cite{Giacalone2021}, a planet candidate is validated if it meets the false-positive probability (FPP) criterion of < 0.015 and the nearby false-positive probability (NFPP) of < $10
^{-3}$. The FPP is defined as 1$-$ ($P_{\rm TP}+P_{\rm PTP}+P_{\rm DTP}+P_{\rm oth}$), where $P_{\rm TP}$ stands for transiting planet probability, $P_{\rm PTP}$ for primary TP probability, and $P_{\rm DTP}$ for diluted TP probability, while NFPP = $P_{\rm NTP}+P_{\rm NEB}+P_{\rm NEBx2P}+P^\prime_{\rm oth}$ refers to a nearby star with probabilities associated with a nearby transiting planet (NTP), nearby eclipsing binary (NEB), and an eclipsing binary with double orbital period. Even though the $P_{\rm oth}$ and $P^\prime_{\rm oth}$ probabilities contribute to adding FPP and NFPP together to make one, they are sums of negligible terms, so they were not included in the following considerations. In our analysis of TOI-5734b, we used \textit{TESS} SPOC PDCSAP light curves from all available sectors, adopting extraction apertures from the light-curve headers. We ran \texttt{TRICERATOPS} 100 times, incorporating contrast curves from NESSI on the 3.5-m WIYN telescope (Kitt Peak National Observatory, USA) to refine our results.
Since FPP varies across runs, we computed its median and 84th percentile to confirm that the initial validation was not an outlier. The final results yielded FPP = $2 \times 10
^{-5}$ and NFPP $< 1 \times 10
^{-6}$, suggesting the planet could be validated. However, the uncertainties (0.7 and 0.01, respectively) warrant caution. 
The ambiguity in the results likely arises from the presence of a nearby star capable of reproducing the observed transit signal (see Fig.\,\ref{fig:tpfplots} the dilution factor estimated in Sect.\,\ref{sec:TESS_data}) and is further supported by the validation test probabilities: 52$\%$ for a transiting planet (TP), 16$\%$ for a primary TP, and 15$\%$ for a diluted TP, which accounts for the possibility of a bound companion or background stars. Moreover, untraceable stellar activity effects may impact the accuracy of the transit fit and, consequently, the final results.
However, we highlight that it is not possible that the transit signal is due to a bound companion, since this option was excluded from both the kinematic analysis of group membership and the \textit{Gaia} DR3 analysis, with the target being a field object with astrometric values consistent with a single star (see Sect.\,\ref{subsec:kinematics}). Also, we excluded the presence of any type of blended eclipsing binary from the analysis of ground-based multi-colour photometry from LCOGT (see Appendix\,\ref{subsec:ground}). In addition, we discarded the possibility for diluted TP, since the in- and out-of-transit analysis and LCOGT photometric observations confirm that the transit signal originates on-target and the fit made on the transit detected by LCOGT reveals a transit depth 1$\sigma$ consistent with that derived in the analysis described in Sect.\,\ref{sec:planet_syst_analy} (see App.\,\ref{subsec:ground} and Fig.\,\ref{TOI-5734_ground_lightcurve}). Finally, by analysing the high-resolution imaging results from the near-infrared \textit{Palomar}-PHARO, and the optical \textit{WIYN}-NESSI and \textit{SAI}-Speckle Polarimeter instruments, we excluded the presence of any contaminant, such as \textit{Gaia} unresolved companions, down to $\Delta mag \sim 5$ within 0.2 – 1.0\,arcsec (see App.\,\ref{app:hri_obs}).
Consequently, combining our validation analysis with the information from ground-based seeing-limited photometry, {\it Gaia} DR3 data and high-resolution imaging data, we have the evidence to consider the planet validated.

\section{Radial velocity and photometry joint analysis}
\label{sec:planet_syst_analy}

TOI-5734 is a young, strongly active star, so the RV measurements are likely affected, hiding planetary signals or injecting spurious ones. Hence, it is fundamental to remove the activity contribution from the RVs. To identify if and how stellar-activity signal influences our RV data, we used the BIS span and the $\log R^\prime_{\rm HK}$ as proxies of the stellar activity. We compared the GLS of the activity indices with that of our data, as shown in Fig.\,\ref{fig:gls_rv_activity}. The BIS span has a periodogram peaking at 11.09\,d suggesting that its signal is related to the stellar rotation period (see Sect.\,\ref{subsec:stellar_rot_period}) and implies the presence of spots on the stellar surface. Instead, the long-term signal introduced by the $\log R^\prime_{\rm HK}$ in the RV data (see Sect.\,\ref{subsec:chrom_act}) should not affect the result of our analysis, so we did not use it to model the RVs for the following analysis. As displayed in Fig.\,\ref{fig:gls_rv_activity} for the GLS of FWHM and for contrast, the long-term trend contribution is visible, together with a significant peak at $\sim$30.55\,d, showing its alias due to the Moon cycle (clearly visible in the window function). In fact, by removing the long-term trend, the 30.55\,d peak goes below the significance thresholds in the GLS of all the activity indicators, as well as the long-term power, and the significance of the rotational period peak strongly increases, as shown in the bottom three panels of Fig.\,\ref{fig:gls_rv_activity}.
We analysed the system to retrieve the planetary parameters using the toolbox {\sc exo-striker}\footnote{\url{https://github.com/3fon3fonov/exostriker}} \citep{exostriker_2019_software}. We performed a joint fit of the RV data and the undetrended transit photometry, which were extracted and cleaned through the {\sc PATHOS} pipeline, and corrected for the dilution factor to consider the flux contamination of the neighbour stars falling inside the photometric aperture (see Sect.\,\ref{sec:TESS_data}).
The transit modelling inside the {\sc exo-striker} toolbox was done through the {\sc batman} package \citep{batman_kreidberg_2015}. We used an algorithm based on GPs to model and remove the signal induced by stellar activity, such as the rotation period signature and the RV long-term trend. Fundamental in this operation was the use of unflattened light curves to exploit the modulation introduced by the stellar rotation. We added the GP regression through the {\sc celerite} package (\citealt{GP_foreman-mackey_2017_celerite1, celerite2_GP}) to simultaneously model the stellar activity in both RV and transit signals. After we approached the analysis with a quasi-periodic kernel, we noticed it could not model the relevant harmonic in the GLS of the residuals at half of the stellar rotation period, which is probably a signature of the flux effect on RV variations due to dark spots or bright faculae (see, e.g. \citealt{lanza_2010_activity_harmonics}). Therefore, we chose a kernel containing the half-period term in its expression; i.e. the dSHO as defined by \cite{GP_foreman-mackey_2017_celerite1} and \cite{celerite2_GP}, because it has two SHO terms that can be used to model stellar rotation and the harmonic at half its period. The kernel hyperparameters are the standard deviation of the process GP$_{\rm dSHO}$ $\sigma$, the primary period of variability GP$_{\rm dSHO}$ $P_{\rm rot}$, the quality factor GP$_{\rm dSHO}$ $Q_0$ (or the quality factor minus one half; this keeps the system underdamped) for the secondary oscillation, and the difference between the quality factors of the first and the second modes GP$_{\rm dSHO}$ $dQ$. This parametrisation (if $dQ$ > 0) ensures that the primary mode is always of higher quality. In the end, GP$_{\rm dSHO}$ $f$ represents the fractional amplitude of the secondary mode compared to the primary. 

We imposed the following as free parameters: the planetary orbital period, $P_{\rm b}$; the RV semi-amplitude, $K$; the orbit eccentricity, $e;$  the argument of periastron, $\omega,$ estimated through the $hk\lambda$ parametrisation from \cite{eastman_2013} ($h$=$e$cos$\omega$, $k$=$e$sin$\omega$, $\lambda$=$\omega$+Ma, with Ma the mean anomaly); the inclination, $i$; the time of the first mid-transit, $t_0$; the scaled planetary radius, $R_{\rm p}/R_\star;$ and the stellar density, $\rho_\star$. We also left the following as free parameters: the quadratic limb-darkening coefficients (LDC; \citealt{kipping_2013_limbdark_param}), the RV offset and jitter, the photometry relative flux offset and noise, and the GP hyperparameters both for the photometry and RV data. Conscious that the GP can influence the RV retrieval and confident of the goodness of our transit estimates, we decided to use them to guide the choice of informative priors. For this reason, since stellar rotation is particularly evident from the light curves, we extracted $P_{\rm rot}$ from the GP$_{\rm dSHO}$ hyperparameter by modelling the light curves and also used it to model the GP on RVs. Since the light curve and RV datasets cover different BJDs, when scanning the system in different moments during the stellar activity evolution, we preferred to estimate the GP$_{\rm dSHO}$ hyperparameters $Q_0$ and $dQ$ separately. The same was done for the GP$_{\rm dSHO}$ $\sigma$ hyperparameters for the RV and photometric data, since they have different units (meters-per-second, m/s and parts-per-million, ppm, respectively). Also, since stellar surface inhomogeneities do not induce the same relative power of the first harmonic in RV and photometric time series, we decided to keep the GP$_{\rm dSHO}$ $f$ hyperparameters independent in our joint model.

We initialised the dynamic nested sampling (NS) algorithm implemented in the {\sc dynesty} package \citep{nested_sampling_speagle_2020} to efficiently explore the parameter space of orbital elements and study their posteriors. We imposed the likelihood maximisation and ran 130 live points for each fitted parameter (130 $\times~ n_{\rm dim}$ total live points, where $n_{\rm dim}$ is the total number of parameters). 
To avoid the presence of biases in the exploration of the parameter space, we imposed uniform priors for most of the used parameters (as reported in Table\,\ref{tab:prior_and_param} and Table\,\ref{tab:prior_and_param_other}), with a very wide prior defined for $K$ to allow a large range of $K$ values to be fitted. Since the planet is transiting, we preferred a Gaussian prior for the orbital inclination, $i,$ around 90$^\circ$, which, being considered as the angle of symmetry of the sampling interval, and hence the likelihood, was also set as the upper limit of the sampling range. Also, for the LDC we adopted a Gaussian prior taking into account the mean values selected from the theoretical LD estimates, specifically for the \textit{TESS} filters; we adopted the values reported in Table\,\ref{tab:stellar_prop} for \feh, \logg,\,and \teff\,. We referred to the estimates tabulated in \cite{claret_2018_ldc_fsm} for \textit{TESS} LDCs.
Given the caution suggested in setting LDC for stars with a \teff\,lower than 5000\,K (see \citealt{Patel_espinoza_2022_ldc}) and a relevant activity (see \citealt{maxted2018_ldc_unc}), such as TOI-5734, we set a sufficiently wide prior, as reported in Table\,\ref{tab:prior_and_param_other}. 
The priors of the RV and photometry GP$_{\rm dSHO}$ hyperparameters $\sigma$ and $f$ were set as wide and uniform. We imposed a Jeffreys prior for the RV jitter as well as for the RV and photometry GP$_{\rm dSHO}$ hyperparameters $Q_0$ and $dQ$ to optimally scan all the orders of magnitude of the imposed ranges. Finally, for the RV and photometry common hyperparameter GP$_{\rm dSHO}$ $P_{\rm rot}$, we set a sufficiently wide Gaussian prior based on the computation of the GLS periodograms in Sect.\,\ref{subsec:stellar_rot_period}.

Through the NS algorithm, we tested models involving zero planets (baseline null model), one planet circular (eccentricity fixed to zero) with $K=0$ (fixed), one planet circular with $K$ variable, and one planet with free eccentricity (elliptical orbit in the $hk\lambda$ parametrisation) without and with a linear or quadratic trend in the RV fit. Initially, we also tried to model the system with a two-planet configuration, but the results were not favoured over the other models, so we did not continue investigating this configuration. For each model, we extracted the logarithm of the Bayesian evidence ($\log Z$) to evaluate the model that better fits the data, consistently with the prescriptions by \cite{trotta2008bayes}. The results are shown in Table\,\ref{tab:dlogz_models_toi5734}. 
After fitting the model with no planets, a significant improvement in the fit quality was derived from the inclusion of transit constraints in the models with one planet. The first test with this model was done by imposing $K=0$, which means assuming that no planetary signal is present in the RV data. This way, the GP modelled the RV data oscillation, considering them to be produced entirely by stellar activity. This test was computed to check if RV data contain a signal from the photometrically detected transiting planet and to ensure that the modelling of the activity-induced RVs with a GP does not bias the mass estimate of the planet. Consequently, we also fitted a planetary signal in the RVs; hence, we left $K$ free to vary. We found a further significant improvement in the quality of the fit, with a $\Delta\log\,Z$ of 8.9 with respect to the $K=0$ model. Following \cite{trotta2008bayes}, this shows strong evidence that the RVs do contain the planet signal, which can be retrieved in our joint transit and RV GP analysis.
Additionally, since the one-planet eccentricity-free model has a negligible difference in Bayesian evidence with the circular one, we adopted the latter model. Also, including a linear trend in the RVs before fitting the models was inconclusive; the results yielded the same conclusion. The $K$-free model showed one planet with a circular orbit as the favoured one. This was expected since the dSHO kernel has the right characteristics to model the peak and the lower frequencies (higher periods), working as a detrending of the long-term activity (see Fig.\,\ref{fig:rv_rhk_bis_timmeseries}). 

\begin{figure}[h]
    \centering
    \includegraphics[width=0.85\linewidth]{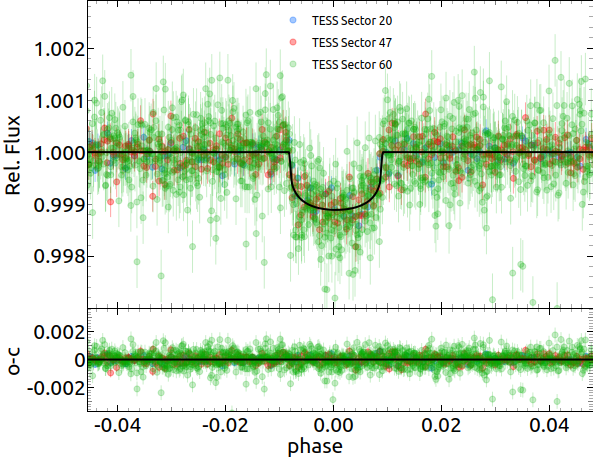}  
    \includegraphics[width=0.85\linewidth]{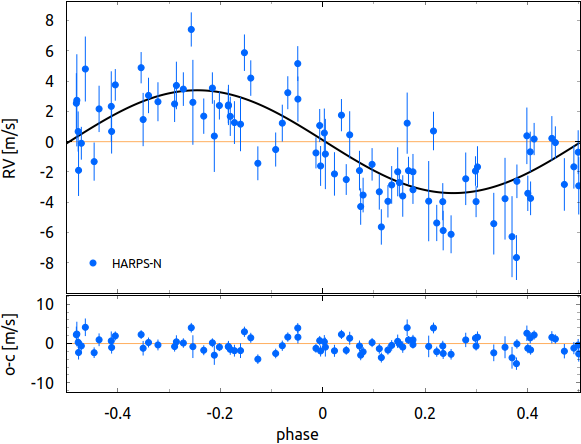}
    \caption{Top: \textit{TESS} phase-folded transits after subtracting the GP-modelled activity, along with the transit model shown with a black line and the residuals below it. Bottom: HARPS-N phased curve of RV data after subtracting the GP-modelled activity, with the planetary RV model (black line) and the residuals plotted below. The RV error bars are plotted without the jitter, the value of which is reported in Table\,\ref{tab:prior_and_param_other}.}
    \label{fig:lc_rv_phase_fold}
\end{figure}

The resulting planetary system parameters are listed in Table\,\ref{tab:prior_and_param} as median, along with the 16$^{th}$ and 84$^{th}$ percentiles. Due to the symmetry of the likelihood around $i_{\rm b}$=90$^\circ$, the inclination posterior is skewed and non-Gaussian; we therefore report the posterior mode and the 68$\%$ highest posterior-density (HPD) interval rather than the mean or median. We did the same for the impact parameter, $b_{\rm b}$, which was directly derived from $i_{\rm b}$. The phase-folded curves of transits and RVs with the GP activity signal subtracted and the residuals of the planetary model are displayed in Fig.\,\ref{fig:lc_rv_phase_fold}. Figure\,\ref{fig:rv_o_c_withgp} shows the RV data modulated by the GP modelling after subtracting the planetary signal. The 68.3$\%$ confidence levels of the NS posterior probability parameter distribution represent the 1$\sigma$ uncertainties of the system parameters. From the fit, we derived parameters scaled on stellar values, so we used the stellar parameters derived in Sect.\,\ref{sec:stellar_analysis} to estimate the final planetary parameters. The corner plots of the posterior distributions of the fitted and derived parameters are shown in Figs.\,\ref{fig:cornerplot_planpar} and \ref{fig:cornerplot_stat_data_fit}. From the adopted joint transit and RV model of TOI-5734b we derived the final estimates of the orbital period of $P_{\rm b}=6.1841876_{-0.0000094}^{+0.0000080}$\,d, which is compatible with the one found in Sect.\,\ref{sec:transit_validation} without the use of the GP to model the stellar activity, the stellar rotation period of $P_{\rm rot} = 11.09_{-0.08}^{+0.07}$\,d, the radius of $R_{\rm b}=2.10^{+0.12}_{-0.12}$\,$R_\oplus$, and the mass of $M_{\rm b}=9.1^{+2.6}_{-2.6}$\,$M_\oplus$. The mass is compatible with that derived through the mass-radius relations (e.g. \citealt{spiegel_2014_mr_relations, muller_2024_mr_relations}). The precisions of our radius and mass
estimates lie below the threshold commonly used to define high-precision estimates of 30$\%$  (they are, respectively, $\sim 2\%$ and $\sim 29\%$ precision), making them reliable estimates. A further analysis of the completeness of our RV time series is detailed in Appendix\,\ref{app:completeness_rv}.
\setlength{\tabcolsep}{1pt}
\begin{table}[h]
\tiny
\caption{Nested sampling priors and posteriors (median and the 16$^{th}$ and 84$^{th}$ percentiles) of TOI-5734b orbital parameters.}
\label{tab:prior_and_param}
\begin{center}
\begin{tabular}{lrrr}
\hline\hline
\noalign{\vskip 1.mm}
Parameter & Unit & Adopted Priors & Posterior \\ \noalign{\vskip 1.mm} 
\hline \noalign{\vskip 1.mm} 
\multicolumn{4}{l}{\textit{Orbital parameters}} \\
\noalign{\vskip 1.mm} 
RV semi-amplitude ($K_{\rm b}$) & m/s & $\mathcal{U}$(0, 50) & $3.9_{-1.1}^{+1.1}$ \\ \noalign{\vskip 1.mm}
Orbital Period ($P_{\rm b}$) & days & $\mathcal{U}$(6.18, 6.19) & $6.1841876_{-0.0000094}^{+0.0000080}$ \\ \noalign{\vskip 1.mm}
Orbital inclination ($i_{\rm b}$) & deg & $\mathcal{N}$(90, 3) [80, 90] & $89.88_{-0.63}^{+0.12}$ \\ \noalign{\vskip 1.mm}
$e\sin(\omega)$ &  & (fixed) & 0 \\ \noalign{\vskip 1.mm}
$e\cos(\omega)$ &  & (fixed) & 0 \\ \noalign{\vskip 1.mm}
Central time of the  & BJD & $\mathcal{U}$(2458842, & $2458842.5107_{-0.0012}^{+ 0.0016}$ \\ 
\,\,first transit ($t_{\rm 0,b}$) &  & 2458843) &  \\ 
\noalign{\vskip 1.mm}
Scaled planetary radius ($R_{\rm b}/R_\star$) &  & $\mathcal{U}$(0.01, 0.12) & $0.03011_{-0.00064}^{+0.00066}$ \\ \noalign{\vskip 1.mm} 
Stellar density & kg/m$^3$ & $\mathcal{N}$(3900, 800) & $3119_{-340}^{+210}$ \\ 
\noalign{\vskip 1.mm} 
\hline \noalign{\vskip 1.mm} 
\multicolumn{4}{l}{\textit{Derived planetary parameters}} \\
\noalign{\vskip 1.mm} 
Semi-major axis ($a_{\rm b}$) & au &  & $ 0.05921_{-0.00024}^{+0.00024}$ \\ \noalign{\vskip 1.mm}
Impact parameter ($b_{\rm b}$) &   &  & $0.04_{-0.04}^{+0.22}$ \\ \noalign{\vskip 1.mm}
Transit duration ($T_{\rm b,\,1,4}$) & days &  & $0.0991_{-0.0070}^{+0.0061}$ \\ \noalign{\vskip 1.mm}
Planet mass ($M_{\rm b}$) & $M_\oplus$ &  & $9.1^{+2.6}_{-2.6}$ \\ \noalign{\vskip 1.mm}
Planet radius ($R_{\rm b}$) & $R_\oplus$ &  & $2.10^{+0.12}_{-0.12}$ \\ \noalign{\vskip 1.mm}
Planet density ($\rho_{\rm b}$) &   $\rho_\oplus$ &  & $0.98_{-0.30}^{+0.36}$ \\ \noalign{\vskip 1.mm}
Equilibrium temperature ($T_{\rm eq}$) & K &  & $688_{-23}^{+23} $ \\ \noalign{\vskip 1.mm}
Surface gravity ($g_{\rm b}$) & m/s$^2$ &  & $20.2_{-6.0}^{+6.6}$ \\
\noalign{\vskip 1.mm} \hline
\end{tabular}
\end{center}
\begin{tablenotes}
    \tiny {
        \item{Notes: The main planetary parameters derived from the analysis are shown at the bottom of the table, while the complementary parameters (e.g. GP hyperparameters) are reported in Table\,\ref{tab:prior_and_param_other}. The inclination, $i_{\rm b}$, and the impact parameter, $b_{\rm b}$, are expressed in terms of posterior mode and the 68$\%$ HPD interval.
        }
    }
\end{tablenotes}
\end{table}
\begin{table}[h]
\begin{center}
\caption{Comparison between joint fit models where the logarithm of the Bayesian evidence $\log Z$ was derived from the NS runs with the 1D--GP activity modelling.}
\label{tab:dlogz_models_toi5734}
\begin{tabular}{l|c}
\hline \hline
$\log Z$ $-$ 81074.18  & RV+LC 1D-GP \\
\hline
TOI-5734 & No trend \\
\hline
GP+0p (null) & $-349.87$  \\
GP+1p circular $K$=0 & 0.00  \\
GP+1p circular $K$-free & $8.90$  \\
GP+1p $hk\lambda$ $K$-free & $9.92$ \\
\hline
\end{tabular}
\end{center}
\begin{tablenotes}
    \tiny{    
        \item{Notes: Rows report the $\log Z$ values extracted from the RV+LC simultaneous fit with one-dimensional GP applied to a model with zero planets (null model), one planet in circular orbit with transit constraints but no planetary RV signal (GP+1p circular $K=0$), one planet circular with RV planetary modelling (GP+1p circular $K$-free), and one planet with free eccentricity; we used the $hk\lambda$ parametrisation (GP+1p $hk\lambda$ $K$-free). For readability, $\log Z=81001.82$ was subtracted from all the models.
    }

 }
\end{tablenotes}
\end{table}

\section{Discussion}
\label{sec:discussion}

\subsection{Mass-radius diagram}

\begin{figure}[h]
    \centering
    \includegraphics[width=0.8\linewidth]{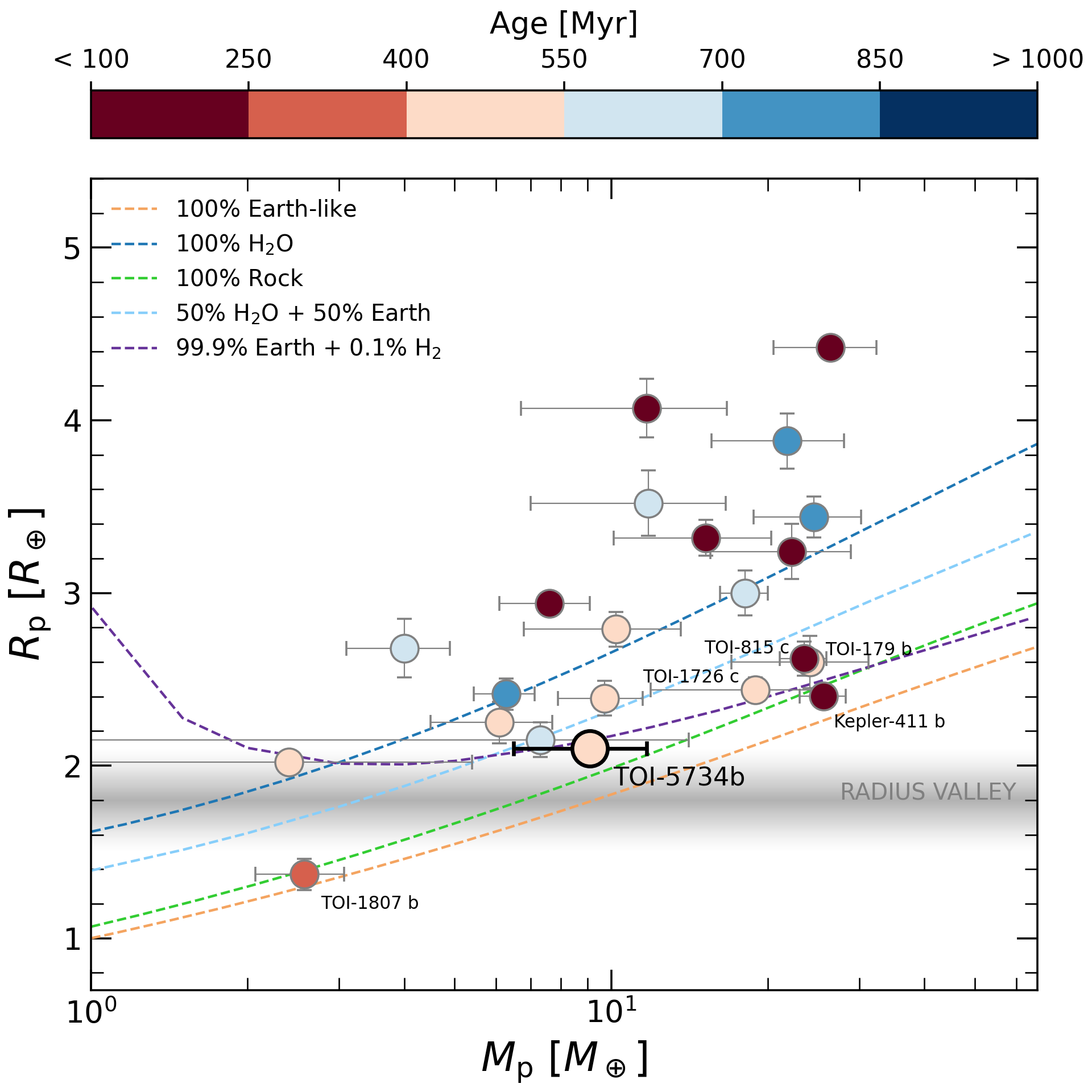}
    \caption{Mass--radius diagram for young ($\leq 
    1$\,Gyr) planets with precise age estimates (relative errors < 50$\%$). Planets are colour-coded based on their age, with a discrete colour bar at the top of the figure. Young planets with similar characteristics to TOI-5734b are labelled with their name. The Zeng et al. (2019) tracks, shown as dashed lines, are reported in the case of planets with an equilibrium temperature of
$T_{\rm eq}=700$\,K, 1\,mbar pressure and the following compositions: 100$\%$ Earth-like, 100$\%$ H$_2$O, 100$\%$ rock (MgSiO$_3$), 50$\%$ Earth+50$\%$ H$_2$O, 99.9$\%$ Earth+0.1$\%$ H$_2$ envelope. The wide, horizontal grey line represents the `radius valley'.}
    \label{fig:mr_plot}
\end{figure}

The values of $M_{\rm b}$ and $R_{\rm b}$ allowed us to derive the planetary mean density of $\rho_{\rm b}= 0.98_{-0.30}^{+0.36}  ~\rho_\oplus = 5.4_{-1.7}^{+2.0}$\,g/cm$^3$ (Table\,\ref{tab:prior_and_param}). The mass-radius diagram is shown in Fig.\,\ref{fig:mr_plot}, where planets younger than 1\,Gyr with accurate (relative errors < 50$\%$) radius, mass, and age estimates are overplotted. We also calculated the planet equilibrium temperature of $T_{\rm eq} = 688_{-23}^{+23}$\,K by assuming an albedo ($A_b$) of 0.3 and a full atmospheric circulation (day-night uniform heat redistribution; $f_{\rm atm}=1$); we did this following the equation derived by the Stephan-Boltzmann law, $T_{\rm eq}=\,$\teff$~\big(\frac{R_\star}{2a}\big)^{1/2} [f_{\rm atm}(1-A_b)]^{1/4}$, where $a$ is the planetary semi-major axis (\citealt{kaltenegger_2011_teq_formula}). 
In the same figure, we also show the planetary composition tracks by \cite{zeng_2019_mr_tracks} chosen by assuming a 1\,mbar surface pressure level and a reference equilibrium temperature of 700\,K, which is close to the value of our target. As shown in the diagram, the planet is mostly compatible with an Earth-like composition (32.5$\%$\,Fe\,+\,67.5$\%$\,MgSiO$_3$) at 99$\%$\,with a 0.1$\%$ H$_2$ atmosphere. Besides this, the planet location is compatible within 1$\sigma$ with a rocky composition (100$\%$\,MgSiO$_3$) and a structure of 50$\%$\,Earth and 50$\%$\,H$_2$O (water world; see \citealt{burn_waterworlds_2024}), suggesting the possible presence of water content of less than 50$\%$.
Given its physical parameters, TOI-5734b is placed on the upper edge of the so-called `radius valley', which is characterised by a scarcity of 1.5--2.0\,$R_\oplus$ planets (e.g. \citealt{petigura_2022}), or it is within its upper bounds if the valley location at the planetary $P_{\rm b}$ is calculated following the relation from \cite{vaneylen_2018_radiusvalley}. In the mass--radius diagram, there are other young planets (labelled with their name) with similar characteristics to those of TOI-5734b. Sorted by increasing radius, they are TOI-1807b (\citealt{nardiello2022}), Kepler-411b (\citealt{kepler411_mrplot}), TOI-1726c (\citealt{Damasso_2023_yo40, Mallorquin2023_toi1726}), TOI-179b (\citealt{Desidera_2023_toi179}), and TOI-815c (\citealt{toi815_psaridi2024}). Both TOI-1807b and Kepler-411b orbit two younger stars with properties (\teff, $M_{\rm \star}$, $R_{\rm \star}$) comparable to those of TOI-5734. They have shorter orbital periods than TOI-5734b (respectively $\sim$0.55\,d and $\sim$3.01\,d for TOI-1807b and Kepler-411b) and low eccentricity (fixed to 0 and 0.15, respectively). TOI-1807b is placed below the `radius valley'; it has an Earth-like density ($\rho_p=5.5\pm1.6$\,g/cm$^3$) consistent with a rocky terrestrial composition (silicates and iron) and no extended H/He envelope, while Kepler-411b, with a $T_{\rm eq}$ of 1138\,K and a density of 10.3$\pm$1.3\,g/cm$^3$, possibly has a 21\% iron-core mass fraction with a rocky mantle. Also, TOI-179b orbits a K2V star with an age of 400\,Myr and a short period of 4.14\,d. It is a mini-Neptune with a high equilibrium temperature of 990\,K with a typical composition of 75\%\,rock\,$+$\,25\%\,water. This planet has lost most of its primordial atmosphere, but its high density ($\rho_{\rm p}=7.5\pm2.5$\,g/cm$^3$) guarantees stability against hydrodynamic evaporation. TOI-1726c and TOI-815c are both mini-Neptunes with longer orbital periods from 20.5 to 35\,d orbiting, respectively, a 400\,Myr-old G2V and a 200\,Myr-old K3V star. They have similar densities of, respectively, $7.1\pm2.7$\,g/cm$^3$ and $7.2\pm1.1$\,g/cm$^3$, and a different $T_{\rm eq}$ (680\,K for TOI-1726c and 469\,K for TOI-815c), suggesting a composition of an Earth-like core and a water layer (<50\% and 10\% H$_2$O, respectively). TOI-1726c has an atmospheric mass fraction that can significantly change during the lifetime of the system, even though it is scarcely affected by atmospheric evaporation. TOI-815c has little or no atmosphere, since it is likely that the planet has lost the majority of its primordial atmosphere. 

\subsection{Atmospheric evolution}\label{subsec:atm_evol}

\begin{figure*}[h!]
\begin{center}
\begin{tabular}{ccc} 
\includegraphics[width=5.5cm]{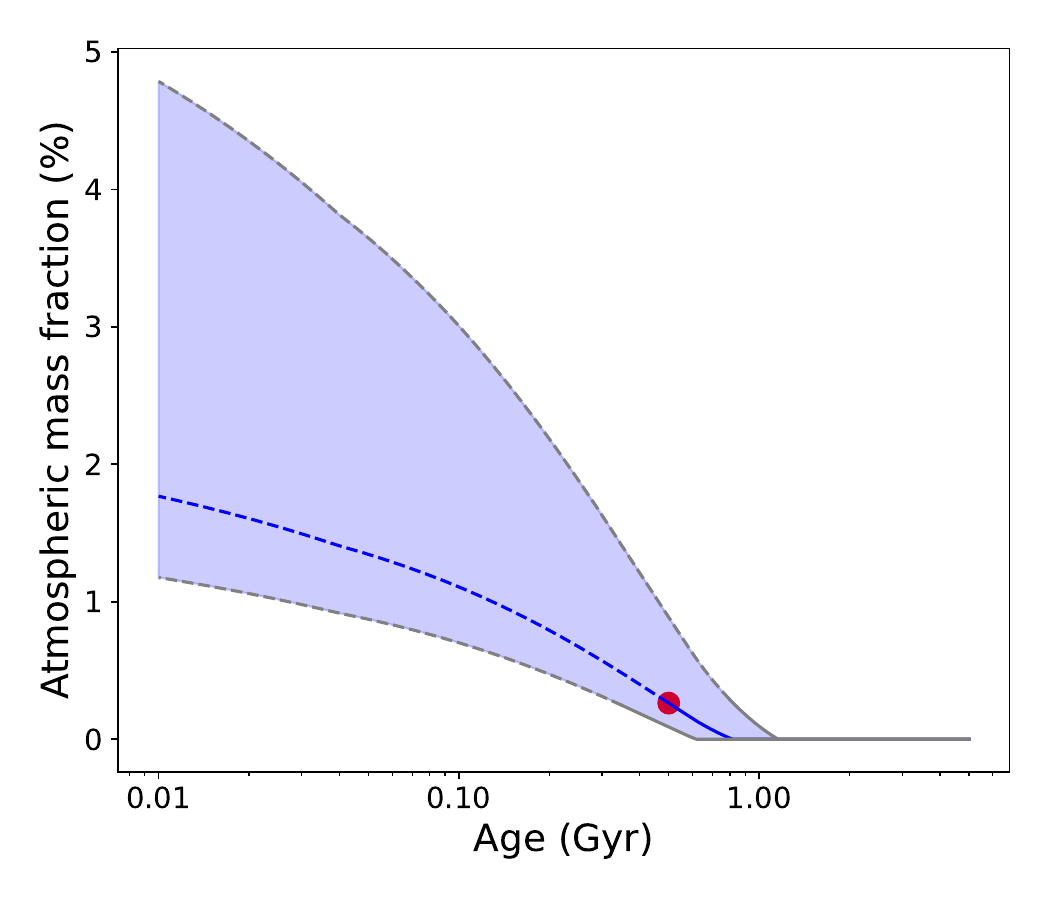} & \includegraphics[width=5.5cm]{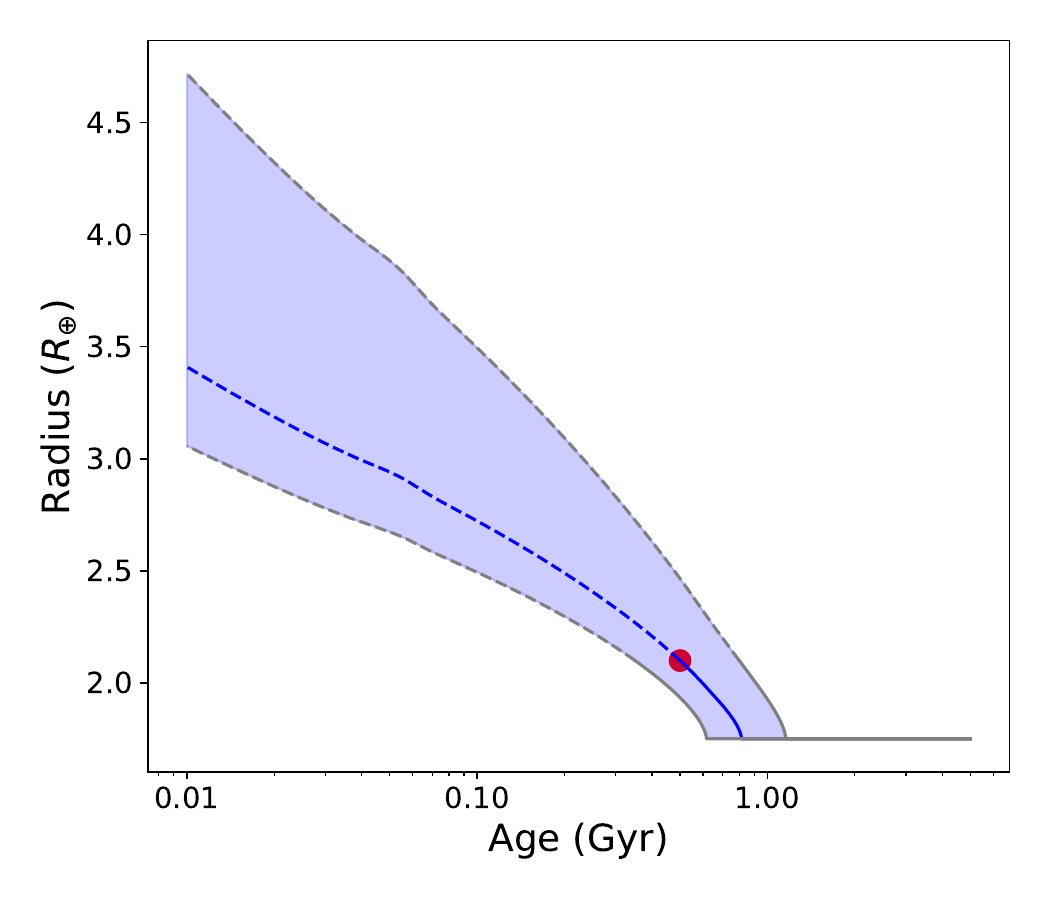} & \includegraphics[width=5.5cm]{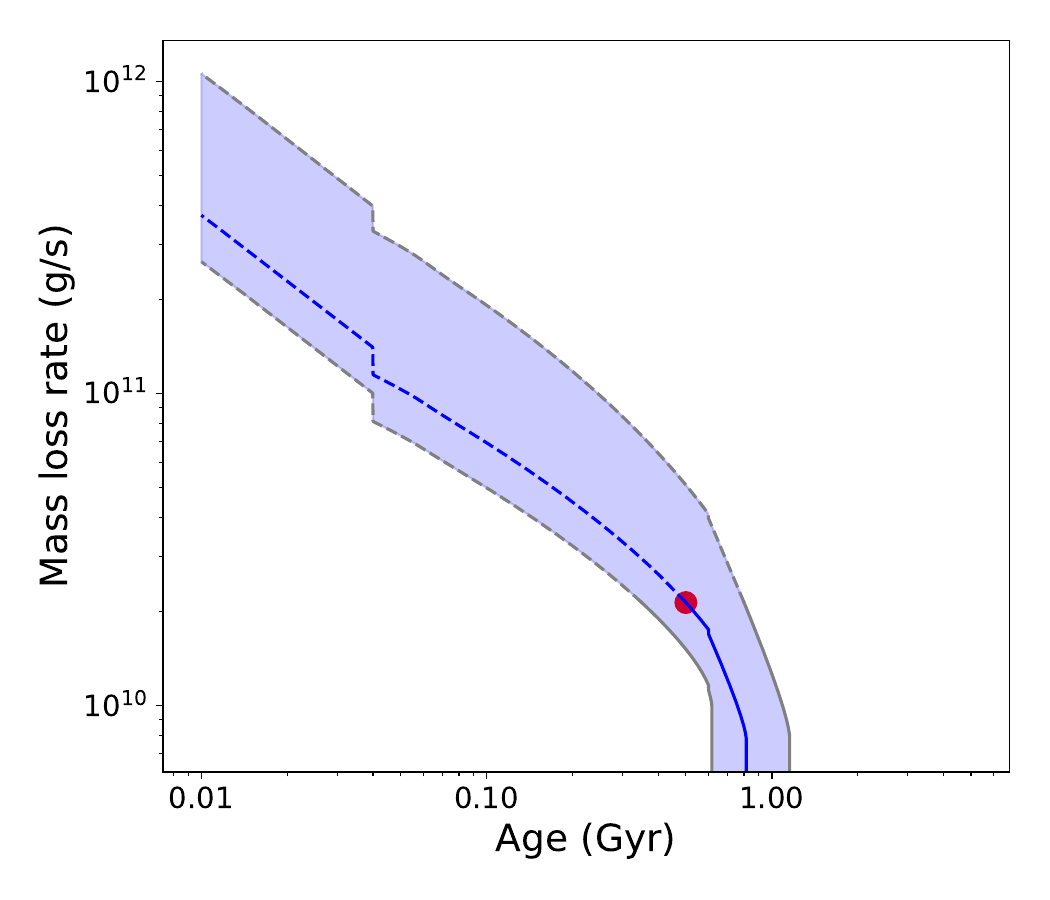} \\
\end{tabular}
\caption{Temporal evolution of mass fraction, radius, and mass-loss rate of TOI-5734b. The panels show the evolution of atmospheric mass fraction (left), radius (middle), and mass-loss rate (right). Solid lines represent the future evolution, whereas dashed lines represent the past one.
The grey lines represent the temporal evolution, considering an age equal to $\pm 1\sigma$ the nominal age. The red circle shows the position of TOI-5734b at the current age.}
\label{fig:evap}
\end{center}
\end{figure*}

Since TOI-5734b is a close-in planet orbiting a young star, it is subjected to the high-energy irradiation emitted by its host. This condition could have a strong impact on the atmospheric evolution of such a low-mass planet, whose gravity is unable to retain volatile chemical species heated by the intense stellar radiation.
To study the atmospheric photo-evaporation over time, we adopted a modelling approach initially proposed by \cite{Locci19} that was refined in subsequent works (see, e.g. \citealt{Mantovan24b}).
In brief, we evaluated the mass-loss rate of the planetary atmosphere by using the analytical approximation derived from the ATES hydrodynamic code \citep{Caldiroli+2021, Caldiroli+2022}, coupled with the planetary core-envelope models by \cite{Fortney2007} and \cite{LopFor14}, the MESA stellar tracks (MIST; \citealt{choi+2016}), the X-ray luminosity time evolution by \citet{Penz08a}, and the X-ray-to-EUV scaling by \citet{SF25}. We underline that we use the term 'core' to refer specifically to all the solid components of the planet.
As a first step, we calculated the planetary structure. Starting from the mass and radius at the current age, we determined the mass and radius of the core, the radius of the gaseous envelope, and the atmospheric mass fraction. To do this, we solved a system of four equations with four unknowns. One equation is from \cite{Fortney2007}, which provides the core radius as a function of core mass; another is from \cite{LopFor14}, which gives the envelope radius as a function of the atmospheric fraction. Finally, we included two closing equations: one that gives the planetary radius as the sum of the envelope and core radius, and another that calculates the planetary mass as the sum of the core and atmospheric mass.
Thus, keeping this planetary structure constant over time, we calculated the mass lost by the planet due to evaporation. We then updated the planetary radius, which changes both due to gravitational contraction and in response to mass loss. Finally, we updated both the bolometric luminosity and the X and EUV luminosities (see Sect.\ref{subsec:star_mass_radius}). Our simulation traces the planet's evolution back in time to 10\,Myr, at which point we assume the stellar disc had dissipated and the planet had reached its final orbit, while the simulation ended at 5\,Gyr.
To perform simulations of the past and future planetary evolution, we assumed an Earth-like core with a rock--iron (67$\%$, 33$\%$) composition. To compute the planetary radius, we considered solar metallicity (\citealt{LopFor14}).
Figure\,\ref{fig:mrcore} shows the values of the core mass, core radius, and atmospheric mass fraction, at the present age, as a function of the planetary radius. We found that the system has a solution for only a very low value of the atmospheric mass fraction. We needed to move to higher values of planetary radii or lower values of the rocky fraction to find a higher value of the atmospheric mass fraction. An ice--rock composition would produce a higher core radius, giving even lower values for the atmospheric mass fraction. However, the system of an ice--rock composition does not find a solution for the nominal values of mass and radius, suggesting that an Earth-like core composition is more appropriate.
At the current age, we determined a core mass and a radius of 9.07\,$M_\oplus$ and 1.75\,$R_\oplus$, respectively. According to our calculations, the planet has only a small atmospheric fraction, $\sim 0.25\%$, suggesting that it has likely already lost most of its primordial envelope entirely. 
We emphasise that this conclusion holds under the assumption of an Earth-like core and a hydrogen-dominated atmosphere. 
As for the future evolution, we found that the planet will completely lose its primary atmosphere in approximately 300\,Myr, and it will end as a Chthonian planet (rocky core remnant of the atmospheric escape process a planet underwent during its evolution; \citealt{chthonian_hebrard_2004}), with a radius and mass equivalent to those of the core (unless it develops a secondary atmosphere).
We also determined that the planet began its evolution with a mass of $\sim 9.24\,M_\oplus$ and a radius of $\sim 3.41\,R_\oplus$. 
In Fig.\,\ref{fig:evap}, we show the temporal evolution of some planetary parameters. The figure also shows the evolutions taking into account different ages for the star; in particular, we considered the values of 350\,Myr and 800\,Myr, i.e. $\pm 1\sigma$ with respect to the nominal age. 
An older age would imply that the planet has undergone the photoevaporation process for a longer time; hence, it would have had a greater mass at 10\,Myr. The opposite argument would apply in the case of a younger age.
Changing the law used to calculate the mass-loss rate does not alter the planet's fate, it just changes the timescale needed for the planet to lose its atmosphere. 
In particular, the timescale is a factor of $\sim 1.8$ longer in the case of the formulation proposed by \cite{kuby+2018a}, because it predicts a lower mass-loss rate with respect to that computed with the analytical formulation of ATES \citep{Caldiroli+2021, Caldiroli+2022}. We conclude that, during its lifetime, the planet has likely moved from the region at larger radii in the mass--radius diagram to its current position near the radius gap. 

\subsection{Orbital evolution}

We estimated the timescale of tidal circularisation of the orbit of TOI-5734b. In this system, orbit circularisation is dominated by tides inside the planet. Assuming a modified tidal-quality factor of $Q^{\prime}_{pl}$ = 1000, which is appropriate for a planet with a rocky internal structure (\citealt{lanza2021_circularization_qlow}), we find a circularisation e-folding timescale e/|de/dt| of 1.1\,Gyr, which is comparable with or slightly longer than the age of the system. 
Therefore, the planet is likely to have formed by migration in a disk rather than through high-eccentricity migration (HEM; \citealt{rasio_ford_1996_hem}) followed by circularisation, because the latter scenario would require a time span longer than the age of the system. Moreover, the semi-major axis of the planet is more than ten times larger than the semi-major axis corresponding to the Roche limit (0.0053\,au for our target), while in the case of HEM, one would expect a semi-major axis around three times the Roche limit (\citealt{Ford_2006_hem}).

\section{Conclusions}
\label{sec:conclusions}

In this paper, we present the dynamical mass determination of a planet transiting the active star TOI-5734, first observed with \textit{TESS}. TOI-5734 is a K3-K4\,V star with a \teff\ of 4750\,K. Based on the stellar $P_{\rm rot}$ ($\sim 11.09$\,d), coronal and chromospheric activity, and lithium $EW$, we find that the star is relatively young, with an age $\sim 500$\,Myr. We combined the \textit{TESS} observations with a dataset of RV measurements collected with HARPS-N at TNG to model stellar activity and planetary signal simultaneously. We used a GP regression to extract the activity signal with a dSHO kernel, which modelled the stellar rotational period and its first harmonic. This approach helped us to constrain the stellar and planetary parameters. We tested several models, and we found that TOI-5734 is orbited by a planet with a mass of $\sim 9.1\,M_\oplus$ and a radius of $\sim 2.10\,R_\oplus$ leading to a density of $\sim 5.4~{\rm g/cm^3}\sim0.98\,\rho_\oplus$.
TOI-5734b is a hot sub-Neptune possibly with a rocky composition and a depleted (almost) primary atmosphere that is located slightly above the upper edge of the `radius valley' in the mass-radius diagram. This is also confirmed by the simulation of the past and future planetary evolution, suggesting that the planet will completely lose its primordial envelope within 300\,Myr, ending its evolution as a Chthonian planet. Despite these results, taking into account its mass and radius uncertainties, the planet location is possibly compatible with a water-world model (composition of 50$\%$ H$_2$O), which cannot be discarded in the absence of other informative data.

Thanks to the JWST and the future launch of \textit{Ariel} in 2029, it will be possible to probe the exoplanet atmospheres of young transiting planets and look at the atmospheric composition and the possible presence of clouds and hazes  in detail. This will also allow us to obtain insight into the formation and migration mechanisms of young exoplanets.
For the particular case of TOI-5734b, we employed the ArielRad radiometric model (\citealt{Mugnai_etal2020}) to evaluate the number of transits needed to achieve the required $S/N$ to detect the presence of a primary atmosphere during the \textit{Ariel} \textit{Tier 1} observation phase. Assuming a cloud-free atmosphere and the planetary and stellar properties obtained in this work, the number of transits is 19.
In particular, the detection of the primary atmosphere for this planet could validate our atmospheric evolution simulations and deepen our knowledge of the evolution of young exoplanetary atmospheres.

\begin{acknowledgements}

This work made use of \texttt{tpfplotter} by J. Lillo-Box (publicly available at www.github.com/jlillo/tpfplotter), which also made use of the Python packages \texttt{astropy}, \texttt{lightkurve}, \texttt{matplotlib} and \texttt{numpy}. This research has made use of the Exoplanet Follow-up Observation Programme (ExoFOP; DOI: 10.26134/ExoFOP5) website, which is operated by the California Institute of Technology, under contract with the National Aeronautics and Space Administration under the Exoplanet Exploration Program.
This work includes data collected with the \textit{TESS} mission, obtained from the MAST data archive at the Space Telescope Science Institute (STScI). This work has made use of data from the European Space Agency (ESA) mission {\it Gaia} (\url{https://www.cosmos.esa.int/gaia}), processed by the {\it Gaia} Data Processing and Analysis Consortium (DPAC, \url{https://www.cosmos.esa.int/web/gaia/dpac/consortium}). Funding for the DPAC has been provided by national institutions, in particular, the institutions participating in the {\it Gaia} Multilateral Agreement. This work makes use of observations from the LCOGT network. Part of the LCOGT telescope time was granted by NOIRLab through the Mid-Scale Innovations Program (MSIP). MSIP is funded by NSF. This work is based on observations obtained at the Hale Telescope, Palomar Observatory, as part of a collaborative agreement between the Caltech Optical Observatories and the Jet Propulsion Laboratory operated by Caltech for NASA. This work has been financially supported by the PRIN-INAF 2019 Planetary systems at young ages (PLATEA), the grant INAF 2022 TRAME@JWST (TRacing the Accretion Metallicity rElationship with NIRSpec@JWST), the ASI-INAF agreement 2021-5-HH.0, the INAF Guest Observer Grant (Large) "GAPS2" and the INAF Guest Observer Grant (Normal) "ArMS: the \textit{Ariel} Masses Survey Large Program at the TNG", according to the INAF Fundamental Astrophysics funding scheme. Part of the research activities described in this paper were carried out with contribution of the Next Generation EU funds within the National Recovery and Resilience Plan (PNRR), Mission 4 - Education and Research, Component 2 - From Research to Business (M4C2), Investment Line 3.1 - Strengthening and creation of Research Infrastructures, Project IR0000034 – “STILES - Strengthening the Italian Leadership in ELT and SKA”. SF thanks the Max Planck Institute for Astronomy (Heidelberg) for the kind hospitality. T.T. acknowledges support by the BNSF program "VIHREN-2021" project No. KP-06-DV/5. LMan acknowledges financial contribution from PRIN MUR 2022 project 2022J4H55R. IAS acknowledges the support of the MV Lomonosov Moscow State University Program of Development. TZi acknowledges support from the CHEOPS ASI-INAF agreement n. 2019-29-HH.0, NVIDIA Academic Hardware Grant Program for the use of the Titan V GPU card and the Italian MUR Departments of Excellence grant 2023-2027 Quantum Frontiers. DRC acknowledges partial support from NASA Grant 18-2XRP18$\_$2-0007. We acknowledge the Italian centre for Astronomical Archives (IA2, \url{https://www.ia2.inaf.it}) for providing technical assistance, services and supporting activities. SF thanks G.S.A.\ Pallavolo\ Ariano\ a.s.d. for its support and wishes to dedicate this work to it.

\end{acknowledgements}

\bibliographystyle{aa}
\bibliography{Biblio.bib}

@ARTICLE{Pizzolato+2003,
       author = {{Pizzolato}, N. and {Maggio}, A. and {Micela}, G. and {Sciortino}, S. and {Ventura}, P.},
        title = "{The stellar activity-rotation relationship revisited: Dependence of saturated and non-saturated X-ray emission regimes on stellar mass for late-type dwarfs}",
      journal = {\aap},
         year = 2003,
        month = jan,
       volume = {397},
        pages = {147-157},
          doi = {10.1051/0004-6361:20021560},
       adsurl = {https://ui.adsabs.harvard.edu/abs/2003A&A...397..147P}
}

@ARTICLE{eRosita2024,
       author = {{Merloni}, A. and {Lamer}, G. and {Liu}, T. and {Ramos-Ceja}, M.~E. and {Brunner}, H. and {Bulbul}, E. and {Dennerl}, K. and {Doroshenko}, V. and {Freyberg}, M.~J. and {Friedrich}, S. and {Gatuzz}, E. and {Georgakakis}, A. and {Haberl}, F. and {Igo}, Z. and {Kreykenbohm}, I. and {Liu}, A. and {Maitra}, C. and {Malyali}, A. and {Mayer}, M.~G.~F. and {Nandra}, K. and {Predehl}, P. and {Robrade}, J. and {Salvato}, M. and {Sanders}, J.~S. and {Stewart}, I. and {Tub{\'\i}n-Arenas}, D. and {Weber}, P. and {Wilms}, J. and {Arcodia}, R. and {Artis}, E. and {Aschersleben}, J. and {Avakyan}, A. and {Aydar}, C. and {Bahar}, Y.~E. and {Balzer}, F. and {Becker}, W. and {Berger}, K. and {Boller}, T. and {Bornemann}, W. and {Br{\"u}ggen}, M. and {Brusa}, M. and {Buchner}, J. and {Burwitz}, V. and {Camilloni}, F. and {Clerc}, N. and {Comparat}, J. and {Coutinho}, D. and {Czesla}, S. and {Dannhauer}, S.~M. and {Dauner}, L. and {Dauser}, T. and {Dietl}, J. and {Dolag}, K. and {Dwelly}, T. and {Egg}, K. and {Ehl}, E. and {Freund}, S. and {Friedrich}, P. and {Gaida}, R. and {Garrel}, C. and {Ghirardini}, V. and {Gokus}, A. and {Gr{\"u}nwald}, G. and {Grandis}, S. and {Grotova}, I. and {Gruen}, D. and {Gueguen}, A. and {H{\"a}mmerich}, S. and {Hamaus}, N. and {Hasinger}, G. and {Haubner}, K. and {Homan}, D. and {Ider Chitham}, J. and {Joseph}, W.~M. and {Joyce}, A. and {K{\"o}nig}, O. and {Kaltenbrunner}, D.~M. and {Khokhriakova}, A. and {Kink}, W. and {Kirsch}, C. and {Kluge}, M. and {Knies}, J. and {Krippendorf}, S. and {Krumpe}, M. and {Kurpas}, J. and {Li}, P. and {Liu}, Z. and {Locatelli}, N. and {Lorenz}, M. and {M{\"u}ller}, S. and {Magaudda}, E. and {Mannes}, C. and {McCall}, H. and {Meidinger}, N. and {Michailidis}, M. and {Migkas}, K. and {Mu{\~n}oz-Giraldo}, D. and {Musiimenta}, B. and {Nguyen-Dang}, N.~T. and {Ni}, Q. and {Olechowska}, A. and {Ota}, N. and {Pacaud}, F. and {Pasini}, T. and {Perinati}, E. and {Pires}, A.~M. and {Pommranz}, C. and {Ponti}, G. and {Poppenhaeger}, K. and {P{\"u}hlhofer}, G. and {Rau}, A. and {Reh}, M. and {Reiprich}, T.~H. and {Roster}, W. and {Saeedi}, S. and {Santangelo}, A. and {Sasaki}, M. and {Schmitt}, J. and {Schneider}, P.~C. and {Schrabback}, T. and {Schuster}, N. and {Schwope}, A. and {Seppi}, R. and {Serim}, M.~M. and {Shreeram}, S. and {Sokolova-Lapa}, E. and {Starck}, H. and {Stelzer}, B. and {Stierhof}, J. and {Suleimanov}, V. and {Tenzer}, C. and {Traulsen}, I. and {Tr{\"u}mper}, J. and {Tsuge}, K. and {Urrutia}, T. and {Veronica}, A. and {Waddell}, S.~G.~H. and {Willer}, R. and {Wolf}, J. and {Yeung}, M.~C.~H. and {Zainab}, A. and {Zangrandi}, F. and {Zhang}, X. and {Zhang}, Y. and {Zheng}, X.},
        title = "{The SRG/eROSITA all-sky survey. First X-ray catalogues and data release of the western Galactic hemisphere}",
      journal = {\aap},
         year = 2024,
        month = feb,
       volume = {682},
          eid = {A34},
        pages = {A34},
          doi = {10.1051/0004-6361/202347165},
archivePrefix = {arXiv},
       eprint = {2401.17274},
 primaryClass = {astro-ph.HE},
       adsurl = {https://ui.adsabs.harvard.edu/abs/2024A&A...682A..34M}
}

@article{vaneylen_2018_radiusvalley,
    author = {Van Eylen, V and Agentoft, Camilla and Lundkvist, M S and Kjeldsen, H and Owen, J E and Fulton, B J and Petigura, E and Snellen, I},
    title = {An asteroseismic view of the radius valley: stripped cores, not born rocky},
    journal = {\mnras},
    volume = {479},
    number = {4},
    pages = {4786-4795},
    year = {2018},
    month = {07},
    issn = {0035-8711},
    doi = {10.1093/mnras/sty1783},
    url = {https://doi.org/10.1093/mnras/sty1783},
    eprint = {https://academic.oup.com/mnras/article-pdf/479/4/4786/25193635/sty1783.pdf}
}

@ARTICLE{lanza2021_circularization_qlow,
       author = {{Lanza}, A.~F.},
        title = "{An internal heating mechanism operating in ultra-short-period planets orbiting magnetically active stars}",
      journal = {\aap},
         year = 2021,
        month = sep,
       volume = {653},
          eid = {A112},
        pages = {A112},
          doi = {10.1051/0004-6361/202140284},
archivePrefix = {arXiv},
       eprint = {2107.03044},
 primaryClass = {astro-ph.EP},
       adsurl = {https://ui.adsabs.harvard.edu/abs/2021A&A...653A.112L}
}

@ARTICLE{Brown:2013,
       author = {{Brown}, T.~M. and {Baliber}, N. and {Bianco}, F.~B. and {Bowman}, M. and {Burleson}, B. and {Conway}, P. and {Crellin}, M. and {Depagne}, {\'E}. and {De Vera}, J. and {Dilday}, B. and {Dragomir}, D. and {Dubberley}, M. and {Eastman}, J.~D. and {Elphick}, M. and {Falarski}, M. and {Foale}, S. and {Ford}, M. and {Fulton}, B.~J. and {Garza}, J. and {Gomez}, E.~L. and {Graham}, M. and {Greene}, R. and {Haldeman}, B. and {Hawkins}, E. and {Haworth}, B. and {Haynes}, R. and {Hidas}, M. and {Hjelstrom}, A.~E. and {Howell}, D.~A. and {Hygelund}, J. and {Lister}, T.~A. and {Lobdill}, R. and {Martinez}, J. and {Mullins}, D.~S. and {Norbury}, M. and {Parrent}, J. and {Paulson}, R. and {Petry}, D.~L. and {Pickles}, A. and {Posner}, V. and {Rosing}, W.~E. and {Ross}, R. and {Sand}, D.~J. and {Saunders}, E.~S. and {Shobbrook}, J. and {Shporer}, A. and {Street}, R.~A. and {Thomas}, D. and {Tsapras}, Y. and {Tufts}, J.~R. and {Valenti}, S. and {Vander Horst}, K. and {Walker}, Z. and {White}, G. and {Willis}, M.},
        title = "{Las Cumbres Observatory Global Telescope Network}",
      journal = {\pasp},
         year = 2013,
        month = sep,
       volume = {125},
       number = {931},
        pages = {1031},
          doi = {10.1086/673168},
archivePrefix = {arXiv},
       eprint = {1305.2437},
 primaryClass = {astro-ph.IM},
       adsurl = {https://ui.adsabs.harvard.edu/abs/2013PASP..125.1031B}
}

@ARTICLE{Collins:2017,
   author = {{Collins}, K.~A. and {Kielkopf}, J.~F. and {Stassun}, K.~G. and
    {Hessman}, F.~V.},
    title = "{AstroImageJ: Image Processing and Photometric Extraction for Ultra-precise Astronomical Light Curves}",
  journal = {\aj},
archivePrefix = "arXiv",
   eprint = {1601.02622},
 primaryClass = "astro-ph.IM",
     year = 2017,
    month = feb,
   volume = 153,
      eid = {77},
    pages = {77},
      doi = {10.3847/1538-3881/153/2/77},
   adsurl = {http://adsabs.harvard.edu/abs/2017AJ....153...77C}
}

@INPROCEEDINGS{McCully:2018,
       author = {{McCully}, Curtis and {Volgenau}, Nikolaus H. and
         {Harbeck}, Daniel-Rolf and {Lister}, Tim A. and {Saunders}, Eric S. and
         {Turner}, Monica L. and {Siiverd}, Robert J. and {Bowman}, Mark},
        title = "{Real-time processing of the imaging data from the network of Las Cumbres Observatory Telescopes using BANZAI}",
    booktitle = {\procspie},
         year = "2018",
       series = {SPIE Conference Series},
       volume = {10707},
        month = "Jul",
          eid = {107070K},
        pages = {107070K},
          doi = {10.1117/12.2314340},
archivePrefix = {arXiv},
       eprint = {1811.04163},
 primaryClass = {astro-ph.IM},
       adsurl = {https://ui.adsabs.harvard.edu/abs/2018SPIE10707E..0KM}
}

@INPROCEEDINGS{collins:2019,
       author = {{Collins}, Karen},
        title = "{TESS Follow-up Observing Program Working Group (TFOP WG) Sub Group 1 (SG1): Ground-based Time-series Photometry}",
    booktitle = {American Astronomical Society Meeting Abstracts \#233},
         year = 2019,
       series = {American Astronomical Society Meeting Abstracts},
       volume = {233},
        month = jan,
          eid = {140.05},
        pages = {140.05},
       adsurl = {https://ui.adsabs.harvard.edu/abs/2019AAS...23314005C}
}

@ARTICLE{Evans+2010,
       author = {{Evans}, Ian N. and {Primini}, Francis A. and {Glotfelty}, Kenny J. and {Anderson}, Craig S. and {Bonaventura}, Nina R. and {Chen}, Judy C. and {Davis}, John E. and {Doe}, Stephen M. and {Evans}, Janet D. and {Fabbiano}, Giuseppina and {Galle}, Elizabeth C. and {Gibbs}, II, Danny G. and {Grier}, John D. and {Hain}, Roger M. and {Hall}, Diane M. and {Harbo}, Peter N. and {He}, Xiangqun Helen and {Houck}, John C. and {Karovska}, Margarita and {Kashyap}, Vinay L. and {Lauer}, Jennifer and {McCollough}, Michael L. and {McDowell}, Jonathan C. and {Miller}, Joseph B. and {Mitschang}, Arik W. and {Morgan}, Douglas L. and {Mossman}, Amy E. and {Nichols}, Joy S. and {Nowak}, Michael A. and {Plummer}, David A. and {Refsdal}, Brian L. and {Rots}, Arnold H. and {Siemiginowska}, Aneta and {Sundheim}, Beth A. and {Tibbetts}, Michael S. and {Van Stone}, David W. and {Winkelman}, Sherry L. and {Zografou}, Panagoula},
        title = "{The Chandra Source Catalog}",
      journal = {\apjs},
         year = 2010,
        month = jul,
       volume = {189},
       number = {1},
        pages = {37-82},
          doi = {10.1088/0067-0049/189/1/37},
archivePrefix = {arXiv},
       eprint = {1005.4665},
 primaryClass = {astro-ph.HE},
       adsurl = {https://ui.adsabs.harvard.edu/abs/2010ApJS..189...37E}
}

@ARTICLE{Naponiello2024,
       author = {{Naponiello}, L. and {Bonomo}, A.~S. and {Mancini}, L. and {Steinmeyer}, M. -L. and {Biazzo}, K. and {Polychroni}, D. and {Dorn}, C. and {Turrini}, D. and {Lanza}, A.~F. and {Sozzetti}, A. and {Desidera}, S. and {Damasso}, M. and {Collins}, K.~A. and {Carleo}, I. and {Collins}, K.~I. and {Colombo}, S. and {D'Arpa}, M.~C. and {Dumusque}, X. and {Gonz{\'a}lez}, M. and {Guilluy}, G. and {Lorenzi}, V. and {Mantovan}, G. and {Nardiello}, D. and {Pinamonti}, M. and {Schwarz}, R.~P. and {Singh}, V. and {Watkins}, C.~N. and {Zingales}, T.},
        title = "{The GAPS programme at TNG: LXIV. An inner eccentric sub-Neptune and an outer sub-Neptune-mass candidate around BD+00 444 (TOI-2443)}",
      journal = {\aap},
         year = 2025,
        month = jan,
       volume = {693},
          eid = {A7},
        pages = {A7},
          doi = {10.1051/0004-6361/202451859},
archivePrefix = {arXiv},
       eprint = {2411.09417},
 primaryClass = {astro-ph.EP},
       adsurl = {https://ui.adsabs.harvard.edu/abs/2025A&A...693A...7N}
}

@article{Piaulet_Ghorayeb_2024,
doi = {10.3847/2041-8213/ad6f00},
url = {https://dx.doi.org/10.3847/2041-8213/ad6f00},
year = {2024},
month = {oct},
publisher = {The American Astronomical Society},
volume = {974},
number = {1},
pages = {L10},
author = {Piaulet-Ghorayeb, Caroline and Benneke, Björn and Radica, Michael and Raul, Eshan and Coulombe, Louis-Philippe and Ahrer, Eva-Maria and Kubyshkina, Daria and Howard, Ward S. and Krissansen-Totton, Joshua and MacDonald, Ryan J. and Roy, Pierre-Alexis and Louca, Amy and Christie, Duncan and Fournier-Tondreau, Marylou and Allart, Romain and Miguel, Yamila and Schlichting, Hilke E. and Welbanks, Luis and Cadieux, Charles and Dorn, Caroline and Evans-Soma, Thomas M. and Fortney, Jonathan J. and Pierrehumbert, Raymond and Lafrenière, David and Acuña, Lorena and Komacek, Thaddeus and Innes, Hamish and Beatty, Thomas G. and Cloutier, Ryan and Doyon, René and Gagnebin, Anna and Gapp, Cyril and Knutson, Heather A.},
title = {JWST/NIRISS Reveals the Water-rich “Steam World” Atmosphere of GJ 9827 d},
journal = {\apjl}
}

@article{luque_palle_2022,
author = {Rafael Luque  and Enric Pallé },
title = {Density, not radius, separates rocky and water-rich small planets orbiting M dwarf stars},
journal = {Science},
volume = {377},
number = {6611},
pages = {1211-1214},
year = {2022},
doi = {10.1126/science.abl7164},
URL = {https://www.science.org/doi/abs/10.1126/science.abl7164},
eprint = {https://www.science.org/doi/pdf/10.1126/science.abl7164}
}

@ARTICLE{toi815_psaridi2024,
       author = {{Psaridi}, Angelica and {Osborn}, Hugh and {Bouchy}, Fran{\c{c}}ois and {Lendl}, Monika and {Parc}, L{\'e}na and {Billot}, Nicolas and {Broeg}, Christopher and {Sousa}, S{\'e}rgio G. and {Adibekyan}, Vardan and {Attia}, Mara and {Bonfanti}, Andrea and {Chakraborty}, Hritam and {Collins}, Karen A. and {Davoult}, Jeanne and {Delgado-Mena}, Elisa and {Grieves}, Nolan and {Guillot}, Tristan and {Heitzmann}, Alexis and {Helled}, Ravit and {Hellier}, Coel and {Jenkins}, Jon M. and {Knierim}, Henrik and {Krenn}, Andreas and {Lissauer}, Jack J. and {Luque}, Rafael and {Rapetti}, David and {Santos}, Nuno C. and {Su{\'a}rez}, Olga and {Venturini}, Julia and {Wilkin}, Francis P. and {Wilson}, Thomas G. and {Winn}, Joshua N. and {Ziegler}, Carl and {Zingales}, Tiziano and {Alibert}, Yann and {Brandeker}, Alexis and {Egger}, Jo Ann and {Gandolfi}, Davide and {Hooton}, Matthew J. and {Tuson}, Amy and {Ulmer-Moll}, Sol{\`e}ne and {Abe}, Lyu and {Allart}, Romain and {Alonso}, Roi and {Anderson}, David R. and {Escud{\'e}}, Guillem Anglada and {B{\'a}rczy}, Tamas and {Barrado}, David and {Barros}, Susana C.~C. and {Baumjohann}, Wolfgang and {Beck}, Mathias and {Beck}, Thomas and {Benz}, Willy and {Bonfils}, Xavier and {Borsato}, Luca and {Bourrier}, Vincent and {Ciardi}, David R. and {Cameron}, Andrew Collier and {Charnoz}, S{\'e}bastien and {Cointepas}, Marion and {Csizmadia}, Szil{\'a}rd and {Cubillos}, Patricio and {Lo Curto}, Gaspare and {Davies}, Melvyn B. and {Daylan}, Tansu and {Deleuil}, Magali and {Deline}, Adrien and {Delrez}, Laetitia and {Demangeon}, Olivier D.~S. and {Demory}, Brice-Olivier and {Dorn}, Caroline and {Dumusque}, Xavier and {Ehrenreich}, David and {Erikson}, Anders and {Lecavelier des Etangs}, Alain and {de Miguel}, Diana and {Fortier}, Andrea and {Fossati}, Luca and {Frensch}, Yolanda G.~C. and {Fridlund}, Malcolm and {Gillon}, Micha{\"e}l and {G{\"u}del}, Manuel and {G{\"u}nther}, Maximilian N. and {Hagelberg}, Janis and {Helling}, Christiane and {Hoyer}, Sergio and {Isaak}, Kate G. and {Kiss}, Laszlo L. and {Lam}, Kristine W.~F. and {Laskar}, Jacques and {Lavie}, Baptiste and {Lovis}, Christophe and {Magrin}, Demetrio and {Marafatto}, Luca and {Maxted}, Pierre and {McDermott}, Scott and {M{\'e}karnia}, Djamel and {Mordasini}, Christoph and {Murgas}, Felipe and {Nascimbeni}, Valerio and {Nielsen}, Louise D. and {Olofsson}, G{\"o}ran and {Ottensamer}, Roland and {Pagano}, Isabella and {Pall{\'e}}, Enric and {Peter}, Gisbert and {Piotto}, Giampaolo and {Pollacco}, Don and {Queloz}, Didier and {Ragazzoni}, Roberto and {Ramos}, Devin and {Rando}, Nicola and {Rauer}, Heike and {Reimers}, Christian and {Ribas}, Ignasi and {Seager}, Sara and {S{\'e}gransan}, Damien and {Scandariato}, Gaetano and {Simon}, Attila E. and {Smith}, Alexis M.~S. and {Stalport}, Manu and {Steller}, Manfred and {Szab{\'o}}, Gyula and {Thomas}, Nicolas and {Pritchard}, Tyler A. and {Udry}, St{\'e}phane and {Corral Van Damme}, Carlos and {Van Grootel}, Val{\'e}rie and {Villaver}, Eva and {Walter}, Ingo and {Walton}, Nicholas and {Watkins}, Cristilyn N. and {West}, Richard G.},
        title = "{Discovery of two warm mini-Neptunes with contrasting densities orbiting the young K3V star TOI-815}",
      journal = {\aap},
         year = 2024,
        month = may,
       volume = {685},
          eid = {A5},
        pages = {A5},
          doi = {10.1051/0004-6361/202348678},
archivePrefix = {arXiv},
       eprint = {2401.15709},
 primaryClass = {astro-ph.EP},
       adsurl = {https://ui.adsabs.harvard.edu/abs/2024A&A...685A...5P}
}

@ARTICLE{Bonomo2023,
       author = {{Bonomo}, A.~S. and {Dumusque}, X. and {Massa}, A. and {Mortier}, A. and {Bongiolatti}, R. and {Malavolta}, L. and {Sozzetti}, A. and {Buchhave}, L.~A. and {Damasso}, M. and {Haywood}, R.~D. and {Morbidelli}, A. and {Latham}, D.~W. and {Molinari}, E. and {Pepe}, F. and {Poretti}, E. and {Udry}, S. and {Affer}, L. and {Boschin}, W. and {Charbonneau}, D. and {Cosentino}, R. and {Cretignier}, M. and {Ghedina}, A. and {Lega}, E. and {L{\'o}pez-Morales}, M. and {Margini}, M. and {Mart{\'\i}nez Fiorenzano}, A.~F. and {Mayor}, M. and {Micela}, G. and {Pedani}, M. and {Pinamonti}, M. and {Rice}, K. and {Sasselov}, D. and {Tronsgaard}, R. and {Vanderburg}, A.},
        title = "{Cold Jupiters and improved masses in 38 Kepler and K2 small planet systems from 3661 HARPS-N radial velocities. No excess of cold Jupiters in small planet systems}",
      journal = {\aap},
         year = 2023,
        month = sep,
       volume = {677},
          eid = {A33},
        pages = {A33},
          doi = {10.1051/0004-6361/202346211},
archivePrefix = {arXiv},
       eprint = {2304.05773},
 primaryClass = {astro-ph.EP},
       adsurl = {https://ui.adsabs.harvard.edu/abs/2023A&A...677A..33B}
}

@ARTICLE{zeng_2019_mr_tracks,
       author = {{Zeng}, Li and {Jacobsen}, Stein B. and {Sasselov}, Dimitar D. and {Petaev}, Michail I. and {Vanderburg}, Andrew and {Lopez-Morales}, Mercedes and {Perez-Mercader}, Juan and {Mattsson}, Thomas R. and {Li}, Gongjie and {Heising}, Matthew Z. and {Bonomo}, Aldo S. and {Damasso}, Mario and {Berger}, Travis A. and {Cao}, Hao and {Levi}, Amit and {Wordsworth}, Robin D.},
        title = "{Growth model interpretation of planet size distribution}",
      journal = {Proceedings of the National Academy of Science},
         year = 2019,
        month = may,
       volume = {116},
       number = {20},
        pages = {9723-9728},
          doi = {10.1073/pnas.1812905116},
archivePrefix = {arXiv},
       eprint = {1906.04253},
 primaryClass = {astro-ph.EP},
       adsurl = {https://ui.adsabs.harvard.edu/abs/2019PNAS..116.9723Z}
}

@ARTICLE{lanza_2010_activity_harmonics,
       author = {{Lanza}, A.~F. and {Bonomo}, A.~S. and {Moutou}, C. and {Pagano}, I. and {Messina}, S. and {Leto}, G. and {Cutispoto}, G. and {Aigrain}, S. and {Alonso}, R. and {Barge}, P. and {Deleuil}, M. and {Auvergne}, M. and {Baglin}, A. and {Collier Cameron}, A.},
        title = "{Photospheric activity, rotation, and radial velocity variations of the planet-hosting star CoRoT-7}",
      journal = {\aap},
         year = 2010,
        month = sep,
       volume = {520},
          eid = {A53},
        pages = {A53},
          doi = {10.1051/0004-6361/201014403},
archivePrefix = {arXiv},
       eprint = {1005.3602},
 primaryClass = {astro-ph.SR},
       adsurl = {https://ui.adsabs.harvard.edu/abs/2010A&A...520A..53L}
}

@ARTICLE{torres2010,
       author = {{Torres}, Guillermo},
        title = "{On the Use of Empirical Bolometric Corrections for Stars}",
      journal = {\aj},
         year = 2010,
        month = nov,
       volume = {140},
       number = {5},
        pages = {1158-1162},
          doi = {10.1088/0004-6256/140/5/1158},
archivePrefix = {arXiv},
       eprint = {1008.3913},
 primaryClass = {astro-ph.SR},
       adsurl = {https://ui.adsabs.harvard.edu/abs/2010AJ....140.1158T}
}

@ARTICLE{Schlegel1998,
       author = {{Schlegel}, David J. and {Finkbeiner}, Douglas P. and {Davis}, Marc},
        title = "{Maps of Dust Infrared Emission for Use in Estimation of Reddening and Cosmic Microwave Background Radiation Foregrounds}",
      journal = {\apj},
         year = 1998,
        month = jun,
       volume = {500},
       number = {2},
        pages = {525-553},
          doi = {10.1086/305772},
archivePrefix = {arXiv},
       eprint = {astro-ph/9710327},
 primaryClass = {astro-ph},
       adsurl = {https://ui.adsabs.harvard.edu/abs/1998ApJ...500..525S}
}

@article{Stassun_2017,
doi = {10.3847/1538-3881/aa5df3},
url = {https://dx.doi.org/10.3847/1538-3881/aa5df3},
year = {2017},
month = {mar},
publisher = {The American Astronomical Society},
volume = {153},
number = {3},
pages = {136},
author = {Stassun, Keivan G. and Collins, Karen A. and Gaudi, B. Scott},
title = {Accurate Empirical Radii and Masses of Planets and Their Host Stars with Gaia Parallaxes},
journal = {\aj}
}

@ARTICLE{Sozzetti2023,
       author = {{Sozzetti}, A.},
        title = "{A dynamical mass for GJ 463 b: A massive super-Jupiter companion beyond the snow line of a nearby M dwarf}",
      journal = {\aap},
         year = 2023,
        month = feb,
       volume = {670},
          eid = {L17},
        pages = {L17},
          doi = {10.1051/0004-6361/202245454},
archivePrefix = {arXiv},
       eprint = {2302.00413},
 primaryClass = {astro-ph.EP},
       adsurl = {https://ui.adsabs.harvard.edu/abs/2023A&A...670L..17S}
}

@ARTICLE{Lindegren2018,
       author = {{Lindegren}, L. and {Hern{\'a}ndez}, J. and {Bombrun}, A. and {Klioner}, S. and {Bastian}, U. and {Ramos-Lerate}, M. and {de Torres}, A. and {Steidelm{\"u}ller}, H. and {Stephenson}, C. and {Hobbs}, D. and {Lammers}, U. and {Biermann}, M. and {Geyer}, R. and {Hilger}, T. and {Michalik}, D. and {Stampa}, U. and {McMillan}, P.~J. and {Casta{\~n}eda}, J. and {Clotet}, M. and {Comoretto}, G. and {Davidson}, M. and {Fabricius}, C. and {Gracia}, G. and {Hambly}, N.~C. and {Hutton}, A. and {Mora}, A. and {Portell}, J. and {van Leeuwen}, F. and {Abbas}, U. and {Abreu}, A. and {Altmann}, M. and {Andrei}, A. and {Anglada}, E. and {Balaguer-N{\'u}{\~n}ez}, L. and {Barache}, C. and {Becciani}, U. and {Bertone}, S. and {Bianchi}, L. and {Bouquillon}, S. and {Bourda}, G. and {Br{\"u}semeister}, T. and {Bucciarelli}, B. and {Busonero}, D. and {Buzzi}, R. and {Cancelliere}, R. and {Carlucci}, T. and {Charlot}, P. and {Cheek}, N. and {Crosta}, M. and {Crowley}, C. and {de Bruijne}, J. and {de Felice}, F. and {Drimmel}, R. and {Esquej}, P. and {Fienga}, A. and {Fraile}, E. and {Gai}, M. and {Garralda}, N. and {Gonz{\'a}lez-Vidal}, J.~J. and {Guerra}, R. and {Hauser}, M. and {Hofmann}, W. and {Holl}, B. and {Jordan}, S. and {Lattanzi}, M.~G. and {Lenhardt}, H. and {Liao}, S. and {Licata}, E. and {Lister}, T. and {L{\"o}ffler}, W. and {Marchant}, J. and {Martin-Fleitas}, J. -M. and {Messineo}, R. and {Mignard}, F. and {Morbidelli}, R. and {Poggio}, E. and {Riva}, A. and {Rowell}, N. and {Salguero}, E. and {Sarasso}, M. and {Sciacca}, E. and {Siddiqui}, H. and {Smart}, R.~L. and {Spagna}, A. and {Steele}, I. and {Taris}, F. and {Torra}, J. and {van Elteren}, A. and {van Reeven}, W. and {Vecchiato}, A.},
        title = "{Gaia Data Release 2. The astrometric solution}",
      journal = {\aap},
         year = 2018,
        month = aug,
       volume = {616},
          eid = {A2},
        pages = {A2},
          doi = {10.1051/0004-6361/201832727},
archivePrefix = {arXiv},
       eprint = {1804.09366},
 primaryClass = {astro-ph.IM},
       adsurl = {https://ui.adsabs.harvard.edu/abs/2018A&A...616A...2L}
}

@ARTICLE{Lindegren2021,
       author = {{Lindegren}, L. and {Klioner}, S.~A. and {Hern{\'a}ndez}, J. and {Bombrun}, A. and {Ramos-Lerate}, M. and {Steidelm{\"u}ller}, H. and {Bastian}, U. and {Biermann}, M. and {de Torres}, A. and {Gerlach}, E. and {Geyer}, R. and {Hilger}, T. and {Hobbs}, D. and {Lammers}, U. and {McMillan}, P.~J. and {Stephenson}, C.~A. and {Casta{\~n}eda}, J. and {Davidson}, M. and {Fabricius}, C. and {Gracia-Abril}, G. and {Portell}, J. and {Rowell}, N. and {Teyssier}, D. and {Torra}, F. and {Bartolom{\'e}}, S. and {Clotet}, M. and {Garralda}, N. and {Gonz{\'a}lez-Vidal}, J.~J. and {Torra}, J. and {Abbas}, U. and {Altmann}, M. and {Anglada Varela}, E. and {Balaguer-N{\'u}{\~n}ez}, L. and {Balog}, Z. and {Barache}, C. and {Becciani}, U. and {Bernet}, M. and {Bertone}, S. and {Bianchi}, L. and {Bouquillon}, S. and {Brown}, A.~G.~A. and {Bucciarelli}, B. and {Busonero}, D. and {Butkevich}, A.~G. and {Buzzi}, R. and {Cancelliere}, R. and {Carlucci}, T. and {Charlot}, P. and {Cioni}, M. -R.~L. and {Crosta}, M. and {Crowley}, C. and {del Peloso}, E.~F. and {del Pozo}, E. and {Drimmel}, R. and {Esquej}, P. and {Fienga}, A. and {Fraile}, E. and {Gai}, M. and {Garcia-Reinaldos}, M. and {Guerra}, R. and {Hambly}, N.~C. and {Hauser}, M. and {Jan{\ss}en}, K. and {Jordan}, S. and {Kostrzewa-Rutkowska}, Z. and {Lattanzi}, M.~G. and {Liao}, S. and {Licata}, E. and {Lister}, T.~A. and {L{\"o}ffler}, W. and {Marchant}, J.~M. and {Masip}, A. and {Mignard}, F. and {Mints}, A. and {Molina}, D. and {Mora}, A. and {Morbidelli}, R. and {Murphy}, C.~P. and {Pagani}, C. and {Panuzzo}, P. and {Pe{\~n}alosa Esteller}, X. and {Poggio}, E. and {Re Fiorentin}, P. and {Riva}, A. and {Sagrist{\`a} Sell{\'e}s}, A. and {Sanchez Gimenez}, V. and {Sarasso}, M. and {Sciacca}, E. and {Siddiqui}, H.~I. and {Smart}, R.~L. and {Souami}, D. and {Spagna}, A. and {Steele}, I.~A. and {Taris}, F. and {Utrilla}, E. and {van Reeven}, W. and {Vecchiato}, A.},
        title = "{Gaia Early Data Release 3. The astrometric solution}",
      journal = {\aap},
         year = 2021,
        month = may,
       volume = {649},
          eid = {A2},
        pages = {A2},
          doi = {10.1051/0004-6361/202039709},
archivePrefix = {arXiv},
       eprint = {2012.03380},
 primaryClass = {astro-ph.IM},
       adsurl = {https://ui.adsabs.harvard.edu/abs/2021A&A...649A...2L}
}

@ARTICLE{stassun2018_parallaxes,
       author = {{Stassun}, Keivan G. and {Torres}, Guillermo},
        title = "{Evidence for a Systematic Offset of -80 {\ensuremath{\mu}}as in the Gaia DR2 Parallaxes}",
      journal = {\apj},
         year = 2018,
        month = jul,
       volume = {862},
       number = {1},
          eid = {61},
        pages = {61},
          doi = {10.3847/1538-4357/aacafc},
archivePrefix = {arXiv},
       eprint = {1805.03526},
 primaryClass = {astro-ph.SR},
       adsurl = {https://ui.adsabs.harvard.edu/abs/2018ApJ...862...61S}
}

@article{Stassun_2016,
doi = {10.3847/2041-8205/831/1/L6},
url = {https://dx.doi.org/10.3847/2041-8205/831/1/L6},
year = {2016},
month = {oct},
publisher = {The American Astronomical Society},
volume = {831},
number = {1},
pages = {L6},
author = {Stassun, Keivan G. and Torres, Guillermo},
title = {EVIDENCE FOR A SYSTEMATIC OFFSET OF −0.25 mas IN THE GAIA DR1 PARALLAXES},
journal = {\apjl}
}

@ARTICLE{stassuntorres:2021,
       author = {{Stassun}, Keivan G. and {Torres}, Guillermo},
        title = "{Parallax Systematics and Photocenter Motions of Benchmark Eclipsing Binaries in Gaia EDR3}",
      journal = {\apjl},
         year = 2021,
        month = feb,
       volume = {907},
       number = {2},
          eid = {L33},
        pages = {L33},
          doi = {10.3847/2041-8213/abdaad},
archivePrefix = {arXiv},
       eprint = {2101.03425},
 primaryClass = {astro-ph.SR},
       adsurl = {https://ui.adsabs.harvard.edu/abs/2021ApJ...907L..33S}
}

@article{Mallorquin2023_toi1726,
	author = {{Mallorquín}, M. and {Béjar}, V. J. S. and {Lodieu}, N. and {Zapatero Osorio, M. R.} and {Tabernero, H.} and {Suárez Mascareño, A.} and {Zechmeister, M.} and {Luque, R.} and {Pallé, E.} and {Montes, D.}},
	title = {Dynamical masses of two young transiting sub-Neptunes orbiting HD 63433},
	DOI= "10.1051/0004-6361/202245397",
	url= "https://doi.org/10.1051/0004-6361/202245397",
	journal = {\aap},
	year = 2023,
	volume = 671,
	pages = "A163"
}

@article{bonomo2025,
	author = {{Bonomo}, A.S. and {Borsato, L.} and {Rajpaul, V. M.} and {Zeng, L.} and {Damasso, M.} and {Hara, N. C.} and {Cretignier, M.} and {Leleu, A.} and {Unger, N.} and {Dumusque, X.} and {Lienhard, F.} and {Mortier, A.} and {Naponiello, L.} and {Malavolta, L.} and {Sozzetti, A.} and {Latham, D. W.} and {Rice, K.} and {Bongiolatti, R.} and {Buchhave, L.} and {Cameron, A. C.} and {Fiorenzano, A. F.} and {Ghedina, A.} and {Haywood, R. D.} and {Lacedelli, G.} and {Massa, A.} and {Pepe, F.} and {Poretti, E.} and {Udry, S.}},
	title = {In-depth characterization of the Kepler-10 three-planet system with HARPS-N radial velocities and Kepler transit timing variations},
	DOI= "10.1051/0004-6361/202453026",
	url= "https://doi.org/10.1051/0004-6361/202453026",
	journal = {\aap},
	year = 2025,
	volume = 696,
	pages = "A233",
}

@article{Ford_2006_hem,
doi = {10.1086/500734},
url = {https://dx.doi.org/10.1086/500734},
year = {2006},
month = {jan},
publisher = {},
volume = {638},
number = {1},
pages = {L45},
author = {Ford, Eric B. and Rasio, Frederic A.},
title = {On the Relation between Hot Jupiters and the Roche Limit},
journal = {\apj}
}

@ARTICLE{burn_waterworlds_2024,
       author = {{Burn}, Remo and {Mordasini}, Christoph and {Mishra}, Lokesh and {Haldemann}, Jonas and {Venturini}, Julia and {Emsenhuber}, Alexandre and {Henning}, Thomas},
        title = "{A radius valley between migrated steam worlds and evaporated rocky cores}",
      journal = {Nature Astronomy},
         year = 2024,
        month = apr,
       volume = {8},
        pages = {463-471},
          doi = {10.1038/s41550-023-02183-7},
archivePrefix = {arXiv},
       eprint = {2401.04380},
 primaryClass = {astro-ph.EP},
       adsurl = {https://ui.adsabs.harvard.edu/abs/2024NatAs...8..463B}
}

@ARTICLE{maxted2018_ldc_unc,
       author = {{Maxted}, P.~F.~L.},
        title = "{Comparison of the power-2 limb-darkening law from the STAGGER-grid to Kepler light curves of transiting exoplanets}",
      journal = {\aap},
         year = 2018,
        month = aug,
       volume = {616},
          eid = {A39},
        pages = {A39},
          doi = {10.1051/0004-6361/201832944},
archivePrefix = {arXiv},
       eprint = {1804.07943},
 primaryClass = {astro-ph.SR},
       adsurl = {https://ui.adsabs.harvard.edu/abs/2018A&A...616A..39M}
}

@article{Patel_espinoza_2022_ldc,
doi = {10.3847/1538-3881/ac5f55},
url = {https://dx.doi.org/10.3847/1538-3881/ac5f55},
year = {2022},
month = {apr},
publisher = {The American Astronomical Society},
volume = {163},
number = {5},
pages = {228},
author = {Jayshil A. Patel and Néstor Espinoza},
title = {Empirical Limb-darkening Coefficients and Transit Parameters of Known Exoplanets from TESS},
journal = {\aj}
}

@ARTICLE{eastman_2013,
       author = {{Eastman}, Jason and {Gaudi}, B. Scott and {Agol}, Eric},
        title = "{EXOFAST: A Fast Exoplanetary Fitting Suite in IDL}",
      journal = {\pasp},
         year = 2013,
        month = jan,
       volume = {125},
       number = {923},
        pages = {83},
          doi = {10.1086/669497},
archivePrefix = {arXiv},
       eprint = {1206.5798},
 primaryClass = {astro-ph.IM},
       adsurl = {https://ui.adsabs.harvard.edu/abs/2013PASP..125...83E}
}

@ARTICLE{filomenoetal2024,
       author = {{Filomeno}, S. and {Biazzo}, K. and {Baratella}, M. and {Benatti}, S. and {D'Orazi}, V. and {Desidera}, S. and {Mancini}, L. and {Messina}, S. and {Polychroni}, D. and {Turrini}, D. and {Cabona}, L. and {Carleo}, I. and {Damasso}, M. and {Malavolta}, L. and {Mantovan}, G. and {Nardiello}, D. and {Scandariato}, G. and {Sozzetti}, A. and {Zingales}, T. and {Andreuzzi}, G. and {Antoniucci}, S. and {Bignamini}, A. and {Bonomo}, A.~S. and {Claudi}, R. and {Cosentino}, R. and {Fiorenzano}, A.~F.~M. and {Fonte}, S. and {Harutyunyan}, A. and {Knapic}, C.},
        title = "{The GAPS Programme at TNG: LXI. Atmospheric parameters and elemental abundances of TESS young exoplanet host stars}",
      journal = {\aap},
         year = 2024,
        month = oct,
       volume = {690},
          eid = {A370},
        pages = {A370},
          doi = {10.1051/0004-6361/202450611},
archivePrefix = {arXiv},
       eprint = {2409.00675},
 primaryClass = {astro-ph.SR},
       adsurl = {https://ui.adsabs.harvard.edu/abs/2024A&A...690A.370F}
}

@article{Rainer_2023,
   title={The GAPS programme at TNG: XLIV. Projected rotational velocities of 273 exoplanet-host stars observed with HARPS-N},
   volume={676},
   ISSN={1432-0746},
   url={http://dx.doi.org/10.1051/0004-6361/202245564},
   DOI={10.1051/0004-6361/202245564},
   journal={A\&A},
   publisher={EDP Sciences},
   author={Rainer, M. and Desidera, S. and Borsa, F. and Barbato, D. and Biazzo, K. and Bonomo, A. and Gratton, R. and Messina, S. and Scandariato, G. and Affer, L. and Benatti, S. and Carleo, I. and Cabona, L. and Covino, E. and Lanza, A. F. and Ligi, R. and Maldonado, J. and Mancini, L. and Nardiello, D. and Sicilia, D. and Sozzetti, A. and Bignamini, A. and Cosentino, R. and Knapic, C. and Martínez Fiorenzano, A. F. and Molinari, E. and Pedani, M. and Poretti, E.},
   year={2023},
   month=aug, 
   pages={A90} 
}

@ARTICLE{zechmeister_2009_gls,
       author = {{Zechmeister}, M. and {K{\"u}rster}, M.},
        title = "{The generalised Lomb-Scargle periodogram. A new formalism for the floating-mean and Keplerian periodograms}",
      journal = {\aap},
         year = 2009,
        month = mar,
       volume = {496},
       number = {2},
        pages = {577-584},
          doi = {10.1051/0004-6361:200811296},
archivePrefix = {arXiv},
       eprint = {0901.2573},
 primaryClass = {astro-ph.IM},
       adsurl = {https://ui.adsabs.harvard.edu/abs/2009A&A...496..577Z}
}

@ARTICLE{pecaut_mamaj_2013_st_type_table,
       author = {{Pecaut}, Mark J. and {Mamajek}, Eric E.},
        title = "{Intrinsic Colors, Temperatures, and Bolometric Corrections of Pre-main-sequence Stars}",
      journal = {\apjs},
         year = 2013,
        month = sep,
       volume = {208},
       number = {1},
          eid = {9},
        pages = {9},
          doi = {10.1088/0067-0049/208/1/9},
archivePrefix = {arXiv},
       eprint = {1307.2657},
 primaryClass = {astro-ph.SR},
       adsurl = {https://ui.adsabs.harvard.edu/abs/2013ApJS..208....9P}
}

@ARTICLE{mamajek_age_est_2008,
       author = {{Mamajek}, Eric E. and {Hillenbrand}, Lynne A.},
        title = "{Improved Age Estimation for Solar-Type Dwarfs Using Activity-Rotation Diagnostics}",
      journal = {\apj},
         year = 2008,
        month = nov,
       volume = {687},
       number = {2},
        pages = {1264-1293},
          doi = {10.1086/591785},
archivePrefix = {arXiv},
       eprint = {0807.1686},
 primaryClass = {astro-ph},
       adsurl = {https://ui.adsabs.harvard.edu/abs/2008ApJ...687.1264M}
}

@INPROCEEDINGS{jenkins_2016_spoc,
       author = {{Jenkins}, Jon M. and {Twicken}, Joseph D. and {McCauliff}, Sean and {Campbell}, Jennifer and {Sanderfer}, Dwight and {Lung}, David and {Mansouri-Samani}, Masoud and {Girouard}, Forrest and {Tenenbaum}, Peter and {Klaus}, Todd and {Smith}, Jeffrey C. and {Caldwell}, Douglas A. and {Chacon}, A.~D. and {Henze}, Christopher and {Heiges}, Cory and {Latham}, David W. and {Morgan}, Edward and {Swade}, Daryl and {Rinehart}, Stephen and {Vanderspek}, Roland},
        title = "{The TESS science processing operations center}",
    booktitle = {Software and Cyberinfrastructure for Astronomy IV},
         year = 2016,
       editor = {{Chiozzi}, Gianluca and {Guzman}, Juan C.},
       series = {SPIE Conference Series},
       volume = {9913},
        month = aug,
          eid = {99133E},
        pages = {99133E},
          doi = {10.1117/12.2233418},
       adsurl = {https://ui.adsabs.harvard.edu/abs/2016SPIE.9913E..3EJ}
}

@ARTICLE{baliunas_1995_mtwilson,
       author = {{Baliunas}, S.~L. and {Donahue}, R.~A. and {Soon}, W.~H. and {Horne}, J.~H. and {Frazer}, J. and {Woodard-Eklund}, L. and {Bradford}, M. and {Rao}, L.~M. and {Wilson}, O.~C. and {Zhang}, Q. and {Bennett}, W. and {Briggs}, J. and {Carroll}, S.~M. and {Duncan}, D.~K. and {Figueroa}, D. and {Lanning}, H.~H. and {Misch}, T. and {Mueller}, J. and {Noyes}, R.~W. and {Poppe}, D. and {Porter}, A.~C. and {Robinson}, C.~R. and {Russell}, J. and {Shelton}, J.~C. and {Soyumer}, T. and {Vaughan}, A.~H. and {Whitney}, J.~H.},
        title = "{Chromospheric Variations in Main-Sequence Stars. II.}",
      journal = {\apj},
         year = 1995,
        month = jan,
       volume = {438},
        pages = {269},
          doi = {10.1086/175072},
       adsurl = {https://ui.adsabs.harvard.edu/abs/1995ApJ...438..269B}
}

@ARTICLE{seval_rvcalc_zechmeister_2018,
       author = {{Zechmeister}, M. and {Reiners}, A. and {Amado}, P.~J. and {Azzaro}, M. and {Bauer}, F.~F. and {B{\'e}jar}, V.~J.~S. and {Caballero}, J.~A. and {Guenther}, E.~W. and {Hagen}, H. -J. and {Jeffers}, S.~V. and {Kaminski}, A. and {K{\"u}rster}, M. and {Launhardt}, R. and {Montes}, D. and {Morales}, J.~C. and {Quirrenbach}, A. and {Reffert}, S. and {Ribas}, I. and {Seifert}, W. and {Tal-Or}, L. and {Wolthoff}, V.},
        title = "{Spectrum radial velocity analyser (SERVAL). High-precision radial velocities and two alternative spectral indicators}",
      journal = {\aap},
         year = 2018,
        month = jan,
       volume = {609},
          eid = {A12},
        pages = {A12},
          doi = {10.1051/0004-6361/201731483},
archivePrefix = {arXiv},
       eprint = {1710.10114},
 primaryClass = {astro-ph.IM},
       adsurl = {https://ui.adsabs.harvard.edu/abs/2018A&A...609A..12Z}
}

@ARTICLE{dutra_ferreira_2016_microturb_rel,
       author = {{Dutra-Ferreira}, L. and {Pasquini}, L. and {Smiljanic}, R. and {Porto de Mello}, G.~F. and {Steffen}, M.},
        title = "{Consistent metallicity scale for cool dwarfs and giants. A benchmark test using the Hyades}",
      journal = {\aap},
         year = 2016,
        month = jan,
       volume = {585},
          eid = {A75},
        pages = {A75},
          doi = {10.1051/0004-6361/201526783},
archivePrefix = {arXiv},
       eprint = {1509.07725},
 primaryClass = {astro-ph.SR},
       adsurl = {https://ui.adsabs.harvard.edu/abs/2016A&A...585A..75D}
}

@ARTICLE{claret_2018_ldc_fsm,
       author = {{Claret}, Antonio},
        title = "{A new method to compute limb-darkening coefficients for stellar atmosphere models with spherical symmetry: the space missions TESS, Kepler, CoRoT, and MOST}",
      journal = {\aap},
         year = 2018,
        month = oct,
       volume = {618},
          eid = {A20},
        pages = {A20},
          doi = {10.1051/0004-6361/201833060},
archivePrefix = {arXiv},
       eprint = {1804.10135},
 primaryClass = {astro-ph.SR},
       adsurl = {https://ui.adsabs.harvard.edu/abs/2018A&A...618A..20C}
}

@ARTICLE{Gaia_dr3_2023_j,
       author = {{Gaia Collaboration} and {Vallenari}, A. and {Brown}, A.~G.~A. and {Prusti}, T. and {de Bruijne}, J.~H.~J. and {Arenou}, F. and {Babusiaux}, C. and {Biermann}, M. and {Creevey}, O.~L. and {Ducourant}, C. and {Evans}, D.~W. and {Eyer}, L. and {Guerra}, R. and {Hutton}, A. and {Jordi}, C. and {Klioner}, S.~A. and {Lammers}, U.~L. and {Lindegren}, L. and {Luri}, X. and {Mignard}, F. and {Panem}, C. and {Pourbaix}, D. and {Randich}, S. and {Sartoretti}, P. and {Soubiran}, C. and {Tanga}, P. and {Walton}, N.~A. and {Bailer-Jones}, C.~A.~L. and {Bastian}, U. and {Drimmel}, R. and {Jansen}, F. and {Katz}, D. and {Lattanzi}, M.~G. and {van Leeuwen}, F. and {Bakker}, J. and {Cacciari}, C. and {Casta{\~n}eda}, J. and {De Angeli}, F. and {Fabricius}, C. and {Fouesneau}, M. and {Fr{\'e}mat}, Y. and {Galluccio}, L. and {Guerrier}, A. and {Heiter}, U. and {Masana}, E. and {Messineo}, R. and {Mowlavi}, N. and {Nicolas}, C. and {Nienartowicz}, K. and {Pailler}, F. and {Panuzzo}, P. and {Riclet}, F. and {Roux}, W. and {Seabroke}, G.~M. and {Sordo}, R. and {Th{\'e}venin}, F. and {Gracia-Abril}, G. and {Portell}, J. and {Teyssier}, D. and {Altmann}, M. and {Andrae}, R. and {Audard}, M. and {Bellas-Velidis}, I. and {Benson}, K. and {Berthier}, J. and {Blomme}, R. and {Burgess}, P.~W. and {Busonero}, D. and {Busso}, G. and {C{\'a}novas}, H. and {Carry}, B. and {Cellino}, A. and {Cheek}, N. and {Clementini}, G. and {Damerdji}, Y. and {Davidson}, M. and {de Teodoro}, P. and {Nu{\~n}ez Campos}, M. and {Delchambre}, L. and {Dell'Oro}, A. and {Esquej}, P. and {Fern{\'a}ndez-Hern{\'a}ndez}, J. and {Fraile}, E. and {Garabato}, D. and {Garc{\'\i}a-Lario}, P. and {Gosset}, E. and {Haigron}, R. and {Halbwachs}, J. -L. and {Hambly}, N.~C. and {Harrison}, D.~L. and {Hern{\'a}ndez}, J. and {Hestroffer}, D. and {Hodgkin}, S.~T. and {Holl}, B. and {Jan{\ss}en}, K. and {Jevardat de Fombelle}, G. and {Jordan}, S. and {Krone-Martins}, A. and {Lanzafame}, A.~C. and {L{\"o}ffler}, W. and {Marchal}, O. and {Marrese}, P.~M. and {Moitinho}, A. and {Muinonen}, K. and {Osborne}, P. and {Pancino}, E. and {Pauwels}, T. and {Recio-Blanco}, A. and {Reyl{\'e}}, C. and {Riello}, M. and {Rimoldini}, L. and {Roegiers}, T. and {Rybizki}, J. and {Sarro}, L.~M. and {Siopis}, C. and {Smith}, M. and {Sozzetti}, A. and {Utrilla}, E. and {van Leeuwen}, M. and {Abbas}, U. and {{\'A}brah{\'a}m}, P. and {Abreu Aramburu}, A. and {Aerts}, C. and {Aguado}, J.~J. and {Ajaj}, M. and {Aldea-Montero}, F. and {Altavilla}, G. and {{\'A}lvarez}, M.~A. and {Alves}, J. and {Anders}, F. and {Anderson}, R.~I. and {Anglada Varela}, E. and {Antoja}, T. and {Baines}, D. and {Baker}, S.~G. and {Balaguer-N{\'u}{\~n}ez}, L. and {Balbinot}, E. and {Balog}, Z. and {Barache}, C. and {Barbato}, D. and {Barros}, M. and {Barstow}, M.~A. and {Bartolom{\'e}}, S. and {Bassilana}, J. -L. and {Bauchet}, N. and {Becciani}, U. and {Bellazzini}, M. and {Berihuete}, A. and {Bernet}, M. and {Bertone}, S. and {Bianchi}, L. and {Binnenfeld}, A. and {Blanco-Cuaresma}, S. and {Blazere}, A. and {Boch}, T. and {Bombrun}, A. and {Bossini}, D. and {Bouquillon}, S. and {Bragaglia}, A. and {Bramante}, L. and {Breedt}, E. and {Bressan}, A. and {Brouillet}, N. and {Brugaletta}, E. and {Bucciarelli}, B. and {Burlacu}, A. and {Butkevich}, A.~G. and {Buzzi}, R. and {Caffau}, E. and {Cancelliere}, R. and {Cantat-Gaudin}, T. and {Carballo}, R. and {Carlucci}, T. and {Carnerero}, M.~I. and {Carrasco}, J.~M. and {Casamiquela}, L. and {Castellani}, M. and {Castro-Ginard}, A. and {Chaoul}, L. and {Charlot}, P. and {Chemin}, L. and {Chiaramida}, V. and {Chiavassa}, A. and {Chornay}, N. and {Comoretto}, G. and {Contursi}, G. and {Cooper}, W.~J. and {Cornez}, T. and {Cowell}, S. and {Crifo}, F. and {Cropper}, M. and {Crosta}, M. and {Crowley}, C. and {Dafonte}, C. and {Dapergolas}, A. and {David}, M. and {David}, P. and {de Laverny}, P. and {De Luise}, F. and {De March}, R. and {De Ridder}, J. and {de Souza}, R. and {de Torres}, A. and {del Peloso}, E.~F. and {del Pozo}, E. and {Delbo}, M. and {Delgado}, A. and {Delisle}, J. -B. and {Demouchy}, C. and {Dharmawardena}, T.~E. and {Di Matteo}, P. and {Diakite}, S. and {Diener}, C. and {Distefano}, E. and {Dolding}, C. and {Edvardsson}, B. and {Enke}, H. and {Fabre}, C. and {Fabrizio}, M. and {Faigler}, S. and {Fedorets}, G. and {Fernique}, P. and {Fienga}, A. and {Figueras}, F. and {Fournier}, Y. and {Fouron}, C. and {Fragkoudi}, F. and {Gai}, M. and {Garcia-Gutierrez}, A. and {Garcia-Reinaldos}, M. and {Garc{\'\i}a-Torres}, M. and {Garofalo}, A. and {Gavel}, A. and {Gavras}, P. and {Gerlach}, E. and {Geyer}, R. and {Giacobbe}, P. and {Gilmore}, G. and {Girona}, S. and {Giuffrida}, G. and {Gomel}, R. and {Gomez}, A. and {Gonz{\'a}lez-N{\'u}{\~n}ez}, J. and {Gonz{\'a}lez-Santamar{\'\i}a}, I. and {Gonz{\'a}lez-Vidal}, J.~J. and {Granvik}, M. and {Guillout}, P. and {Guiraud}, J. and {Guti{\'e}rrez-S{\'a}nchez}, R. and {Guy}, L.~P. and {Hatzidimitriou}, D. and {Hauser}, M. and {Haywood}, M. and {Helmer}, A. and {Helmi}, A. and {Sarmiento}, M.~H. and {Hidalgo}, S.~L. and {Hilger}, T. and {H{\l}adczuk}, N. and {Hobbs}, D. and {Holland}, G. and {Huckle}, H.~E. and {Jardine}, K. and {Jasniewicz}, G. and {Jean-Antoine Piccolo}, A. and {Jim{\'e}nez-Arranz}, {\'O}. and {Jorissen}, A. and {Juaristi Campillo}, J. and {Julbe}, F. and {Karbevska}, L. and {Kervella}, P. and {Khanna}, S. and {Kontizas}, M. and {Kordopatis}, G. and {Korn}, A.~J. and {K{\'o}sp{\'a}l}, {\'A}. and {Kostrzewa-Rutkowska}, Z. and {Kruszy{\'n}ska}, K. and {Kun}, M. and {Laizeau}, P. and {Lambert}, S. and {Lanza}, A.~F. and {Lasne}, Y. and {Le Campion}, J. -F. and {Lebreton}, Y. and {Lebzelter}, T. and {Leccia}, S. and {Leclerc}, N. and {Lecoeur-Taibi}, I. and {Liao}, S. and {Licata}, E.~L. and {Lindstr{\o}m}, H.~E.~P. and {Lister}, T.~A. and {Livanou}, E. and {Lobel}, A. and {Lorca}, A. and {Loup}, C. and {Madrero Pardo}, P. and {Magdaleno Romeo}, A. and {Managau}, S. and {Mann}, R.~G. and {Manteiga}, M. and {Marchant}, J.~M. and {Marconi}, M. and {Marcos}, J. and {Marcos Santos}, M.~M.~S. and {Mar{\'\i}n Pina}, D. and {Marinoni}, S. and {Marocco}, F. and {Marshall}, D.~J. and {Martin Polo}, L. and {Mart{\'\i}n-Fleitas}, J.~M. and {Marton}, G. and {Mary}, N. and {Masip}, A. and {Massari}, D. and {Mastrobuono-Battisti}, A. and {Mazeh}, T. and {McMillan}, P.~J. and {Messina}, S. and {Michalik}, D. and {Millar}, N.~R. and {Mints}, A. and {Molina}, D. and {Molinaro}, R. and {Moln{\'a}r}, L. and {Monari}, G. and {Mongui{\'o}}, M. and {Montegriffo}, P. and {Montero}, A. and {Mor}, R. and {Mora}, A. and {Morbidelli}, R. and {Morel}, T. and {Morris}, D. and {Muraveva}, T. and {Murphy}, C.~P. and {Musella}, I. and {Nagy}, Z. and {Noval}, L. and {Oca{\~n}a}, F. and {Ogden}, A. and {Ordenovic}, C. and {Osinde}, J.~O. and {Pagani}, C. and {Pagano}, I. and {Palaversa}, L. and {Palicio}, P.~A. and {Pallas-Quintela}, L. and {Panahi}, A. and {Payne-Wardenaar}, S. and {Pe{\~n}alosa Esteller}, X. and {Penttil{\"a}}, A. and {Pichon}, B. and {Piersimoni}, A.~M. and {Pineau}, F. -X. and {Plachy}, E. and {Plum}, G. and {Poggio}, E. and {Pr{\v{s}}a}, A. and {Pulone}, L. and {Racero}, E. and {Ragaini}, S. and {Rainer}, M. and {Raiteri}, C.~M. and {Rambaux}, N. and {Ramos}, P. and {Ramos-Lerate}, M. and {Re Fiorentin}, P. and {Regibo}, S. and {Richards}, P.~J. and {Rios Diaz}, C. and {Ripepi}, V. and {Riva}, A. and {Rix}, H. -W. and {Rixon}, G. and {Robichon}, N. and {Robin}, A.~C. and {Robin}, C. and {Roelens}, M. and {Rogues}, H.~R.~O. and {Rohrbasser}, L. and {Romero-G{\'o}mez}, M. and {Rowell}, N. and {Royer}, F. and {Ruz Mieres}, D. and {Rybicki}, K.~A. and {Sadowski}, G. and {S{\'a}ez N{\'u}{\~n}ez}, A. and {Sagrist{\`a} Sell{\'e}s}, A. and {Sahlmann}, J. and {Salguero}, E. and {Samaras}, N. and {Sanchez Gimenez}, V. and {Sanna}, N. and {Santove{\~n}a}, R. and {Sarasso}, M. and {Schultheis}, M. and {Sciacca}, E. and {Segol}, M. and {Segovia}, J.~C. and {S{\'e}gransan}, D. and {Semeux}, D. and {Shahaf}, S. and {Siddiqui}, H.~I. and {Siebert}, A. and {Siltala}, L. and {Silvelo}, A. and {Slezak}, E. and {Slezak}, I. and {Smart}, R.~L. and {Snaith}, O.~N. and {Solano}, E. and {Solitro}, F. and {Souami}, D. and {Souchay}, J. and {Spagna}, A. and {Spina}, L. and {Spoto}, F. and {Steele}, I.~A. and {Steidelm{\"u}ller}, H. and {Stephenson}, C.~A. and {S{\"u}veges}, M. and {Surdej}, J. and {Szabados}, L. and {Szegedi-Elek}, E. and {Taris}, F. and {Taylor}, M.~B. and {Teixeira}, R. and {Tolomei}, L. and {Tonello}, N. and {Torra}, F. and {Torra}, J. and {Torralba Elipe}, G. and {Trabucchi}, M. and {Tsounis}, A.~T. and {Turon}, C. and {Ulla}, A. and {Unger}, N. and {Vaillant}, M.~V. and {van Dillen}, E. and {van Reeven}, W. and {Vanel}, O. and {Vecchiato}, A. and {Viala}, Y. and {Vicente}, D. and {Voutsinas}, S. and {Weiler}, M. and {Wevers}, T. and {Wyrzykowski}, {\L}. and {Yoldas}, A. and {Yvard}, P. and {Zhao}, H. and {Zorec}, J. and {Zucker}, S. and {Zwitter}, T.},
        title = "{Gaia Data Release 3. Summary of the content and survey properties}",
      journal = {\aap},
         year = 2023,
        month = jun,
       volume = {674},
          eid = {A1},
        pages = {A1},
          doi = {10.1051/0004-6361/202243940},
archivePrefix = {arXiv},
       eprint = {2208.00211},
 primaryClass = {astro-ph.GA},
       adsurl = {https://ui.adsabs.harvard.edu/abs/2023A&A...674A...1G}
}

@ARTICLE{espinoza_2019,
       author = {{Espinoza}, N{\'e}stor and {Kossakowski}, Diana and {Brahm}, Rafael},
        title = "{juliet: a versatile modelling tool for transiting and non-transiting exoplanetary systems}",
      journal = {\mnras},
         year = 2019,
        month = dec,
       volume = {490},
       number = {2},
        pages = {2262-2283},
          doi = {10.1093/mnras/stz2688},
archivePrefix = {arXiv},
       eprint = {1812.08549},
 primaryClass = {astro-ph.EP},
       adsurl = {https://ui.adsabs.harvard.edu/abs/2019MNRAS.490.2262E}
}

@ARTICLE{benatti_2021_intro,
       author = {{Benatti}, S. and {Damasso}, M. and {Borsa}, F. and {Locci}, D. and {Pillitteri}, I. and {Desidera}, S. and {Maggio}, A. and {Micela}, G. and {Wolk}, S. and {Claudi}, R. and {Malavolta}, L. and {Modirrousta-Galian}, D.},
        title = "{Constraints on the mass and on the atmospheric composition and evolution of the low-density young planet DS Tucanae A b}",
      journal = {\aap},
         year = 2021,
        month = jun,
       volume = {650},
          eid = {A66},
        pages = {A66},
          doi = {10.1051/0004-6361/202140416},
archivePrefix = {arXiv},
       eprint = {2103.12922},
 primaryClass = {astro-ph.EP},
       adsurl = {https://ui.adsabs.harvard.edu/abs/2021A&A...650A..66B}
}

@INPROCEEDINGS{chthonian_hebrard_2004,
       author = {{H{\'e}brard}, G. and {Lecavelier Des {\'E}tangs}, A. and {Vidal-Madjar}, A. and {D{\'e}sert}, J. -M. and {Ferlet}, R.},
        title = "{Evaporation Rate of Hot Jupiters and Formation of Chthonian Planets}",
    booktitle = {Extrasolar Planets: Today and Tomorrow},
         year = 2004,
       editor = {{Beaulieu}, J. and {Lecavelier Des Etangs}, A. and {Terquem}, C.},
       series = {Astronomical Society of the Pacific Conference Series},
       volume = {321},
        month = dec,
        pages = {203},
          doi = {10.48550/arXiv.astro-ph/0312384},
archivePrefix = {arXiv},
       eprint = {astro-ph/0312384},
 primaryClass = {astro-ph},
       adsurl = {https://ui.adsabs.harvard.edu/abs/2004ASPC..321..203H}
}

@ARTICLE{Nardiello+_2020_pathos,
       author = {{Nardiello}, D. and {Piotto}, G. and {Deleuil}, M. and {Malavolta}, L. and {Montalto}, M. and {Bedin}, L.~R. and {Borsato}, L. and {Granata}, V. and {Libralato}, M. and {Manthopoulou}, E.~E.},
        title = "{A PSF-based Approach to TESS High quality data Of Stellar clusters (PATHOS) - II. Search for exoplanets in open clusters of the Southern ecliptic hemisphere and their frequency}",
      journal = {\mnras},
         year = 2020,
        month = jul,
       volume = {495},
       number = {4},
        pages = {4924-4942},
          doi = {10.1093/mnras/staa1465},
archivePrefix = {arXiv},
       eprint = {2005.12281},
 primaryClass = {astro-ph.EP},
       adsurl = {https://ui.adsabs.harvard.edu/abs/2020MNRAS.495.4924N}
}

@ARTICLE{nardiello+_2021_pathos,
       author = {{Nardiello}, D. and {Deleuil}, M. and {Mantovan}, G. and {Malavolta}, L. and {Lacedelli}, G. and {Libralato}, M. and {Bedin}, L.~R. and {Borsato}, L. and {Granata}, V. and {Piotto}, G.},
        title = "{A PSF-based Approach to TESS High quality data Of Stellar clusters (PATHOS) - IV. Candidate exoplanets around stars in open clusters: frequency and age-planetary radius distribution}",
      journal = {\mnras},
         year = 2021,
        month = aug,
       volume = {505},
       number = {3},
        pages = {3767-3784},
          doi = {10.1093/mnras/stab1497},
archivePrefix = {arXiv},
       eprint = {2105.09952},
 primaryClass = {astro-ph.EP},
       adsurl = {https://ui.adsabs.harvard.edu/abs/2021MNRAS.505.3767N}
}

@ARTICLE{nardiello+_2019_pathos,
       author = {{Nardiello}, D. and {Borsato}, L. and {Piotto}, G. and {Colombo}, L.~S. and {Manthopoulou}, E.~E. and {Bedin}, L.~R. and {Granata}, V. and {Lacedelli}, G. and {Libralato}, M. and {Malavolta}, L. and {Montalto}, M. and {Nascimbeni}, V.},
        title = "{A PSF-based Approach to TESS High quality data Of Stellar clusters (PATHOS) - I. Search for exoplanets and variable stars in the field of 47 Tuc}",
      journal = {\mnras},
         year = 2019,
        month = dec,
       volume = {490},
       number = {3},
        pages = {3806-3823},
          doi = {10.1093/mnras/stz2878},
archivePrefix = {arXiv},
       eprint = {1910.03592},
 primaryClass = {astro-ph.SR},
       adsurl = {https://ui.adsabs.harvard.edu/abs/2019MNRAS.490.3806N}
}

@article{nardiello_2020_pathos,
    author = {Nardiello, D},
    title = "{A PSF-based Approach to TESS High quality data Of Stellar clusters (PATHOS) – III. Exploring the properties of young associations through their variables, dippers, and candidate exoplanets}",
    journal = {\mnras},
    volume = {498},
    number = {4},
    pages = {5972-5989},
    year = {2020},
    month = {09},
    issn = {0035-8711},
    doi = {10.1093/mnras/staa2745},
    url = {https://doi.org/10.1093/mnras/staa2745},
    eprint = {https://academic.oup.com/mnras/article-pdf/498/4/5972/33838922/staa2745.pdf},
}

@ARTICLE{kipping_2013_limbdark_param,
       author = {{Kipping}, David M.},
        title = "{Efficient, uninformative sampling of limb darkening coefficients for two-parameter laws}",
      journal = {\mnras},
         year = 2013,
        month = nov,
       volume = {435},
       number = {3},
        pages = {2152-2160},
          doi = {10.1093/mnras/stt1435},
archivePrefix = {arXiv},
       eprint = {1308.0009},
 primaryClass = {astro-ph.SR},
       adsurl = {https://ui.adsabs.harvard.edu/abs/2013MNRAS.435.2152K}
}

@ARTICLE{batman_kreidberg_2015,
       author = {{Kreidberg}, Laura},
        title = "{batman: BAsic Transit Model cAlculatioN in Python}",
      journal = {\pasp},
         year = 2015,
        month = nov,
       volume = {127},
       number = {957},
        pages = {1161},
          doi = {10.1086/683602},
archivePrefix = {arXiv},
       eprint = {1507.08285},
 primaryClass = {astro-ph.EP},
       adsurl = {https://ui.adsabs.harvard.edu/abs/2015PASP..127.1161K}
}

@ARTICLE{nested_sampling_speagle_2020,
       author = {{Speagle}, Joshua S.},
        title = "{DYNESTY: a dynamic nested sampling package for estimating Bayesian posteriors and evidences}",
      journal = {\mnras},
         year = 2020,
        month = apr,
       volume = {493},
       number = {3},
        pages = {3132-3158},
          doi = {10.1093/mnras/staa278},
archivePrefix = {arXiv},
       eprint = {1904.02180},
 primaryClass = {astro-ph.IM},
       adsurl = {https://ui.adsabs.harvard.edu/abs/2020MNRAS.493.3132S}
}

@ARTICLE{barragan2022_intro,
       author = {{Barrag{\'a}n}, O. and {Armstrong}, D.~J. and {Gandolfi}, D. and {Carleo}, I. and {Vidotto}, A.~A. and {Villarreal D'Angelo}, C. and {Oklop{\v{c}}i{\'c}}, A. and {Isaacson}, H. and {Oddo}, D. and {Collins}, K. and {Fridlund}, M. and {Sousa}, S.~G. and {Persson}, C.~M. and {Hellier}, C. and {Howell}, S. and {Howard}, A. and {Redfield}, S. and {Eisner}, N. and {Georgieva}, I.~Y. and {Dragomir}, D. and {Bayliss}, D. and {Nielsen}, L.~D. and {Klein}, B. and {Aigrain}, S. and {Zhang}, M. and {Teske}, J. and {Twicken}, J.~D. and {Jenkins}, J. and {Esposito}, M. and {Van Eylen}, V. and {Rodler}, F. and {Adibekyan}, V. and {Alarcon}, J. and {Anderson}, D.~R. and {Akana Murphy}, J.~M. and {Barrado}, D. and {Barros}, S.~C.~C. and {Benneke}, B. and {Bouchy}, F. and {Bryant}, E.~M. and {Butler}, R.~P. and {Burt}, J. and {Cabrera}, J. and {Casewell}, S. and {Chaturvedi}, P. and {Cloutier}, R. and {Cochran}, W.~D. and {Crane}, J. and {Crossfield}, I. and {Crouzet}, N. and {Collins}, K.~I. and {Dai}, F. and {Deeg}, H.~J. and {Deline}, A. and {Demangeon}, O.~D.~S. and {Dumusque}, X. and {Figueira}, P. and {Furlan}, E. and {Gnilka}, C. and {Goad}, M.~R. and {Goffo}, E. and {Guti{\'e}rrez-Canales}, F. and {Hadjigeorghiou}, A. and {Hartman}, Z. and {Hatzes}, A.~P. and {Harris}, M. and {Henderson}, B. and {Hirano}, T. and {Hojjatpanah}, S. and {Hoyer}, S. and {Kab{\'a}th}, P. and {Korth}, J. and {Lillo-Box}, J. and {Luque}, R. and {Marmier}, M. and {Mo{\v{c}}nik}, T. and {Muresan}, A. and {Murgas}, F. and {Nagel}, E. and {Osborne}, H.~L.~M. and {Osborn}, A. and {Osborn}, H.~P. and {Palle}, E. and {Raimbault}, M. and {Ricker}, G.~R. and {Rubenzahl}, R.~A. and {Stockdale}, C. and {Santos}, N.~C. and {Scott}, N. and {Schwarz}, R.~P. and {Shectman}, S. and {Raimbault}, M. and {Seager}, S. and {S{\'e}gransan}, D. and {Serrano}, L.~M. and {Skarka}, M. and {Smith}, A.~M.~S. and {{\v{S}}ubjak}, J. and {Tan}, T.~G. and {Udry}, S. and {Watson}, C. and {Wheatley}, P.~J. and {West}, R. and {Winn}, J.~N. and {Wang}, S.~X. and {Wolfgang}, A. and {Ziegler}, C.},
        title = "{The young HD 73583 (TOI-560) planetary system: two 10-M$_{{\ensuremath{\oplus}}}$ mini-Neptunes transiting a 500-Myr-old, bright, and active K dwarf}",
      journal = {\mnras},
         year = 2022,
        month = aug,
       volume = {514},
       number = {2},
        pages = {1606-1627},
          doi = {10.1093/mnras/stac638},
archivePrefix = {arXiv},
       eprint = {2110.13069},
 primaryClass = {astro-ph.EP},
       adsurl = {https://ui.adsabs.harvard.edu/abs/2022MNRAS.514.1606B}
}

@book{Rasmussen2006GP,
  author = {Rasmussen, Carl Edward and Williams, Christopher K. I.},
  publisher = {The MIT Press},
  title = {Gaussian Processes for Machine Learning},
  year = 2006
}

@article{GP_foreman-mackey_2017_celerite1,
   author = {{Foreman-Mackey}, D. and {Agol}, E. and {Ambikasaran}, S. and
            {Angus}, R.},
    title = "{Fast and Scalable Gaussian Process Modeling with Applications to
              Astronomical Time Series}",
  journal = {\aj},
     year = 2017,
    month = dec,
   volume = 154,
    pages = {220},
      doi = {10.3847/1538-3881/aa9332},
   adsurl = {http://adsabs.harvard.edu/abs/2017AJ....154..220F}
}

@article{celerite2_GP,
   author = {{Foreman-Mackey}, D.},
    title = "{Scalable Backpropagation for Gaussian Processes using Celerite}",
  journal = {Research Notes of the American Astronomical Society},
     year = 2018,
    month = feb,
   volume = 2,
   number = 1,
    pages = {31},
      doi = {10.3847/2515-5172/aaaf6c},
   adsurl = {http://adsabs.harvard.edu/abs/2018RNAAS...2a..31F}
}

@MISC{exostriker_2019_software,
       author = {{Trifonov}, Trifon},
        title = "{The Exo-Striker: Transit and radial velocity interactive fitting tool for orbital analysis and N-body simulations}",
 howpublished = {Astrophysics Source Code Library, record ascl:1906.004},
         year = 2019,
        month = jun,
          eid = {ascl:1906.004},
        pages = {ascl:1906.004},
archivePrefix = {ascl},
       eprint = {1906.004},
       adsurl = {https://ui.adsabs.harvard.edu/abs/2019ascl.soft06004T}
}

@ARTICLE{tls_hippke_heller_2019,
       author = {{Hippke}, Michael and {Heller}, Ren{\'e}},
        title = "{Optimized transit detection algorithm to search for periodic transits of small planets}",
      journal = {\aap},
         year = "2019",
        month = "Mar",
       volume = {623},
          eid = {A39},
        pages = {A39},
          doi = {10.1051/0004-6361/201834672},
archivePrefix = {arXiv},
       eprint = {1901.02015},
 primaryClass = {astro-ph.EP},
       adsurl = {https://ui.adsabs.harvard.edu/\#abs/2019A&A...623A..39H}
}

@ARTICLE{wotan_hippke_2019,
       author = {{Hippke}, Michael and {David}, Trevor J. and {Mulders}, Gijs D. and {Heller}, Ren{\'e}},
        title = "{W{\={o}}tan: Comprehensive Time-series Detrending in Python}",
      journal = {\aj},
         year = 2019,
        month = oct,
       volume = {158},
       number = {4},
          eid = {143},
        pages = {143},
          doi = {10.3847/1538-3881/ab3984},
archivePrefix = {arXiv},
       eprint = {1906.00966},
 primaryClass = {astro-ph.EP},
       adsurl = {https://ui.adsabs.harvard.edu/abs/2019AJ....158..143H}
}

@ARTICLE{TICv8.2_paegert_2022,
       author = {{Paegert}, M. and {Stassun}, K.~G. and {Collins}, K.~A. and {Pepper}, J. and {Torres}, G. and {Jenkins}, J. and {Twicken}, J.~D. and {Latham}, D.~W.},
        title = "{VizieR Online Data Catalog: TESS Input Catalog version 8.2 (TIC v8.2) (Paegert+, 2021)}",
      journal = {VizieR Online Data Catalog},
         year = 2022,
        month = feb,
          eid = {IV/39},
        pages = {IV/39},
       adsurl = {https://ui.adsabs.harvard.edu/abs/2022yCat.4039....0P}
}

@ARTICLE{tpfplotter_aller_2020,
       author = {{Aller}, A. and {Lillo-Box}, J. and {Jones}, D. and {Miranda}, L.~F. and {Barcel{\'o} Forteza}, S.},
        title = "{Planetary nebulae seen with TESS: Discovery of new binary central star candidates from Cycle 1}",
      journal = {\aap},
         year = 2020,
        month = mar,
       volume = {635},
          eid = {A128},
        pages = {A128},
          doi = {10.1051/0004-6361/201937118},
archivePrefix = {arXiv},
       eprint = {1911.09991},
 primaryClass = {astro-ph.SR},
       adsurl = {https://ui.adsabs.harvard.edu/abs/2020A&A...635A.128A}
}

@INPROCEEDINGS{ricker_2014,
       author = {{Ricker}, George R. and {Winn}, Joshua N. and {Vanderspek}, Roland and {Latham}, David W. and {Bakos}, G{\'a}sp{\'a}r. {\'A}. and {Bean}, Jacob L. and {Berta-Thompson}, Zachory K. and {Brown}, Timothy M. and {Buchhave}, Lars and {Butler}, Nathaniel R. and {Butler}, R. Paul and {Chaplin}, William J. and {Charbonneau}, David and {Christensen-Dalsgaard}, J{\o}rgen and {Clampin}, Mark and {Deming}, Drake and {Doty}, John and {De Lee}, Nathan and {Dressing}, Courtney and {Dunham}, E.~W. and {Endl}, Michael and {Fressin}, Francois and {Ge}, Jian and {Henning}, Thomas and {Holman}, Matthew J. and {Howard}, Andrew W. and {Ida}, Shigeru and {Jenkins}, Jon and {Jernigan}, Garrett and {Johnson}, John A. and {Kaltenegger}, Lisa and {Kawai}, Nobuyuki and {Kjeldsen}, Hans and {Laughlin}, Gregory and {Levine}, Alan M. and {Lin}, Douglas and {Lissauer}, Jack J. and {MacQueen}, Phillip and {Marcy}, Geoffrey and {McCullough}, P.~R. and {Morton}, Timothy D. and {Narita}, Norio and {Paegert}, Martin and {Palle}, Enric and {Pepe}, Francesco and {Pepper}, Joshua and {Quirrenbach}, Andreas and {Rinehart}, S.~A. and {Sasselov}, Dimitar and {Sato}, Bun'ei and {Seager}, Sara and {Sozzetti}, Alessandro and {Stassun}, Keivan G. and {Sullivan}, Peter and {Szentgyorgyi}, Andrew and {Torres}, Guillermo and {Udry}, Stephane and {Villasenor}, Joel},
        title = "{Transiting Exoplanet Survey Satellite (TESS)}",
    booktitle = {Space Telescopes and Instrumentation 2014: Optical, Infrared, and Millimeter Wave},
         year = 2014,
       editor = {{Oschmann}, Jacobus M., Jr. and {Clampin}, Mark and {Fazio}, Giovanni G. and {MacEwen}, Howard A.},
       series = {SPIE Conference Series},
       volume = {9143},
        month = aug,
          eid = {914320},
        pages = {914320},
          doi = {10.1117/12.2063489},
archivePrefix = {arXiv},
       eprint = {1406.0151},
 primaryClass = {astro-ph.EP},
       adsurl = {https://ui.adsabs.harvard.edu/abs/2014SPIE.9143E..20R}
}

@ARTICLE{dasilva_2006,
       author = {{da Silva}, L. and {Girardi}, L. and {Pasquini}, L. and {Setiawan}, J. and {von der L{\"u}he}, O. and {de Medeiros}, J.~R. and {Hatzes}, A. and {D{\"o}llinger}, M.~P. and {Weiss}, A.},
        title = "{Basic physical parameters of a selected sample of evolved stars}",
      journal = {\aap},
         year = 2006,
        month = nov,
       volume = {458},
       number = {2},
        pages = {609-623},
          doi = {10.1051/0004-6361:20065105},
archivePrefix = {arXiv},
       eprint = {astro-ph/0608160},
 primaryClass = {astro-ph},
       adsurl = {https://ui.adsabs.harvard.edu/abs/2006A&A...458..609D}
}

@article{carleo2020,
	author = {{Carleo}, I. and {Malavolta, L.} and {Lanza, A. F.} and {Damasso, M.} and {Desidera, S.} and {Borsa, F.} and {Mallonn, M.} and {Pinamonti, M.} and {Gratton, R.} and {Alei, E.} and {Benatti, S.} and {Mancini, L.} and {Maldonado, J.} and {Biazzo, K.} and {Esposito, M.} and {Frustagli, G.} and {Gonz\'alez-\'Alvarez, E.} and {Micela, G.} and {Scandariato, G.} and {Sozzetti, A.} and {Affer, L.} and {Bignamini, A.} and {Bonomo, A. S.} and {Claudi, R.} and {Cosentino, R.} and {Covino, E.} and {Fiorenzano, A. F. M.} and {Giacobbe, P.} and {Harutyunyan, A.} and {Leto, G.} and {Maggio, A.} and {Molinari, E.} and {Nascimbeni, V.} and {Pagano, I.} and {Pedani, M.} and {Piotto, G.} and {Poretti, E.} and {Rainer, M.} and {Redfield, S.} and {Baffa, C.} and {Baruffolo, A.} and {Buchschacher, N.} and {Billotti, V.} and {Cecconi, M.} and {Falcini, G.} and {Fantinel, D.} and {Fini, L.} and {Galli, A.} and {Ghedina, A.} and {Ghinassi, F.} and {Giani, E.} and {Gonzalez, C.} and {Gonzalez, M.} and {Guerra, J.} and {Hernandez Diaz, M.} and {Hernandez, N.} and {Iuzzolino, M.} and {Lodi, M.} and {Oliva, E.} and {Origlia, L.} and {Perez Ventura, H.} and {Puglisi, A.} and {Riverol, C.} and {Riverol, L.} and {San Juan, J.} and {Sanna, N.} and {Scuderi, S.} and {Seemann, U.} and {Sozzi, M.} and {Tozzi, A.}},
	title = {The GAPS Programme at TNG - XXI. A GIARPS case study of known young planetary candidates: confirmation of HD 285507 b and refutation of AD Leonis b},
	DOI= "10.1051/0004-6361/201937369",
	url= "https://doi.org/10.1051/0004-6361/201937369",
	journal = {A\&A},
	year = 2020,
	volume = 638,
	pages = "A5",
}

@inproceedings{cosentino2014harps,
  title={HARPS-N@ TNG, two year harvesting data: performances and results},
  author={Cosentino, Rosario and Lovis, Christophe and Pepe, Francesco and Cameron, Andrew Collier and Latham, David W and Molinari, Emilio and Udry, Stephane and Bezawada, Naidu and Buchschacher, Nicolas and Figueira, Pedro and others},
  booktitle={Ground-based and Airborne Instrumentation for Astronomy V},
  volume={9147},
  pages={2658--2669},
  year={2014},
  organization={SPIE}
}

@ARTICLE{sousa_2015_ares_v2,
       author = {{Sousa}, S.~G. and {Santos}, N.~C. and {Adibekyan}, V. and {Delgado-Mena}, E. and {Israelian}, G.},
        title = "{ARES v2: new features and improved performance}",
      journal = {\aap},
         year = 2015,
        month = may,
       volume = {577},
          eid = {A67},
        pages = {A67},
          doi = {10.1051/0004-6361/201425463},
archivePrefix = {arXiv},
       eprint = {1504.02725},
 primaryClass = {astro-ph.IM},
       adsurl = {https://ui.adsabs.harvard.edu/abs/2015A&A...577A..67S}
}

@INPROCEEDINGS{adamow_2017_pymoogi,
       author = {{Adamow}, Monika M.},
        title = "{pyMOOGi - python wrapper for MOOG}",
    booktitle = {American Astronomical Society Meeting Abstracts \#230},
         year = 2017,
       series = {American Astronomical Society Meeting Abstracts},
       volume = {230},
        month = jun,
          eid = {216.07},
        pages = {216.07},
       adsurl = {https://ui.adsabs.harvard.edu/abs/2017AAS...23021607A}
}

@ARTICLE{Damasso_2023_yo40,
       author = {{Damasso}, M. and {Locci}, D. and {Benatti}, S. and {Maggio}, A. and {Nardiello}, D. and {Baratella}, M. and {Biazzo}, K. and {Bonomo}, A.~S. and {Desidera}, S. and {D'Orazi}, V. and {Mallonn}, M. and {Lanza}, A.~F. and {Sozzetti}, A. and {Marzari}, F. and {Borsa}, F. and {Maldonado}, J. and {Mancini}, L. and {Poretti}, E. and {Scandariato}, G. and {Bignamini}, A. and {Borsato}, L. and {Capuzzo Dolcetta}, R. and {Cecconi}, M. and {Claudi}, R. and {Cosentino}, R. and {Covino}, E. and {Fiorenzano}, A. and {Harutyunyan}, A. and {Mann}, A.~W. and {Micela}, G. and {Molinari}, E. and {Molinaro}, M. and {Pagano}, I. and {Pedani}, M. and {Pinamonti}, M. and {Piotto}, G. and {Stoev}, H.},
        title = "{The GAPS Programme at TNG. XLII. A characterisation study of the multi-planet system around the 400 Myr-old star HD 63433 (TOI-1726)}",
      journal = {\aap},
         year = 2023,
        month = apr,
       volume = {672},
          eid = {A126},
        pages = {A126},
          doi = {10.1051/0004-6361/202245391},
archivePrefix = {arXiv},
       eprint = {2303.15242},
 primaryClass = {astro-ph.EP},
       adsurl = {https://ui.adsabs.harvard.edu/abs/2023A&A...672A.126D}
}

@ARTICLE{Desidera_2023_toi179,
       author = {{Desidera}, S. and {Damasso}, M. and {Gratton}, R. and {Benatti}, S. and {Nardiello}, D. and {D'Orazi}, V. and {Lanza}, A.~F. and {Locci}, D. and {Marzari}, F. and {Mesa}, D. and {Messina}, S. and {Pillitteri}, I. and {Sozzetti}, A. and {Girard}, J. and {Maggio}, A. and {Micela}, G. and {Malavolta}, L. and {Nascimbeni}, V. and {Pinamonti}, M. and {Squicciarini}, V. and {Alcal{\'a}}, J. and {Biazzo}, K. and {Bohn}, A. and {Bonavita}, M. and {Brooks}, K. and {Chauvin}, G. and {Covino}, E. and {Delorme}, P. and {Hagelberg}, J. and {Janson}, M. and {Lagrange}, A. -M. and {Lazzoni}, C.},
        title = "{TOI-179: A young system with a transiting compact Neptune-mass planet and a low-mass companion in outer orbit}",
      journal = {\aap},
         year = 2023,
        month = jul,
       volume = {675},
          eid = {A158},
        pages = {A158},
          doi = {10.1051/0004-6361/202244611},
archivePrefix = {arXiv},
       eprint = {2210.07933},
 primaryClass = {astro-ph.EP},
       adsurl = {https://ui.adsabs.harvard.edu/abs/2023A&A...675A.158D}
}

@article{jeffries_2023_eagles,
    author = {Jeffries, R D and Jackson, R J and Wright, Nicholas J and Weaver, G and Gilmore, G and Randich, S and Bragaglia, A and Korn, A J and Smiljanic, R and Biazzo, K and Casey, A R and Frasca, A and Gonneau, A and Guiglion, G and Morbidelli, L and Prisinzano, L and Sacco, G G and Tautvaišienė, G and Worley, C C and Zaggia, S},
    title = "{The Gaia-ESO Survey: empirical estimates of stellar ages from lithium equivalent widths (eagles)}",
    journal = {\mnras},
    volume = {523},
    number = {1},
    pages = {802-824},
    year = {2023},
    month = {04},
    issn = {0035-8711},
    doi = {10.1093/mnras/stad1293},
    url = {https://doi.org/10.1093/mnras/stad1293},
    eprint = {https://academic.oup.com/mnras/article-pdf/523/1/802/50456187/stad1293.pdf},
}

@ARTICLE{Lind_2009_LiI_nlte,
       author = {{Lind}, K. and {Asplund}, M. and {Barklem}, P.~S.},
        title = "{Departures from LTE for neutral Li in late-type stars}",
      journal = {\aap},
         year = 2009,
        month = aug,
       volume = {503},
       number = {2},
        pages = {541-544},
          doi = {10.1051/0004-6361/200912221},
archivePrefix = {arXiv},
       eprint = {0906.0899},
 primaryClass = {astro-ph.SR},
       adsurl = {https://ui.adsabs.harvard.edu/abs/2009A&A...503..541L}
}

@INPROCEEDINGS{castelli_kurukz_2003,
       author = {{Castelli}, F. and {Kurucz}, R.~L.},
        title = "{New Grids of ATLAS9 Model Atmospheres}",
    booktitle = {Modelling of Stellar Atmospheres},
         year = 2003,
       editor = {{Piskunov}, N. and {Weiss}, W.~W. and {Gray}, D.~F.},
       volume = {210},
        month = jan,
        pages = {A20},
          doi = {10.48550/arXiv.astro-ph/0405087},
archivePrefix = {arXiv},
       eprint = {astro-ph/0405087},
 primaryClass = {astro-ph},
       adsurl = {https://ui.adsabs.harvard.edu/abs/2003IAUS..210P.A20C}
}

@ARTICLE{baratellaetal2020b,
       author = {{Baratella}, M. and {D'Orazi}, V. and {Biazzo}, K. and {Desidera}, S. and {Gratton}, R. and {Benatti}, S. and {Bignamini}, A. and {Carleo}, I. and {Cecconi}, M. and {Claudi}, R. and {Cosentino}, R. and {Ghedina}, A. and {Harutyunyan}, A. and {Lanza}, A.~F. and {Malavolta}, L. and {Maldonado}, J. and {Mallonn}, M. and {Messina}, S. and {Micela}, G. and {Molinari}, E. and {Poretti}, E. and {Scandariato}, G. and {Sozzetti}, A.},
        title = "{The GAPS Programme at TNG. XXV. Stellar atmospheric parameters and chemical composition through GIARPS optical and near-infrared spectra}",
      journal = {\aap},
         year = 2020,
        month = aug,
       volume = {640},
          eid = {A123},
        pages = {A123},
          doi = {10.1051/0004-6361/202038511},
archivePrefix = {arXiv},
       eprint = {2007.00475},
 primaryClass = {astro-ph.SR},
       adsurl = {https://ui.adsabs.harvard.edu/abs/2020A&A...640A.123B}
}

@ARTICLE{Biazzoetal2022,
       author = {{Biazzo}, K. and {D'Orazi}, V. and {Desidera}, S. and {Turrini}, D. and {Benatti}, S. and {Gratton}, R. and {Magrini}, L. and {Sozzetti}, A. and {Baratella}, M. and {Bonomo}, A.~S. and {Borsa}, F. and {Claudi}, R. and {Covino}, E. and {Damasso}, M. and {Di Mauro}, M.~P. and {Lanza}, A.~F. and {Maggio}, A. and {Malavolta}, L. and {Maldonado}, J. and {Marzari}, F. and {Micela}, G. and {Poretti}, E. and {Vitello}, F. and {Affer}, L. and {Bignamini}, A. and {Carleo}, I. and {Cosentino}, R. and {Fiorenzano}, A.~F.~M. and {Giacobbe}, P. and {Harutyunyan}, A. and {Leto}, G. and {Mancini}, L. and {Molinari}, E. and {Molinaro}, M. and {Nardiello}, D. and {Nascimbeni}, V. and {Pagano}, I. and {Pedani}, M. and {Piotto}, G. and {Rainer}, M. and {Scandariato}, G.},
        title = "{The GAPS Programme at TNG. XXXV. Fundamental properties of transiting exoplanet host stars}",
      journal = {\aap},
         year = 2022,
        month = aug,
       volume = {664},
          eid = {A161},
        pages = {A161},
          doi = {10.1051/0004-6361/202243467},
archivePrefix = {arXiv},
       eprint = {2205.15796},
 primaryClass = {astro-ph.SR},
       adsurl = {https://ui.adsabs.harvard.edu/abs/2022A&A...664A.161B}
}

@ARTICLE{sneden1973,
       author = {{Sneden}, C.},
        title = "{The nitrogen abundance of the very metal-poor star HD 122563.}",
      journal = {\apj},
         year = 1973,
        month = sep,
       volume = {184},
        pages = {839},
          doi = {10.1086/152374},
       adsurl = {https://ui.adsabs.harvard.edu/abs/1973ApJ...184..839S}
}

@ARTICLE{gagneetal2018,
       author = {{Gagn{\'e}}, Jonathan and {Mamajek}, Eric E. and {Malo}, Lison and {Riedel}, Adric and {Rodriguez}, David and {Lafreni{\`e}re}, David and {Faherty}, Jacqueline K. and {Roy-Loubier}, Olivier and {Pueyo}, Laurent and {Robin}, Annie C. and {Doyon}, Ren{\'e}},
        title = "{BANYAN. XI. The BANYAN {\ensuremath{\Sigma}} Multivariate Bayesian Algorithm to Identify Members of Young Associations with 150 pc}",
      journal = {\apj},
         year = 2018,
        month = mar,
       volume = {856},
       number = {1},
          eid = {23},
        pages = {23},
          doi = {10.3847/1538-4357/aaae09},
archivePrefix = {arXiv},
       eprint = {1801.09051},
 primaryClass = {astro-ph.SR},
       adsurl = {https://ui.adsabs.harvard.edu/abs/2018ApJ...856...23G}
}

@ARTICLE{gaiacollaborationetal2016,
       author = {{Gaia Collaboration} and {Prusti}, T. and {de Bruijne}, J.~H.~J. and {Brown}, A.~G.~A. and {Vallenari}, A. and {Babusiaux}, C. and {Bailer-Jones}, C.~A.~L. and {Bastian}, U. and {Biermann}, M. and {Evans}, D.~W. and {Eyer}, L. and {Jansen}, F. and {Jordi}, C. and {Klioner}, S.~A. and {Lammers}, U. and {Lindegren}, L. and {Luri}, X. and {Mignard}, F. and {Milligan}, D.~J. and {Panem}, C. and {Poinsignon}, V. and {Pourbaix}, D. and {Randich}, S. and {Sarri}, G. and {Sartoretti}, P. and {Siddiqui}, H.~I. and {Soubiran}, C. and {Valette}, V. and {van Leeuwen}, F. and {Walton}, N.~A. and {Aerts}, C. and {Arenou}, F. and {Cropper}, M. and {Drimmel}, R. and {H{\o}g}, E. and {Katz}, D. and {Lattanzi}, M.~G. and {O'Mullane}, W. and {Grebel}, E.~K. and {Holland}, A.~D. and {Huc}, C. and {Passot}, X. and {Bramante}, L. and {Cacciari}, C. and {Casta{\~n}eda}, J. and {Chaoul}, L. and {Cheek}, N. and {De Angeli}, F. and {Fabricius}, C. and {Guerra}, R. and {Hern{\'a}ndez}, J. and {Jean-Antoine-Piccolo}, A. and {Masana}, E. and {Messineo}, R. and {Mowlavi}, N. and {Nienartowicz}, K. and {Ord{\'o}{\~n}ez-Blanco}, D. and {Panuzzo}, P. and {Portell}, J. and {Richards}, P.~J. and {Riello}, M. and {Seabroke}, G.~M. and {Tanga}, P. and {Th{\'e}venin}, F. and {Torra}, J. and {Els}, S.~G. and {Gracia-Abril}, G. and {Comoretto}, G. and {Garcia-Reinaldos}, M. and {Lock}, T. and {Mercier}, E. and {Altmann}, M. and {Andrae}, R. and {Astraatmadja}, T.~L. and {Bellas-Velidis}, I. and {Benson}, K. and {Berthier}, J. and {Blomme}, R. and {Busso}, G. and {Carry}, B. and {Cellino}, A. and {Clementini}, G. and {Cowell}, S. and {Creevey}, O. and {Cuypers}, J. and {Davidson}, M. and {De Ridder}, J. and {de Torres}, A. and {Delchambre}, L. and {Dell'Oro}, A. and {Ducourant}, C. and {Fr{\'e}mat}, Y. and {Garc{\'\i}a-Torres}, M. and {Gosset}, E. and {Halbwachs}, J. -L. and {Hambly}, N.~C. and {Harrison}, D.~L. and {Hauser}, M. and {Hestroffer}, D. and {Hodgkin}, S.~T. and {Huckle}, H.~E. and {Hutton}, A. and {Jasniewicz}, G. and {Jordan}, S. and {Kontizas}, M. and {Korn}, A.~J. and {Lanzafame}, A.~C. and {Manteiga}, M. and {Moitinho}, A. and {Muinonen}, K. and {Osinde}, J. and {Pancino}, E. and {Pauwels}, T. and {Petit}, J. -M. and {Recio-Blanco}, A. and {Robin}, A.~C. and {Sarro}, L.~M. and {Siopis}, C. and {Smith}, M. and {Smith}, K.~W. and {Sozzetti}, A. and {Thuillot}, W. and {van Reeven}, W. and {Viala}, Y. and {Abbas}, U. and {Abreu Aramburu}, A. and {Accart}, S. and {Aguado}, J.~J. and {Allan}, P.~M. and {Allasia}, W. and {Altavilla}, G. and {{\'A}lvarez}, M.~A. and {Alves}, J. and {Anderson}, R.~I. and {Andrei}, A.~H. and {Anglada Varela}, E. and {Antiche}, E. and {Antoja}, T. and {Ant{\'o}n}, S. and {Arcay}, B. and {Atzei}, A. and {Ayache}, L. and {Bach}, N. and {Baker}, S.~G. and {Balaguer-N{\'u}{\~n}ez}, L. and {Barache}, C. and {Barata}, C. and {Barbier}, A. and {Barblan}, F. and {Baroni}, M. and {Barrado y Navascu{\'e}s}, D. and {Barros}, M. and {Barstow}, M.~A. and {Becciani}, U. and {Bellazzini}, M. and {Bellei}, G. and {Bello Garc{\'\i}a}, A. and {Belokurov}, V. and {Bendjoya}, P. and {Berihuete}, A. and {Bianchi}, L. and {Bienaym{\'e}}, O. and {Billebaud}, F. and {Blagorodnova}, N. and {Blanco-Cuaresma}, S. and {Boch}, T. and {Bombrun}, A. and {Borrachero}, R. and {Bouquillon}, S. and {Bourda}, G. and {Bouy}, H. and {Bragaglia}, A. and {Breddels}, M.~A. and {Brouillet}, N. and {Br{\"u}semeister}, T. and {Bucciarelli}, B. and {Budnik}, F. and {Burgess}, P. and {Burgon}, R. and {Burlacu}, A. and {Busonero}, D. and {Buzzi}, R. and {Caffau}, E. and {Cambras}, J. and {Campbell}, H. and {Cancelliere}, R. and {Cantat-Gaudin}, T. and {Carlucci}, T. and {Carrasco}, J.~M. and {Castellani}, M. and {Charlot}, P. and {Charnas}, J. and {Charvet}, P. and {Chassat}, F. and {Chiavassa}, A. and {Clotet}, M. and {Cocozza}, G. and {Collins}, R.~S. and {Collins}, P. and {Costigan}, G. and {Crifo}, F. and {Cross}, N.~J.~G. and {Crosta}, M. and {Crowley}, C. and {Dafonte}, C. and {Damerdji}, Y. and {Dapergolas}, A. and {David}, P. and {David}, M. and {De Cat}, P. and {de Felice}, F. and {de Laverny}, P. and {De Luise}, F. and {De March}, R. and {de Martino}, D. and {de Souza}, R. and {Debosscher}, J. and {del Pozo}, E. and {Delbo}, M. and {Delgado}, A. and {Delgado}, H.~E. and {di Marco}, F. and {Di Matteo}, P. and {Diakite}, S. and {Distefano}, E. and {Dolding}, C. and {Dos Anjos}, S. and {Drazinos}, P. and {Dur{\'a}n}, J. and {Dzigan}, Y. and {Ecale}, E. and {Edvardsson}, B. and {Enke}, H. and {Erdmann}, M. and {Escolar}, D. and {Espina}, M. and {Evans}, N.~W. and {Eynard Bontemps}, G. and {Fabre}, C. and {Fabrizio}, M. and {Faigler}, S. and {Falc{\~a}o}, A.~J. and {Farr{\`a}s Casas}, M. and {Faye}, F. and {Federici}, L. and {Fedorets}, G. and {Fern{\'a}ndez-Hern{\'a}ndez}, J. and {Fernique}, P. and {Fienga}, A. and {Figueras}, F. and {Filippi}, F. and {Findeisen}, K. and {Fonti}, A. and {Fouesneau}, M. and {Fraile}, E. and {Fraser}, M. and {Fuchs}, J. and {Furnell}, R. and {Gai}, M. and {Galleti}, S. and {Galluccio}, L. and {Garabato}, D. and {Garc{\'\i}a-Sedano}, F. and {Gar{\'e}}, P. and {Garofalo}, A. and {Garralda}, N. and {Gavras}, P. and {Gerssen}, J. and {Geyer}, R. and {Gilmore}, G. and {Girona}, S. and {Giuffrida}, G. and {Gomes}, M. and {Gonz{\'a}lez-Marcos}, A. and {Gonz{\'a}lez-N{\'u}{\~n}ez}, J. and {Gonz{\'a}lez-Vidal}, J.~J. and {Granvik}, M. and {Guerrier}, A. and {Guillout}, P. and {Guiraud}, J. and {G{\'u}rpide}, A. and {Guti{\'e}rrez-S{\'a}nchez}, R. and {Guy}, L.~P. and {Haigron}, R. and {Hatzidimitriou}, D. and {Haywood}, M. and {Heiter}, U. and {Helmi}, A. and {Hobbs}, D. and {Hofmann}, W. and {Holl}, B. and {Holland}, G. and {Hunt}, J.~A.~S. and {Hypki}, A. and {Icardi}, V. and {Irwin}, M. and {Jevardat de Fombelle}, G. and {Jofr{\'e}}, P. and {Jonker}, P.~G. and {Jorissen}, A. and {Julbe}, F. and {Karampelas}, A. and {Kochoska}, A. and {Kohley}, R. and {Kolenberg}, K. and {Kontizas}, E. and {Koposov}, S.~E. and {Kordopatis}, G. and {Koubsky}, P. and {Kowalczyk}, A. and {Krone-Martins}, A. and {Kudryashova}, M. and {Kull}, I. and {Bachchan}, R.~K. and {Lacoste-Seris}, F. and {Lanza}, A.~F. and {Lavigne}, J. -B. and {Le Poncin-Lafitte}, C. and {Lebreton}, Y. and {Lebzelter}, T. and {Leccia}, S. and {Leclerc}, N. and {Lecoeur-Taibi}, I. and {Lemaitre}, V. and {Lenhardt}, H. and {Leroux}, F. and {Liao}, S. and {Licata}, E. and {Lindstr{\o}m}, H.~E.~P. and {Lister}, T.~A. and {Livanou}, E. and {Lobel}, A. and {L{\"o}ffler}, W. and {L{\'o}pez}, M. and {Lopez-Lozano}, A. and {Lorenz}, D. and {Loureiro}, T. and {MacDonald}, I. and {Magalh{\~a}es Fernandes}, T. and {Managau}, S. and {Mann}, R.~G. and {Mantelet}, G. and {Marchal}, O. and {Marchant}, J.~M. and {Marconi}, M. and {Marie}, J. and {Marinoni}, S. and {Marrese}, P.~M. and {Marschalk{\'o}}, G. and {Marshall}, D.~J. and {Mart{\'\i}n-Fleitas}, J.~M. and {Martino}, M. and {Mary}, N. and {Matijevi{\v{c}}}, G. and {Mazeh}, T. and {McMillan}, P.~J. and {Messina}, S. and {Mestre}, A. and {Michalik}, D. and {Millar}, N.~R. and {Miranda}, B.~M.~H. and {Molina}, D. and {Molinaro}, R. and {Molinaro}, M. and {Moln{\'a}r}, L. and {Moniez}, M. and {Montegriffo}, P. and {Monteiro}, D. and {Mor}, R. and {Mora}, A. and {Morbidelli}, R. and {Morel}, T. and {Morgenthaler}, S. and {Morley}, T. and {Morris}, D. and {Mulone}, A.~F. and {Muraveva}, T. and {Musella}, I. and {Narbonne}, J. and {Nelemans}, G. and {Nicastro}, L. and {Noval}, L. and {Ord{\'e}novic}, C. and {Ordieres-Mer{\'e}}, J. and {Osborne}, P. and {Pagani}, C. and {Pagano}, I. and {Pailler}, F. and {Palacin}, H. and {Palaversa}, L. and {Parsons}, P. and {Paulsen}, T. and {Pecoraro}, M. and {Pedrosa}, R. and {Pentik{\"a}inen}, H. and {Pereira}, J. and {Pichon}, B. and {Piersimoni}, A.~M. and {Pineau}, F. -X. and {Plachy}, E. and {Plum}, G. and {Poujoulet}, E. and {Pr{\v{s}}a}, A. and {Pulone}, L. and {Ragaini}, S. and {Rago}, S. and {Rambaux}, N. and {Ramos-Lerate}, M. and {Ranalli}, P. and {Rauw}, G. and {Read}, A. and {Regibo}, S. and {Renk}, F. and {Reyl{\'e}}, C. and {Ribeiro}, R.~A. and {Rimoldini}, L. and {Ripepi}, V. and {Riva}, A. and {Rixon}, G. and {Roelens}, M. and {Romero-G{\'o}mez}, M. and {Rowell}, N. and {Royer}, F. and {Rudolph}, A. and {Ruiz-Dern}, L. and {Sadowski}, G. and {Sagrist{\`a} Sell{\'e}s}, T. and {Sahlmann}, J. and {Salgado}, J. and {Salguero}, E. and {Sarasso}, M. and {Savietto}, H. and {Schnorhk}, A. and {Schultheis}, M. and {Sciacca}, E. and {Segol}, M. and {Segovia}, J.~C. and {Segransan}, D. and {Serpell}, E. and {Shih}, I. -C. and {Smareglia}, R. and {Smart}, R.~L. and {Smith}, C. and {Solano}, E. and {Solitro}, F. and {Sordo}, R. and {Soria Nieto}, S. and {Souchay}, J. and {Spagna}, A. and {Spoto}, F. and {Stampa}, U. and {Steele}, I.~A. and {Steidelm{\"u}ller}, H. and {Stephenson}, C.~A. and {Stoev}, H. and {Suess}, F.~F. and {S{\"u}veges}, M. and {Surdej}, J. and {Szabados}, L. and {Szegedi-Elek}, E. and {Tapiador}, D. and {Taris}, F. and {Tauran}, G. and {Taylor}, M.~B. and {Teixeira}, R. and {Terrett}, D. and {Tingley}, B. and {Trager}, S.~C. and {Turon}, C. and {Ulla}, A. and {Utrilla}, E. and {Valentini}, G. and {van Elteren}, A. and {Van Hemelryck}, E. and {van Leeuwen}, M. and {Varadi}, M. and {Vecchiato}, A. and {Veljanoski}, J. and {Via}, T. and {Vicente}, D. and {Vogt}, S. and {Voss}, H. and {Votruba}, V. and {Voutsinas}, S. and {Walmsley}, G. and {Weiler}, M. and {Weingrill}, K. and {Werner}, D. and {Wevers}, T. and {Whitehead}, G. and {Wyrzykowski}, {\L}. and {Yoldas}, A. and {{\v{Z}}erjal}, M. and {Zucker}, S. and {Zurbach}, C. and {Zwitter}, T. and {Alecu}, A. and {Allen}, M. and {Allende Prieto}, C. and {Amorim}, A. and {Anglada-Escud{\'e}}, G. and {Arsenijevic}, V. and {Azaz}, S. and {Balm}, P. and {Beck}, M. and {Bernstein}, H. -H. and {Bigot}, L. and {Bijaoui}, A. and {Blasco}, C. and {Bonfigli}, M. and {Bono}, G. and {Boudreault}, S. and {Bressan}, A. and {Brown}, S. and {Brunet}, P. -M. and {Bunclark}, P. and {Buonanno}, R. and {Butkevich}, A.~G. and {Carret}, C. and {Carrion}, C. and {Chemin}, L. and {Ch{\'e}reau}, F. and {Corcione}, L. and {Darmigny}, E. and {de Boer}, K.~S. and {de Teodoro}, P. and {de Zeeuw}, P.~T. and {Delle Luche}, C. and {Domingues}, C.~D. and {Dubath}, P. and {Fodor}, F. and {Fr{\'e}zouls}, B. and {Fries}, A. and {Fustes}, D. and {Fyfe}, D. and {Gallardo}, E. and {Gallegos}, J. and {Gardiol}, D. and {Gebran}, M. and {Gomboc}, A. and {G{\'o}mez}, A. and {Grux}, E. and {Gueguen}, A. and {Heyrovsky}, A. and {Hoar}, J. and {Iannicola}, G. and {Isasi Parache}, Y. and {Janotto}, A. -M. and {Joliet}, E. and {Jonckheere}, A. and {Keil}, R. and {Kim}, D. -W. and {Klagyivik}, P. and {Klar}, J. and {Knude}, J. and {Kochukhov}, O. and {Kolka}, I. and {Kos}, J. and {Kutka}, A. and {Lainey}, V. and {LeBouquin}, D. and {Liu}, C. and {Loreggia}, D. and {Makarov}, V.~V. and {Marseille}, M.~G. and {Martayan}, C. and {Martinez-Rubi}, O. and {Massart}, B. and {Meynadier}, F. and {Mignot}, S. and {Munari}, U. and {Nguyen}, A. -T. and {Nordlander}, T. and {Ocvirk}, P. and {O'Flaherty}, K.~S. and {Olias Sanz}, A. and {Ortiz}, P. and {Osorio}, J. and {Oszkiewicz}, D. and {Ouzounis}, A. and {Palmer}, M. and {Park}, P. and {Pasquato}, E. and {Peltzer}, C. and {Peralta}, J. and {P{\'e}turaud}, F. and {Pieniluoma}, T. and {Pigozzi}, E. and {Poels}, J. and {Prat}, G. and {Prod'homme}, T. and {Raison}, F. and {Rebordao}, J.~M. and {Risquez}, D. and {Rocca-Volmerange}, B. and {Rosen}, S. and {Ruiz-Fuertes}, M.~I. and {Russo}, F. and {Sembay}, S. and {Serraller Vizcaino}, I. and {Short}, A. and {Siebert}, A. and {Silva}, H. and {Sinachopoulos}, D. and {Slezak}, E. and {Soffel}, M. and {Sosnowska}, D. and {Strai{\v{z}}ys}, V. and {ter Linden}, M. and {Terrell}, D. and {Theil}, S. and {Tiede}, C. and {Troisi}, L. and {Tsalmantza}, P. and {Tur}, D. and {Vaccari}, M. and {Vachier}, F. and {Valles}, P. and {Van Hamme}, W. and {Veltz}, L. and {Virtanen}, J. and {Wallut}, J. -M. and {Wichmann}, R. and {Wilkinson}, M.~I. and {Ziaeepour}, H. and {Zschocke}, S.},
        title = "{The Gaia mission}",
      journal = {\aap},
         year = 2016,
        month = nov,
       volume = {595},
          eid = {A1},
        pages = {A1},
          doi = {10.1051/0004-6361/201629272},
archivePrefix = {arXiv},
       eprint = {1609.04153},
 primaryClass = {astro-ph.IM},
       adsurl = {https://ui.adsabs.harvard.edu/abs/2016A&A...595A...1G}
}

@ARTICLE{baratellaetal2020a,
       author = {{Baratella}, M. and {D'Orazi}, V. and {Carraro}, G. and {Desidera}, S. and {Randich}, S. and {Magrini}, L. and {Adibekyan}, V. and {Smiljanic}, R. and {Spina}, L. and {Tsantaki}, M. and {Tautvai{\v{s}}ien{\.{e}}}, G. and {Sousa}, S.~G. and {Jofr{\'e}}, P. and {Jim{\'e}nez-Esteban}, F.~M. and {Delgado-Mena}, E. and {Martell}, S. and {Van der Swaelmen}, M. and {Roccatagliata}, V. and {Gilmore}, G. and {Alfaro}, E.~J. and {Bayo}, A. and {Bensby}, T. and {Bragaglia}, A. and {Franciosini}, E. and {Gonneau}, A. and {Heiter}, U. and {Hourihane}, A. and {Jeffries}, R.~D. and {Koposov}, S.~E. and {Morbidelli}, L. and {Prisinzano}, L. and {Sacco}, G. and {Sbordone}, L. and {Worley}, C. and {Zaggia}, S. and {Lewis}, J.},
        title = "{The Gaia-ESO Survey: a new approach to chemically characterising young open clusters. I. Stellar parameters, and iron-peak, {\ensuremath{\alpha}}-, and proton-capture elements}",
      journal = {\aap},
         year = 2020,
        month = feb,
       volume = {634},
          eid = {A34},
        pages = {A34},
          doi = {10.1051/0004-6361/201937055},
archivePrefix = {arXiv},
       eprint = {2001.03179},
 primaryClass = {astro-ph.SR},
       adsurl = {https://ui.adsabs.harvard.edu/abs/2020A&A...634A..34B}
}

@article{husser2013_phoenix_cond,
	author = {{Husser}, T.O. and {Wende-von Berg}, S. and {Dreizler}, S. and {Homeier}, D. and {Reiners}, A. and {Barman}, T. and {Hauschildt}, P. H.},
	title = {A new extensive library of PHOENIX stellar atmospheres and
          synthetic spectra},
	DOI= "10.1051/0004-6361/201219058",
	url= "https://doi.org/10.1051/0004-6361/201219058",
	journal = {\aap},
	year = 2013,
	volume = 553,
	pages = "A6",
	month = ""
}

@ARTICLE{lallement_2018_e_bv,
       author = {{Lallement}, R. and {Capitanio}, L. and {Ruiz-Dern}, L. and {Danielski}, C. and {Babusiaux}, C. and {Vergely}, L. and {Elyajouri}, M. and {Arenou}, F. and {Leclerc}, N.},
        title = "{Three-dimensional maps of interstellar dust in the Local Arm: using Gaia, 2MASS, and APOGEE-DR14}",
      journal = {\aap},
         year = 2018,
        month = aug,
       volume = {616},
          eid = {A132},
        pages = {A132},
          doi = {10.1051/0004-6361/201832832},
archivePrefix = {arXiv},
       eprint = {1804.06060},
 primaryClass = {astro-ph.GA},
       adsurl = {https://ui.adsabs.harvard.edu/abs/2018A&A...616A.132L}
}

@ARTICLE{newton_2019_intro,
       author = {{Newton}, Elisabeth R. and {Mann}, Andrew W. and {Tofflemire}, Benjamin M. and {Pearce}, Logan and {Rizzuto}, Aaron C. and {Vanderburg}, Andrew and {Martinez}, Raquel A. and {Wang}, Jason J. and {Ruffio}, Jean-Baptiste and {Kraus}, Adam L. and {Johnson}, Marshall C. and {Thao}, Pa Chia and {Wood}, Mackenna L. and {Rampalli}, Rayna and {Nielsen}, Eric L. and {Collins}, Karen A. and {Dragomir}, Diana and {Hellier}, Coel and {Anderson}, D.~R. and {Barclay}, Thomas and {Brown}, Carolyn and {Feiden}, Gregory and {Hart}, Rhodes and {Isopi}, Giovanni and {Kielkopf}, John F. and {Mallia}, Franco and {Nelson}, Peter and {Rodriguez}, Joseph E. and {Stockdale}, Chris and {Waite}, Ian A. and {Wright}, Duncan J. and {Lissauer}, Jack J. and {Ricker}, George R. and {Vanderspek}, Roland and {Latham}, David W. and {Seager}, S. and {Winn}, Joshua N. and {Jenkins}, Jon M. and {Bouma}, Luke G. and {Burke}, Christopher J. and {Davies}, Misty and {Fausnaugh}, Michael and {Li}, Jie and {Morris}, Robert L. and {Mukai}, Koji and {Villase{\~n}or}, Joel and {Villeneuva}, Steven and {De Rosa}, Robert J. and {Macintosh}, Bruce and {Mengel}, Matthew W. and {Okumura}, Jack and {Wittenmyer}, Robert A.},
        title = "{TESS Hunt for Young and Maturing Exoplanets (THYME): A Planet in the 45 Myr Tucana-Horologium Association}",
      journal = {\apjl},
         year = 2019,
        month = jul,
       volume = {880},
       number = {1},
          eid = {L17},
        pages = {L17},
          doi = {10.3847/2041-8213/ab2988},
archivePrefix = {arXiv},
       eprint = {1906.10703},
 primaryClass = {astro-ph.EP},
       adsurl = {https://ui.adsabs.harvard.edu/abs/2019ApJ...880L..17N}
}

@ARTICLE{benatti_2019_intro,
       author = {{Benatti}, S. and {Nardiello}, D. and {Malavolta}, L. and {Desidera}, S. and {Borsato}, L. and {Nascimbeni}, V. and {Damasso}, M. and {D'Orazi}, V. and {Mesa}, D. and {Messina}, S. and {Esposito}, M. and {Bignamini}, A. and {Claudi}, R. and {Covino}, E. and {Lovis}, C. and {Sabotta}, S.},
        title = "{A possibly inflated planet around the bright young star DS Tucanae A}",
      journal = {\aap},
         year = 2019,
        month = oct,
       volume = {630},
          eid = {A81},
        pages = {A81},
          doi = {10.1051/0004-6361/201935598},
archivePrefix = {arXiv},
       eprint = {1904.01591},
 primaryClass = {astro-ph.SR},
       adsurl = {https://ui.adsabs.harvard.edu/abs/2019A&A...630A..81B}
}

@ARTICLE{nardiello2022,
       author = {{Nardiello}, D. and {Malavolta}, L. and {Desidera}, S. and {Baratella}, M. and {D'Orazi}, V. and {Messina}, S. and {Biazzo}, K. and {Benatti}, S. and {Damasso}, M. and {Rajpaul}, V.~M. and {Bonomo}, A.~S. and {Capuzzo Dolcetta}, R. and {Mallonn}, M. and {Cale}, B. and {Plavchan}, P. and {El Mufti}, M. and {Bignamini}, A. and {Borsa}, F. and {Carleo}, I. and {Claudi}, R. and {Covino}, E. and {Lanza}, A.~F. and {Maldonado}, J. and {Mancini}, L. and {Micela}, G. and {Molinari}, E. and {Pinamonti}, M. and {Piotto}, G. and {Poretti}, E. and {Scandariato}, G. and {Sozzetti}, A. and {Andreuzzi}, G. and {Boschin}, W. and {Cosentino}, R. and {Fiorenzano}, A.~F.~M. and {Harutyunyan}, A. and {Knapic}, C. and {Pedani}, M. and {Affer}, L. and {Maggio}, A. and {Rainer}, M.},
        title = "{The GAPS Programme at TNG. XXXVII. A precise density measurement of the young ultra-short period planet TOI-1807 b}",
      journal = {\aap},
         year = 2022,
        month = aug,
       volume = {664},
          eid = {A163},
        pages = {A163},
          doi = {10.1051/0004-6361/202243743},
archivePrefix = {arXiv},
       eprint = {2206.03496},
 primaryClass = {astro-ph.EP},
       adsurl = {https://ui.adsabs.harvard.edu/abs/2022A&A...664A.163N}
}

@article{kepler411_mrplot,
	author = {{Sun}, L. and {Ioannidis}, P. and {Gu}, S. and {Schmitt}, J. H. M. M. and {Wang}, X. and {Kouwenhoven}, M. B. N.},
	title = {Kepler-411: a four-planet system with an active host star},
	DOI= "10.1051/0004-6361/201834275",
	url= "https://doi.org/10.1051/0004-6361/201834275",
	journal = {\aap},
	year = 2019,
	volume = 624,
	pages = "A15"
}

@ARTICLE{Tinettietal2018_ariel,
       author = {{Tinetti}, Giovanna and {Drossart}, Pierre and {Eccleston}, Paul and {Hartogh}, Paul and {Heske}, Astrid and {Leconte}, J{\'e}r{\'e}my and {Micela}, Giusi and {Ollivier}, Marc and {Pilbratt}, G{\"o}ran and {Puig}, Ludovic and {Turrini}, Diego and {Vandenbussche}, Bart and {Wolkenberg}, Paulina and {Beaulieu}, Jean-Philippe and {Buchave}, Lars A. and {Ferus}, Martin and {Griffin}, Matt and {Guedel}, Manuel and {Justtanont}, Kay and {Lagage}, Pierre-Olivier and {Machado}, Pedro and {Malaguti}, Giuseppe and {Min}, Michiel and {N{\o}rgaard-Nielsen}, Hans Ulrik and {Rataj}, Mirek and {Ray}, Tom and {Ribas}, Ignasi and {Swain}, Mark and {Szabo}, Robert and {Werner}, Stephanie and {Barstow}, Joanna and {Burleigh}, Matt and {Cho}, James and {Coud{\'e} du Foresto}, Vincent and {Coustenis}, Athena and {Decin}, Leen and {Encrenaz}, Therese and {Galand}, Marina and {Gillon}, Michael and {Helled}, Ravit and {Morales}, Juan Carlos and {Garc{\'\i}a Mu{\~n}oz}, Antonio and {Moneti}, Andrea and {Pagano}, Isabella and {Pascale}, Enzo and {Piccioni}, Giuseppe and {Pinfield}, David and {Sarkar}, Subhajit and {Selsis}, Franck and {Tennyson}, Jonathan and {Triaud}, Amaury and {Venot}, Olivia and {Waldmann}, Ingo and {Waltham}, David and {Wright}, Gillian and {Amiaux}, Jerome and {Augu{\`e}res}, Jean-Louis and {Berth{\'e}}, Michel and {Bezawada}, Naidu and {Bishop}, Georgia and {Bowles}, Neil and {Coffey}, Deirdre and {Colom{\'e}}, Josep and {Crook}, Martin and {Crouzet}, Pierre-Elie and {Da Peppo}, Vania and {Sanz}, Isabel Escudero and {Focardi}, Mauro and {Frericks}, Martin and {Hunt}, Tom and {Kohley}, Ralf and {Middleton}, Kevin and {Morgante}, Gianluca and {Ottensamer}, Roland and {Pace}, Emanuele and {Pearson}, Chris and {Stamper}, Richard and {Symonds}, Kate and {Rengel}, Miriam and {Renotte}, Etienne and {Ade}, Peter and {Affer}, Laura and {Alard}, Christophe and {Allard}, Nicole and {Altieri}, Francesca and {Andr{\'e}}, Yves and {Arena}, Claudio and {Argyriou}, Ioannis and {Aylward}, Alan and {Baccani}, Cristian and {Bakos}, Gaspar and {Banaszkiewicz}, Marek and {Barlow}, Mike and {Batista}, Virginie and {Bellucci}, Giancarlo and {Benatti}, Serena and {Bernardi}, Pernelle and {B{\'e}zard}, Bruno and {Blecka}, Maria and {Bolmont}, Emeline and {Bonfond}, Bertrand and {Bonito}, Rosaria and {Bonomo}, Aldo S. and {Brucato}, John Robert and {Brun}, Allan Sacha and {Bryson}, Ian and {Bujwan}, Waldemar and {Casewell}, Sarah and {Charnay}, Bejamin and {Pestellini}, Cesare Cecchi and {Chen}, Guo and {Ciaravella}, Angela and {Claudi}, Riccardo and {Cl{\'e}dassou}, Rodolphe and {Damasso}, Mario and {Damiano}, Mario and {Danielski}, Camilla and {Deroo}, Pieter and {Di Giorgio}, Anna Maria and {Dominik}, Carsten and {Doublier}, Vanessa and {Doyle}, Simon and {Doyon}, Ren{\'e} and {Drummond}, Benjamin and {Duong}, Bastien and {Eales}, Stephen and {Edwards}, Billy and {Farina}, Maria and {Flaccomio}, Ettore and {Fletcher}, Leigh and {Forget}, Fran{\c{c}}ois and {Fossey}, Steve and {Fr{\"a}nz}, Markus and {Fujii}, Yuka and {Garc{\'\i}a-Piquer}, {\'A}lvaro and {Gear}, Walter and {Geoffray}, Herv{\'e} and {G{\'e}rard}, Jean Claude and {Gesa}, Lluis and {Gomez}, H. and {Graczyk}, Rafa{\l} and {Griffith}, Caitlin and {Grodent}, Denis and {Guarcello}, Mario Giuseppe and {Gustin}, Jacques and {Hamano}, Keiko and {Hargrave}, Peter and {Hello}, Yann and {Heng}, Kevin and {Herrero}, Enrique and {Hornstrup}, Allan and {Hubert}, Benoit and {Ida}, Shigeru and {Ikoma}, Masahiro and {Iro}, Nicolas and {Irwin}, Patrick and {Jarchow}, Christopher and {Jaubert}, Jean and {Jones}, Hugh and {Julien}, Queyrel and {Kameda}, Shingo and {Kerschbaum}, Franz and {Kervella}, Pierre and {Koskinen}, Tommi and {Krijger}, Matthijs and {Krupp}, Norbert and {Lafarga}, Marina and {Landini}, Federico and {Lellouch}, Emanuel and {Leto}, Giuseppe and {Luntzer}, A. and {Rank-L{\"u}ftinger}, Theresa and {Maggio}, Antonio and {Maldonado}, Jesus and {Maillard}, Jean-Pierre and {Mall}, Urs and {Marquette}, Jean-Baptiste and {Mathis}, Stephane and {Maxted}, Pierre and {Matsuo}, Taro and {Medvedev}, Alexander and {Miguel}, Yamila and {Minier}, Vincent and {Morello}, Giuseppe and {Mura}, Alessandro and {Narita}, Norio and {Nascimbeni}, Valerio and {Nguyen Tong}, N. and {Noce}, Vladimiro and {Oliva}, Fabrizio and {Palle}, Enric and {Palmer}, Paul and {Pancrazzi}, Maurizio and {Papageorgiou}, Andreas and {Parmentier}, Vivien and {Perger}, Manuel and {Petralia}, Antonino and {Pezzuto}, Stefano and {Pierrehumbert}, Ray and {Pillitteri}, Ignazio},
        title = "{A chemical survey of exoplanets with ARIEL}",
      journal = {Exp. Astron.},
         year = 2018,
        month = nov,
       volume = {46},
       number = {1},
        pages = {135-209},
          doi = {10.1007/s10686-018-9598-x},
       adsurl = {https://ui.adsabs.harvard.edu/abs/2018ExA....46..135T}
}

@BOOK{2mass,
       author = {{Cutri}, R.~M. and {Skrutskie}, M.~F. and {van Dyk}, S. and {Beichman}, C.~A. and {Carpenter}, J.~M. and {Chester}, T. and {Cambresy}, L. and {Evans}, T. and {Fowler}, J. and {Gizis}, J. and {Howard}, E. and {Huchra}, J. and {Jarrett}, T. and {Kopan}, E.~L. and {Kirkpatrick}, J.~D. and {Light}, R.~M. and {Marsh}, K.~A. and {McCallon}, H. and {Schneider}, S. and {Stiening}, R. and {Sykes}, M. and {Weinberg}, M. and {Wheaton}, W.~A. and {Wheelock}, S. and {Zacarias}, N.},
        title = "{2MASS All Sky Catalog of point sources.}",
         year = 2003,
       adsurl = {https://ui.adsabs.harvard.edu/abs/2003tmc..book.....C}
}

@ARTICLE{2021casagrande,
       author = {{Casagrande}, Luca and {Lin}, Jane and {Rains}, Adam D. and {Liu}, Fan and {Buder}, Sven and {Horner}, Jonathan and {Asplund}, Martin and {Lewis}, Geraint F. and {Martell}, Sarah L. and {Nordlander}, Thomas and {Stello}, Dennis and {Ting}, Yuan-Sen and {Wittenmyer}, Robert A. and {Bland-Hawthorn}, Joss and {Casey}, Andrew R. and {De Silva}, Gayandhi M. and {D'Orazi}, Valentina and {Freeman}, Ken C. and {Hayden}, Michael R. and {Kos}, Janez and {Lind}, Karin and {Schlesinger}, Katharine J. and {Sharma}, Sanjib and {Simpson}, Jeffrey D. and {Zucker}, Daniel B. and {Zwitter}, Toma{\v{z}}},
        title = "{The GALAH survey: effective temperature calibration from the InfraRed Flux Method in the Gaia system}",
      journal = {\mnras},
         year = 2021,
        month = oct,
       volume = {507},
       number = {2},
        pages = {2684-2696},
          doi = {10.1093/mnras/stab2304},
archivePrefix = {arXiv},
       eprint = {2011.02517},
 primaryClass = {astro-ph.SR},
       adsurl = {https://ui.adsabs.harvard.edu/abs/2021MNRAS.507.2684C}
}

@ARTICLE{Fortney2007,
       author = {{Fortney}, J.~J. and {Marley}, M.~S. and {Barnes}, J.~W.},
        title = "{Planetary Radii across Five Orders of Magnitude in Mass and Stellar Insolation: Application to Transits}",
      journal = {\apj},
         year = 2007,
        month = apr,
       volume = {659},
       number = {2},
        pages = {1661-1672},
          doi = {10.1086/512120},
archivePrefix = {arXiv},
       eprint = {astro-ph/0612671},
 primaryClass = {astro-ph},
       adsurl = {https://ui.adsabs.harvard.edu/abs/2007ApJ...659.1661F}
}

@ARTICLE{LopFor14,
       author = {{Lopez}, Eric D. and {Fortney}, Jonathan J.},
        title = "{Understanding the Mass-Radius Relation for Sub-neptunes: Radius as a Proxy for Composition}",
      journal = {\apj},
         year = 2014,
        month = sep,
       volume = {792},
       number = {1},
          eid = {1},
        pages = {1},
          doi = {10.1088/0004-637X/792/1/1},
archivePrefix = {arXiv},
       eprint = {1311.0329},
 primaryClass = {astro-ph.EP},
       adsurl = {https://ui.adsabs.harvard.edu/abs/2014ApJ...792....1L}
}

@ARTICLE{kuby+2018a,
       author = {{Kubyshkina}, D. and {Fossati}, L. and {Erkaev}, N.~V. and {Cubillos}, P.~E. and {Johnstone}, C.~P. and {Kislyakova}, K.~G. and {Lammer}, H. and {Lendl}, M. and {Odert}, P.},
        title = "{Overcoming the Limitations of the Energy-limited Approximation for Planet Atmospheric Escape}",
      journal = {\apjl},
         year = 2018,
        month = oct,
       volume = {866},
       number = {2},
          eid = {L18},
        pages = {L18},
          doi = {10.3847/2041-8213/aae586},
archivePrefix = {arXiv},
       eprint = {1810.06920},
 primaryClass = {astro-ph.EP},
       adsurl = {https://ui.adsabs.harvard.edu/abs/2018ApJ...866L..18K}
}

@ARTICLE{choi+2016,
       author = {{Choi}, Jieun and {Dotter}, Aaron and {Conroy}, Charlie and {Cantiello}, Matteo and {Paxton}, Bill and {Johnson}, Benjamin D.},
        title = "{Mesa Isochrones and Stellar Tracks (MIST). I. Solar-scaled Models}",
      journal = {\apj},
         year = 2016,
        month = jun,
       volume = {823},
       number = {2},
          eid = {102},
        pages = {102},
          doi = {10.3847/0004-637X/823/2/102},
archivePrefix = {arXiv},
       eprint = {1604.08592},
 primaryClass = {astro-ph.SR},
       adsurl = {https://ui.adsabs.harvard.edu/abs/2016ApJ...823..102C}
}

@ARTICLE{Caldiroli+2021,
       author = {{Caldiroli}, Andrea and {Haardt}, Francesco and {Gallo}, Elena and {Spinelli}, Riccardo and {Malsky}, Isaac and {Rauscher}, Emily},
        title = "{Irradiation-driven escape of primordial planetary atmospheres. I. The ATES photoionization hydrodynamics code}",
      journal = {\aap},
         year = 2021,
        month = nov,
       volume = {655},
          eid = {A30},
        pages = {A30},
          doi = {10.1051/0004-6361/202141497},
archivePrefix = {arXiv},
       eprint = {2106.10294},
 primaryClass = {astro-ph.EP},
       adsurl = {https://ui.adsabs.harvard.edu/abs/2021A&A...655A..30C}
}

@ARTICLE{Caldiroli+2022,
       author = {{Caldiroli}, Andrea and {Haardt}, Francesco and {Gallo}, Elena and {Spinelli}, Riccardo and {Malsky}, Isaac and {Rauscher}, Emily},
        title = "{Irradiation-driven escape of primordial planetary atmospheres. II. Evaporation efficiency of sub-Neptunes through hot Jupiters}",
      journal = {\aap},
         year = 2022,
        month = jul,
       volume = {663},
          eid = {A122},
        pages = {A122},
          doi = {10.1051/0004-6361/202142763},
archivePrefix = {arXiv},
       eprint = {2112.00744},
 primaryClass = {astro-ph.EP},
       adsurl = {https://ui.adsabs.harvard.edu/abs/2022A&A...663A.122C}
}

@ARTICLE{Penz08a,
       author = {{Penz}, T. and {Micela}, G. and {Lammer}, H.},
        title = "{Influence of the evolving stellar X-ray luminosity distribution on exoplanetary mass loss}",
      journal = {\aap},
         year = 2008,
        month = jan,
       volume = {477},
       number = {1},
        pages = {309-314},
          doi = {10.1051/0004-6361:20078364},
       adsurl = {https://ui.adsabs.harvard.edu/abs/2008A&A...477..309P}
}

@ARTICLE{SF25,
       author = {{Sanz-Forcada}, J. and {L{\'o}pez-Puertas}, M. and {Lamp{\'o}n}, M. and {Czesla}, S. and {Nortmann}, L. and {Caballero}, J.~A. and {Zapatero Osorio}, M.~R. and {Amado}, P.~J. and {Murgas}, F. and {Orell-Miquel}, J. and {Pall{\'e}}, E. and {Quirrenbach}, A. and {Reiners}, A. and {Ribas}, I. and {S{\'a}nchez-L{\'o}pez}, A. and {Solano}, E.},
        title = "{Connection between planetary He I {\ensuremath{\lambda}}10 830 {\r{A}} absorption and extreme-ultraviolet emission of planet-host stars}",
      journal = {\aap},
         year = 2025,
        month = jan,
       volume = {693},
          eid = {A285},
        pages = {A285},
          doi = {10.1051/0004-6361/202451680},
archivePrefix = {arXiv},
       eprint = {2501.03716},
 primaryClass = {astro-ph.EP},
       adsurl = {https://ui.adsabs.harvard.edu/abs/2025A&A...693A.285S}
}

@ARTICLE{Locci19,
       author = {{Locci}, Daniele and {Cecchi-Pestellini}, Cesare and {Micela}, Giuseppina},
        title = "{Photo-evaporation of close-in gas giants orbiting around G and M stars}",
      journal = {\aap},
         year = 2019,
        month = apr,
       volume = {624},
          eid = {A101},
        pages = {A101},
          doi = {10.1051/0004-6361/201834491},
archivePrefix = {arXiv},
       eprint = {1903.10911},
 primaryClass = {astro-ph.EP},
       adsurl = {https://ui.adsabs.harvard.edu/abs/2019A&A...624A.101L}
}

@ARTICLE{Mantovan+2024,
       author = {{Mantovan}, G. and {Malavolta}, L. and {Desidera}, S. and {Zingales}, T. and {Borsato}, L. and {Piotto}, G. and {Maggio}, A. and {Locci}, D. and {Polychroni}, D. and {Turrini}, D. and {Baratella}, M. and {Biazzo}, K. and {Nardiello}, D. and {Stassun}, K. and {Nascimbeni}, V. and {Benatti}, S. and {Anna John}, A. and {Watkins}, C. and {Bieryla}, A. and {Lissauer}, J.~J. and {Twicken}, J.~D. and {Lanza}, A.~F. and {Winn}, J.~N. and {Messina}, S. and {Montalto}, M. and {Sozzetti}, A. and {Boffin}, H. and {Cheryasov}, D. and {Strakhov}, I. and {Murgas}, F. and {D'Arpa}, M. and {Barkaoui}, K. and {Benni}, P. and {Bignamini}, A. and {Bonomo}, A.~S. and {Borsa}, F. and {Cabona}, L. and {Cameron}, A.~C. and {Claudi}, R. and {Cochran}, W. and {Collins}, K.~A. and {Damasso}, M. and {Dong}, J. and {Endl}, M. and {Fukui}, A. and {F{\H{u}}r{\'e}sz}, G. and {Gandolfi}, D. and {Ghedina}, A. and {Jenkins}, J. and {Kab{\'a}th}, P. and {Latham}, D.~W. and {Lorenzi}, V. and {Luque}, R. and {Maldonado}, J. and {McLeod}, K. and {Molinaro}, M. and {Narita}, N. and {Nowak}, G. and {Orell-Miquel}, J. and {Pall{\'e}}, E. and {Parviainen}, H. and {Pedani}, M. and {Quinn}, S.~N. and {Relles}, H. and {Rowden}, P. and {Scandariato}, G. and {Schwarz}, R. and {Seager}, S. and {Shporer}, A. and {Vanderburg}, A. and {Wilson}, T.~G.},
        title = "{The GAPS programme at TNG. XLIX. TOI-5398, the youngest compact multi-planet system composed of an inner sub-Neptune and an outer warm Saturn}",
      journal = {\aap},
         year = 2024,
        month = feb,
       volume = {682},
          eid = {A129},
        pages = {A129},
          doi = {10.1051/0004-6361/202347472},
archivePrefix = {arXiv},
       eprint = {2310.16888},
 primaryClass = {astro-ph.EP},
       adsurl = {https://ui.adsabs.harvard.edu/abs/2024A&A...682A.129M}
}

@ARTICLE{Mantovan24b,
       author = {{Mantovan}, G. and {Malavolta}, L. and {Locci}, D. and {Polychroni}, D. and {Turrini}, D. and {Maggio}, A. and {Desidera}, S. and {Spinelli}, R. and {Benatti}, S. and {Piotto}, G. and {Lanza}, A.~F. and {Marzari}, F. and {Sozzetti}, A. and {Damasso}, M. and {Nardiello}, D. and {Cabona}, L. and {D'Arpa}, M. and {Guilluy}, G. and {Mancini}, L. and {Micela}, G. and {Nascimbeni}, V. and {Zingales}, T.},
        title = "{Orbital obliquity of the young planet TOI-5398 b and the evolutionary history of the system}",
      journal = {\aap},
         year = 2024,
        month = apr,
       volume = {684},
          eid = {L17},
        pages = {L17},
          doi = {10.1051/0004-6361/202449769},
archivePrefix = {arXiv},
       eprint = {2404.02969},
 primaryClass = {astro-ph.EP},
       adsurl = {https://ui.adsabs.harvard.edu/abs/2024A&A...684L..17M}
}

@ARTICLE{lovis_2011_sindex_to_rhk,
       author = {{Lovis}, C. and {Dumusque}, X. and {Santos}, N.~C. and {Bouchy}, F. and {Mayor}, M. and {Pepe}, F. and {Queloz}, D. and {S{\'e}gransan}, D. and {Udry}, S.},
        title = "{The HARPS search for southern extra-solar planets. XXXI. Magnetic activity cycles in solar-type stars: statistics and impact on precise radial velocities}",
      journal = {arXiv e-prints},
         year = 2011,
        month = jul,
          eid = {arXiv:1107.5325},
        pages = {arXiv:1107.5325},
          doi = {10.48550/arXiv.1107.5325},
archivePrefix = {arXiv},
       eprint = {1107.5325},
 primaryClass = {astro-ph.SR},
       adsurl = {https://ui.adsabs.harvard.edu/abs/2011arXiv1107.5325L}
}

@ARTICLE{covino_2013_gapsI,
       author = {{Covino}, E. and {Esposito}, M. and {Barbieri}, M. and {Mancini}, L. and {Nascimbeni}, V. and {Claudi}, R. and {Desidera}, S. and {Gratton}, R. and {Lanza}, A.~F. and {Sozzetti}, A. and {Biazzo}, K. and {Affer}, L. and {Gandolfi}, D. and {Munari}, U. and {Pagano}, I. and {Bonomo}, A.~S. and {Collier Cameron}, A. and {H{\'e}brard}, G. and {Maggio}, A. and {Messina}, S. and {Micela}, G. and {Molinari}, E. and {Pepe}, F. and {Piotto}, G. and {Ribas}, I. and {Santos}, N.~C. and {Southworth}, J. and {Shkolnik}, E. and {Triaud}, A.~H.~M.~J. and {Bedin}, L. and {Benatti}, S. and {Boccato}, C. and {Bonavita}, M. and {Borsa}, F. and {Borsato}, L. and {Brown}, D. and {Carolo}, E. and {Ciceri}, S. and {Cosentino}, R. and {Damasso}, M. and {Faedi}, F. and {Mart{\'\i}nez Fiorenzano}, A.~F. and {Latham}, D.~W. and {Lovis}, C. and {Mordasini}, C. and {Nikolov}, N. and {Poretti}, E. and {Rainer}, M. and {Rebolo L{\'o}pez}, R. and {Scandariato}, G. and {Silvotti}, R. and {Smareglia}, R. and {Alcal{\'a}}, J.~M. and {Cunial}, A. and {Di Fabrizio}, L. and {Di Mauro}, M.~P. and {Giacobbe}, P. and {Granata}, V. and {Harutyunyan}, A. and {Knapic}, C. and {Lattanzi}, M. and {Leto}, G. and {Lodato}, G. and {Malavolta}, L. and {Marzari}, F. and {Molinaro}, M. and {Nardiello}, D. and {Pedani}, M. and {Prisinzano}, L. and {Turrini}, D.},
        title = "{The GAPS programme with HARPS-N at TNG. I. Observations of the Rossiter-McLaughlin effect and characterisation of the transiting system Qatar-1}",
      journal = {\aap},
         year = 2013,
        month = jun,
       volume = {554},
          eid = {A28},
        pages = {A28},
          doi = {10.1051/0004-6361/201321298},
archivePrefix = {arXiv},
       eprint = {1304.0005},
 primaryClass = {astro-ph.SR},
       adsurl = {https://ui.adsabs.harvard.edu/abs/2013A&A...554A..28C}
}

@ARTICLE{nardiello2025_toi1430,
       author = {{Nardiello}, D. and {Akana Murphy}, J.~M. and {Spinelli}, R. and {Baratella}, M. and {Desidera}, S. and {Nascimbeni}, V. and {Malavolta}, L. and {Biazzo}, K. and {Maggio}, A. and {Locci}, D. and {Benatti}, S. and {Batalha}, N.~M. and {D'Orazi}, V. and {Borsato}, L. and {Piotto}, G. and {Oelkers}, R.~J. and {Mallonn}, M. and {Sozzetti}, A. and {Bedin}, L.~R. and {Mantovan}, G. and {Zingales}, T. and {Affer}, L. and {Bignamini}, A. and {Bonomo}, A.~S. and {Cabona}, L. and {Collins}, K.~A. and {Damasso}, M. and {Filomeno}, S. and {Ghedina}, A. and {Harutyunyan}, A. and {Lanza}, A.~F. and {Mancini}, L. and {Rainer}, M. and {Scandariato}, G. and {Schwarz}, R.~P. and {Sefako}, R. and {Srdoc}, G.},
        title = "{The GAPS Programme at TNG: LXV. Precise density measurement of TOI-1430 b, a young planet with an evaporating atmosphere}",
      journal = {\aap},
         year = 2025,
        month = jan,
       volume = {693},
          eid = {A32},
        pages = {A32},
          doi = {10.1051/0004-6361/202452236},
archivePrefix = {arXiv},
       eprint = {2411.12795},
 primaryClass = {astro-ph.EP},
       adsurl = {https://ui.adsabs.harvard.edu/abs/2025A&A...693A..32N}
}

@ARTICLE{petigura_2022,
       author = {{Petigura}, Erik A. and {Rogers}, James G. and {Isaacson}, Howard and {Owen}, James E. and {Kraus}, Adam L. and {Winn}, Joshua N. and {MacDougall}, Mason G. and {Howard}, Andrew W. and {Fulton}, Benjamin and {Kosiarek}, Molly R. and {Weiss}, Lauren M. and {Behmard}, Aida and {Blunt}, Sarah},
        title = "{The California-Kepler Survey. X. The Radius Gap as a Function of Stellar Mass, Metallicity, and Age}",
      journal = {\aj},
         year = 2022,
        month = apr,
       volume = {163},
       number = {4},
          eid = {179},
        pages = {179},
          doi = {10.3847/1538-3881/ac51e3},
archivePrefix = {arXiv},
       eprint = {2201.10020},
 primaryClass = {astro-ph.EP},
       adsurl = {https://ui.adsabs.harvard.edu/abs/2022AJ....163..179P}
}

\begin{appendix}

\section{Additional tables and figures}\label{app:tab_fig}

\begin{figure}[h]
    \centering
    \includegraphics[width=0.8\linewidth]{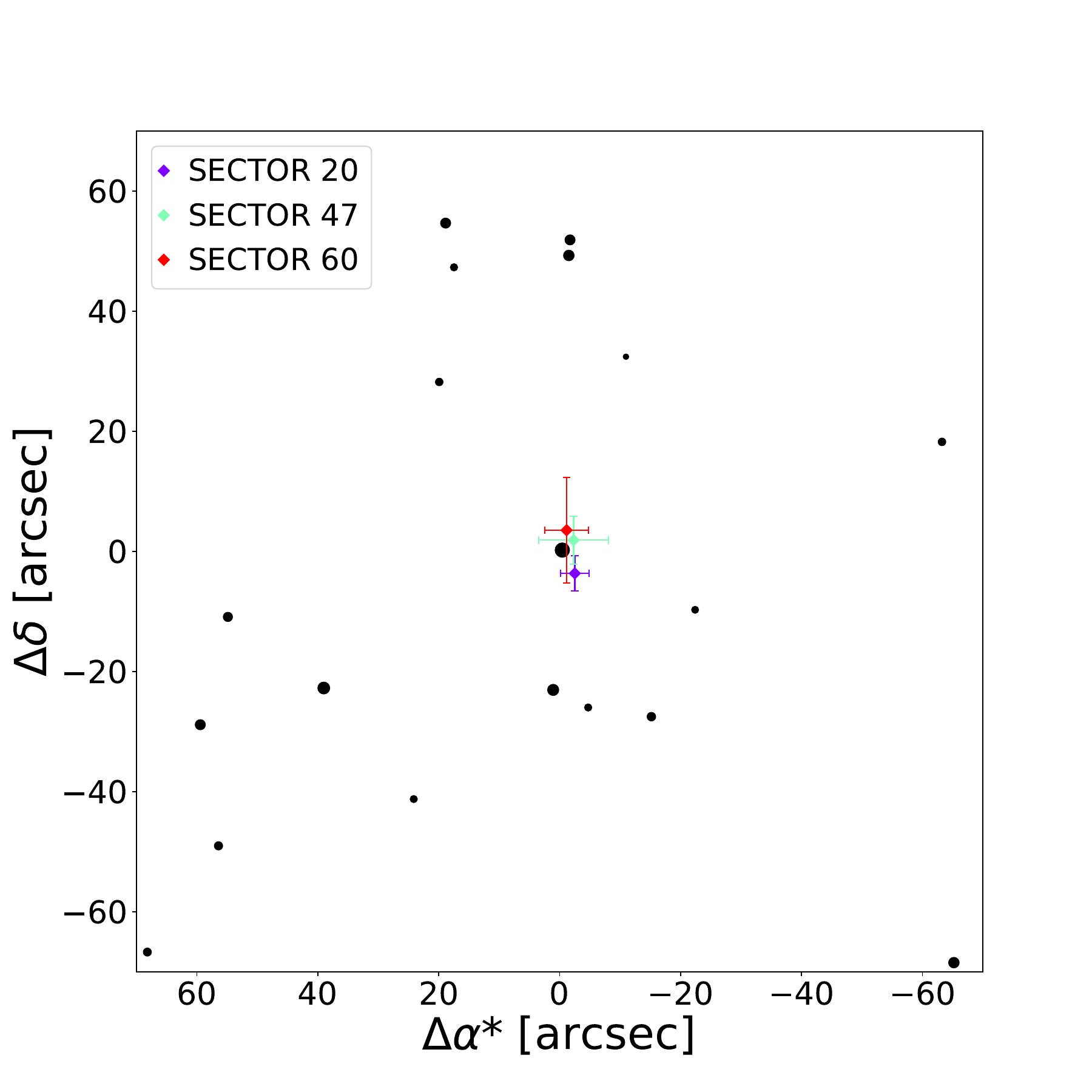}
    \caption{In- and out-of-transit centroid of TOI-5734. Within the errors, the centroid coordinates coincide with the target position (located in (0,0)).}
    \label{fig:inout_centroid}
\end{figure}

\begin{figure}[h]
    \centering
    \includegraphics[width=\linewidth]{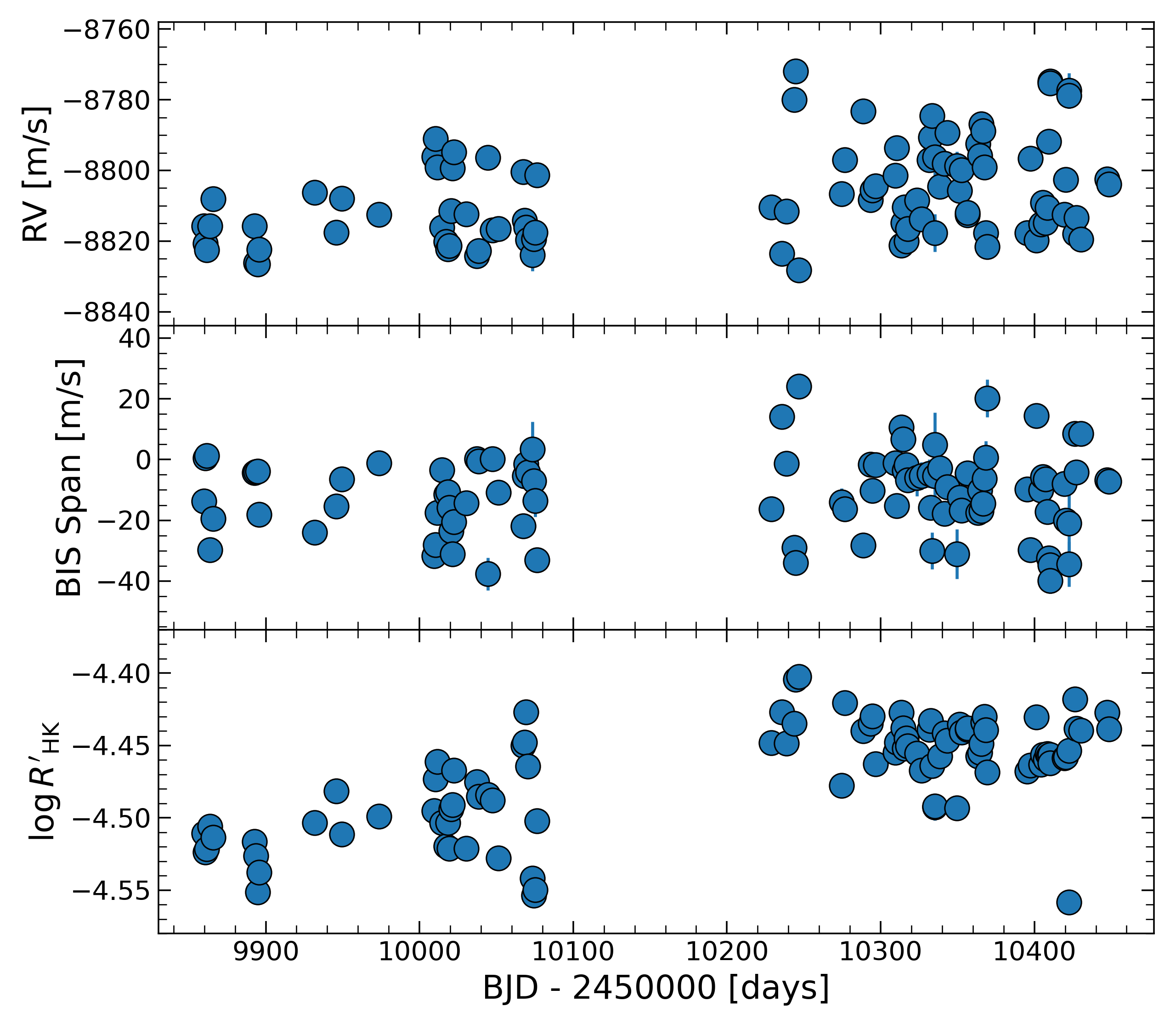}
    \caption{HARPS-N spectroscopic time series used in this work. From top to bottom, the RV, the BIS span and the $\log R^\prime_{\rm HK}$ time series.}
    \label{fig:rv_rhk_bis_timmeseries}
\end{figure}

\begin{figure}[h!]
    \centering
    \includegraphics[width=\linewidth]{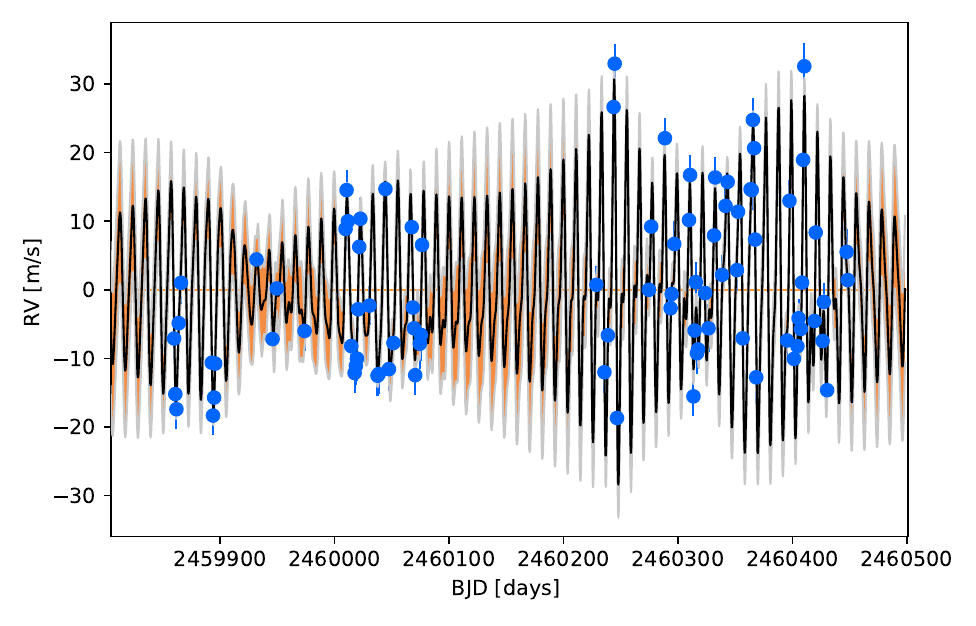}
    \caption{RV data with the GP model component after subtracting the planetary component.}
    \label{fig:rv_o_c_withgp}
\end{figure}

\begin{figure*}[h]
    \centering
    \includegraphics[width=0.75\linewidth]{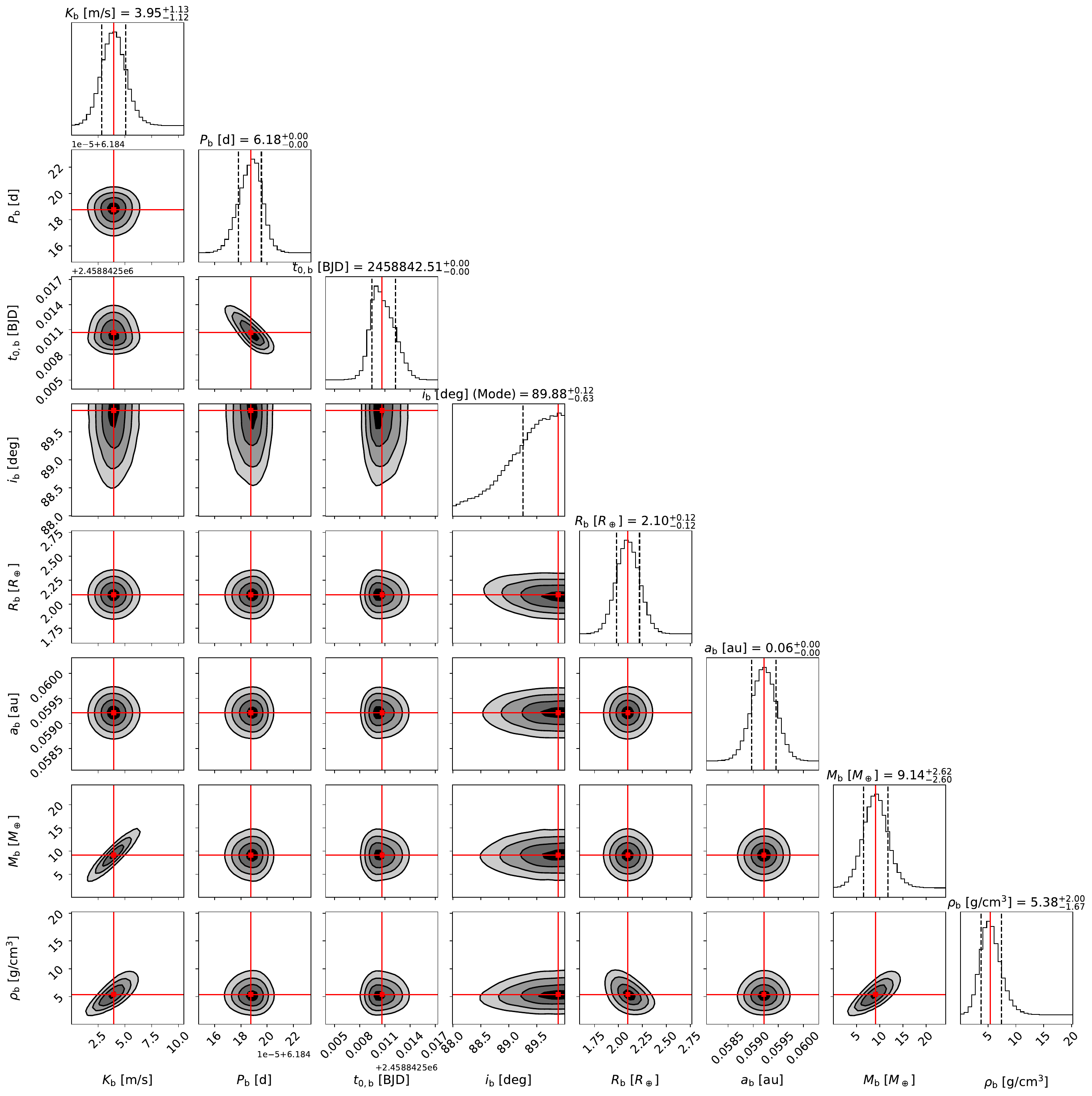}
    \caption{Corner plot of the posterior distribution of the main planetary parameters resulting from the \textit{TESS} and HARPS-N analysis done with a NS scheme for TOI-5734b. The dark contours on the panels represent the 1$\sigma$, 2$\sigma$, and 3$\sigma$ confidence levels of the overall NS. The red line represents the median value of the posterior distribution (the mode in the case of $i_{\rm b}$).}
    \label{fig:cornerplot_planpar}
\end{figure*}

\begin{figure*}[h]
    \centering
    \includegraphics[width=0.8\linewidth]{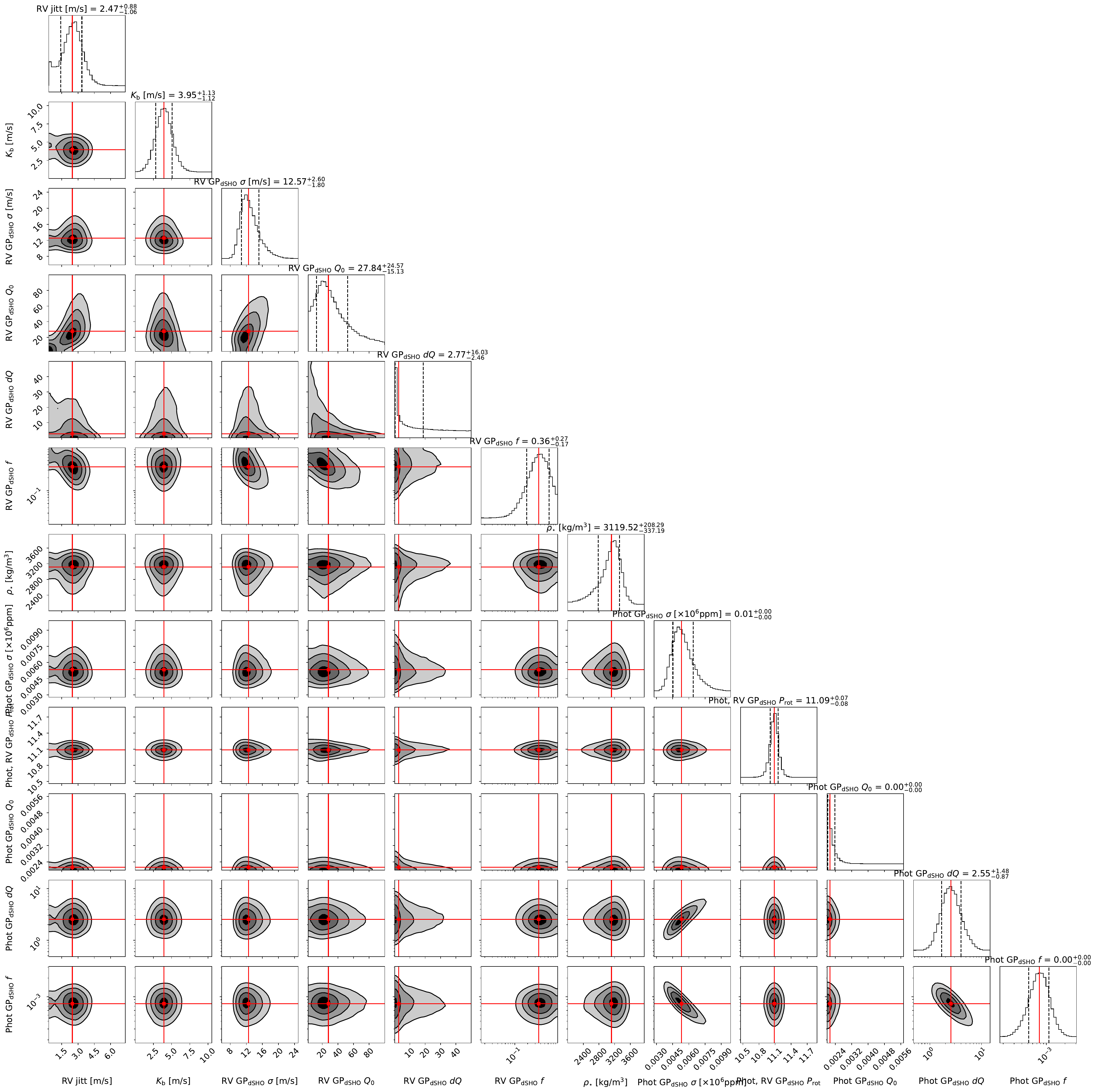}
    \caption{Corner plot of the posterior distribution of the global data resulting from the \textit{TESS} and HARPS-N analysis with the GP rotational kernel for TOI-5734b. The dark contours on the panels represent the 1$\sigma$, 2$\sigma$, and 3$\sigma$ confidence levels of the overall NS. The red line represents the median value of the posterior distribution.}
    \label{fig:cornerplot_stat_data_fit}
\end{figure*}

\begin{figure*}[h!]
\begin{center}
\begin{tabular}{ccc} 
\includegraphics[width=5.5cm]{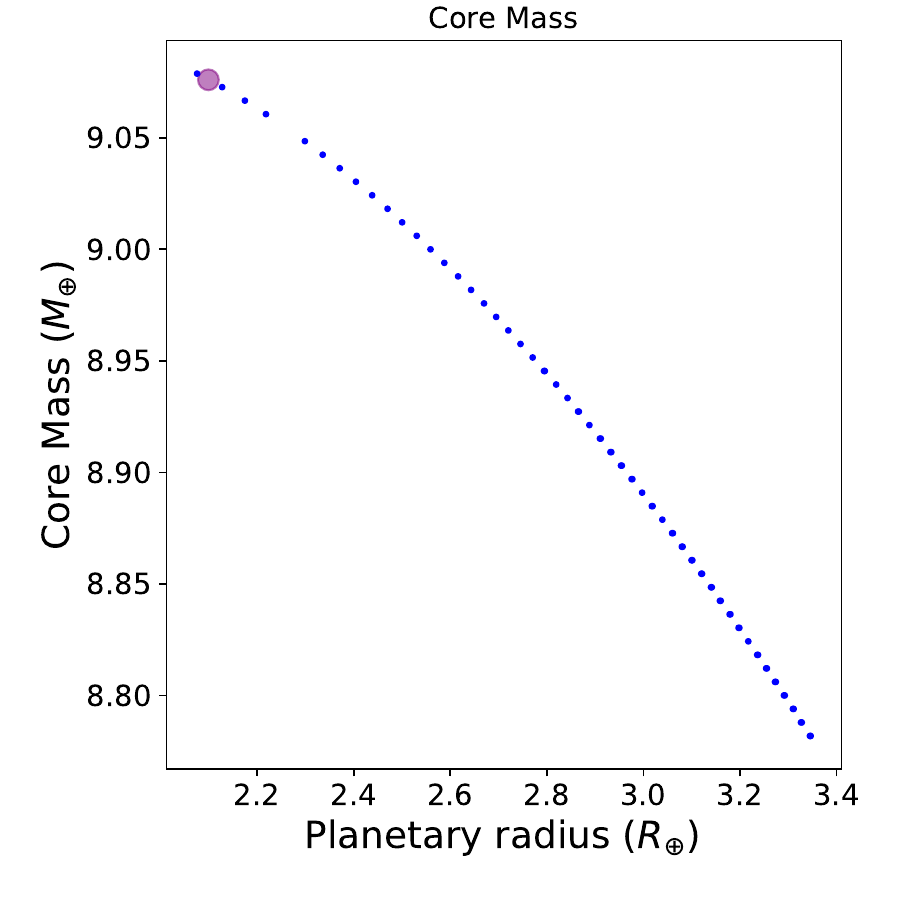} &\includegraphics[width=5.5cm]{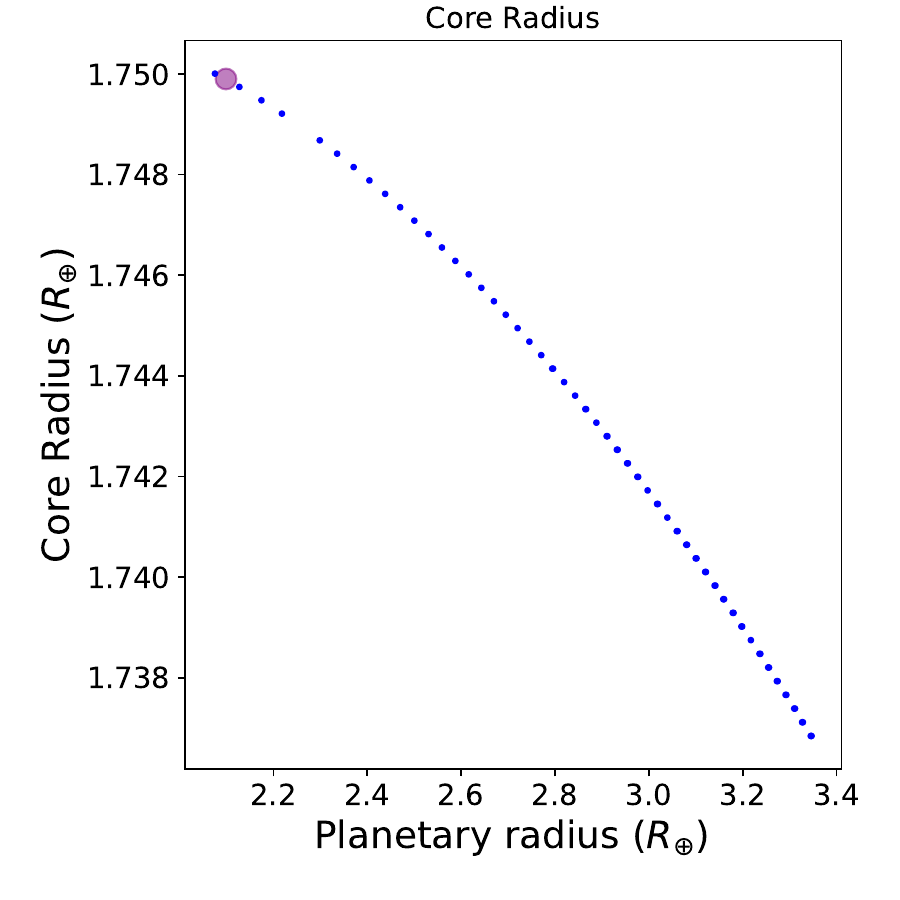} & \includegraphics[width=5.5cm]{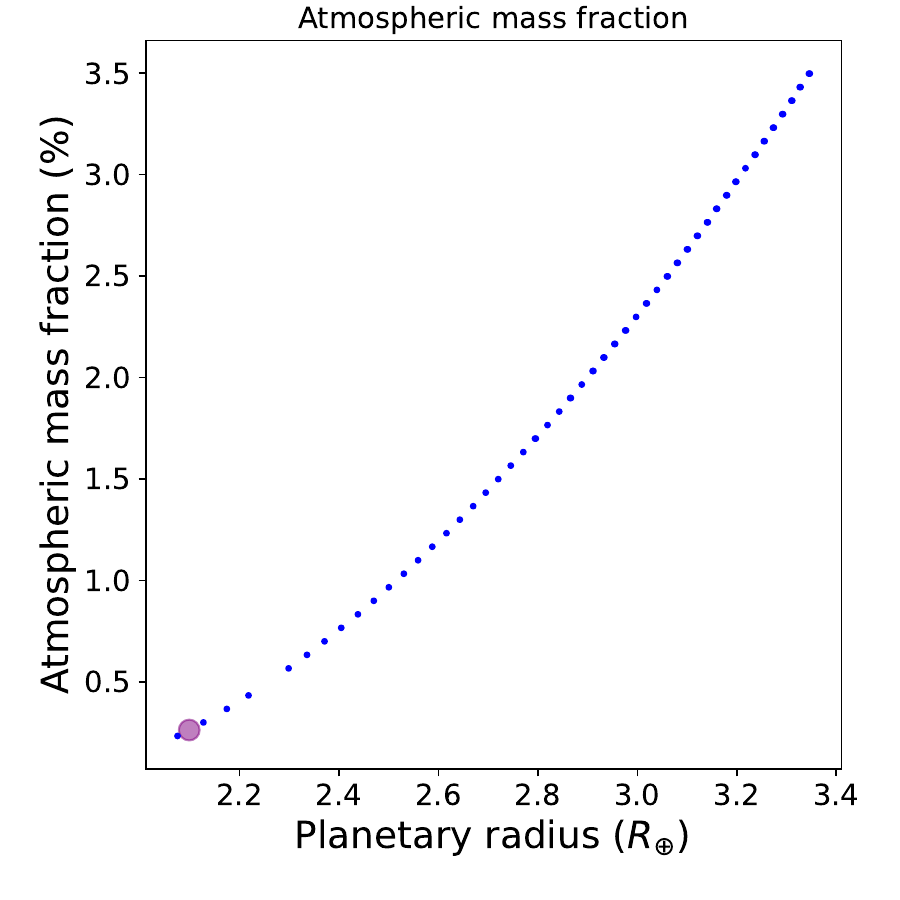} \\
\end{tabular}
\caption{Solutions of core-envelope models for planet TOI-5734b with total mass fixed at the measured value, considering rock-iron core composition with solar metallicity. The left panel shows the core mass versus the planetary radius, the middle panel shows the core radius, and the right panel shows the atmospheric mass fraction. The purple circle shows the position in the diagram of TOI-5734b at the current age.}
\label{fig:mrcore}
\end{center}
\end{figure*}

The study exploited the light curves from three \textit{TESS} sectors (20, 47, 60). We checked that the detected transits originate on our target and are not associated with a neighbour source using the procedure in \cite{Nardiello+_2020_pathos}. This was done by looking at the differences between the mean in- and out-of-transit centroids of each sector, where the centroid and the target (at (0,0) coordinates) show compatible positions, within the errors, as reported in Fig.\,\ref{fig:inout_centroid}.
The analysis was conducted with spectroscopic data extracted from 97 HARPS-N high-resolution spectra. The time series of RVs and the BIS span, and $\log R^\prime_{\rm HK}$ activity indices are tabulated in Table\,\ref{tab:toi_5734_rv_act_time_series} and displayed in Fig.\,\ref{fig:rv_rhk_bis_timmeseries}.
The bottom panel of Fig.\,\ref{fig:rv_rhk_bis_timmeseries} shows an increasing trend of the $\log R^\prime_{\rm HK}$. This trend implies a higher level of activity variability in the second observational season, which leads to more scattered radial-velocity measurements as shown in the top panel of Fig.\,\ref{fig:rv_rhk_bis_timmeseries}.
Fig.\,\ref{fig:rv_o_c_withgp} shows the RV time series fitted with the GP model activity component after subtracting the planetary signal. As shown in the figure, the higher scattering of RV data in the second observational season is completely absorbed by the GP model.
The best fit parameters of our simultaneous RV and light curve model with the GP hyperparameters used to remove the stellar contribution are presented in the corner plots of Fig.\,\ref{fig:cornerplot_planpar} and Fig.\,\ref{fig:cornerplot_stat_data_fit}, showing respectively the posterior distribution of the main planetary parameters and that of the global fit data, such as jitter, GP hyperparameters and LDC, each with the 1\,$\sigma$, 2\,$\sigma$, and 3\,$\sigma$ confidence levels. Nested sampling priors and posteriors (median and the 16$^{th}$ and 84$^{th}$ percentiles) of the GP hyperparameters and the other system parameters used in the analysis are reported in Table\,\ref{tab:prior_and_param_other}.
Following the discussion on atmospheric evolution simulations in Sect.\,\ref{subsec:atm_evol}, we report in Figure\,\ref{fig:mrcore} the solutions of the core-envelope model assuming a rock-iron core composition, where the core mass, core radius, and atmospheric mass fraction, at the present age, are plotted as a function of the planetary radius.

\setlength{\tabcolsep}{3pt}
\begin{table*}[t]
\centering
\tiny
\caption{Time series of TOI-5734 from HARPS-N data showing Barycentric Julian Dates (BJD), radial velocities (RV) with uncertainties and the BIS span and $R'_{\rm HK}$ activity indices with their uncertainties.}
\label{tab:toi_5734_rv_act_time_series}
\begin{minipage}[t]{0.5\textwidth}
\vspace{0pt}
\begin{tabular}{lrrrrrr}
\hline\hline
Epoch & RV & $\sigma_{RV}$  & BIS span  & $\sigma_{BIS}$  & $\log R^\prime_{\rm HK}$ & $\sigma_{\log R^\prime_{\rm HK}}$ \\
\hline
(BJD) & (m/s) & (m/s) & (m/s) & (m/s) & (dex) & (dex) \\
\hline
2459859.718602 & $-$8815.68 & 1.42 & $-$13.74 & 2.83 & $-$4.5107 & 0.0012 \\
2459860.759271 & $-$8820.67 & 1.14 & 0.44 & 2.27 & $-$4.5238 & 0.0008 \\
2459861.737958 & $-$8822.52 & 1.26 & 1.18 & 2.52 & $-$4.5215 & 0.0010 \\
2459863.762178 & $-$8815.72 & 0.87 & $-$29.73 & 1.75 & $-$4.5060 & 0.0005 \\
2459865.737363 & $-$8808.12 & 1.70 & $-$19.47 & 3.41 & $-$4.5137 & 0.0016 \\
2459892.635310 & $-$8815.71 & 1.11 & $-$4.36 & 2.21 & $-$4.5164 & 0.0008 \\
2459893.684689 & $-$8826.11 & 0.99 & $-$4.07 & 1.98 & $-$4.5264 & 0.0007 \\
2459894.663053 & $-$8826.53 & 1.59 & $-$3.86 & 3.17 & $-$4.5515 & 0.0014 \\
2459895.601393 & $-$8822.32 & 0.99 & $-$18.10 & 1.99 & $-$4.5378 & 0.0007 \\
2459931.696051 & $-$8806.23 & 1.21 & $-$24.03 & 2.42 & $-$4.5034 & 0.0009 \\
2459945.727279 & $-$8817.56 & 1.08 & $-$15.34 & 2.16 & $-$4.4815 & 0.0007 \\
2459949.442528 & $-$8807.93 & 1.58 & $-$6.40 & 3.16 & $-$4.5115 & 0.0015 \\
2459973.676069 & $-$8812.49 & 1.19 & $-$1.20 & 2.37 & $-$4.4991 & 0.0010 \\
2460009.513359 & $-$8796.13 & 1.54 & $-$31.71 & 3.08 & $-$4.4951 & 0.0014 \\
2460010.403723 & $-$8791.00 & 1.14 & $-$28.05 & 2.28 & $-$4.4733 & 0.0008 \\
2460011.546551 & $-$8799.04 & 0.98 & $-$17.50 & 1.97 & $-$4.4611 & 0.0006 \\
2460014.526314 & $-$8816.08 & 1.66 & $-$3.46 & 3.32 & $-$4.5035 & 0.0016 \\
2460017.423717 & $-$8820.10 & 1.28 & $-$11.47 & 2.57 & $-$4.5197 & 0.0010 \\
2460018.529524 & $-$8822.28 & 0.90 & $-$10.75 & 1.81 & $-$4.5033 & 0.0006 \\
2460019.496884 & $-$8821.35 & 2.00 & $-$15.76 & 4.01 & $-$4.5211 & 0.0023 \\
2460020.513919 & $-$8811.40 & 1.86 & $-$23.59 & 3.72 & $-$4.4939 & 0.0019 \\
2460021.483588 & $-$8799.35 & 1.73 & $-$31.11 & 3.45 & $-$4.4912 & 0.0016 \\
2460022.504791 & $-$8794.74 & 1.30 & $-$20.46 & 2.59 & $-$4.4674 & 0.0010 \\
2460030.408504 & $-$8812.31 & 1.04 & $-$14.38 & 2.08 & $-$4.5212 & 0.0007 \\
2460037.438803 & $-$8824.13 & 1.19 & 0.14 & 2.38 & $-$4.4751 & 0.0009 \\
2460038.456733 & $-$8822.75 & 1.14 & $-$0.60 & 2.27 & $-$4.4854 & 0.0008 \\
2460044.368221 & $-$8796.40 & 2.67 & $-$37.67 & 5.33 & $-$4.4840 & 0.0030 \\
2460047.385986 & $-$8816.87 & 1.75 & 0.22 & 3.50 & $-$4.4879 & 0.0016 \\
2460051.376628 & $-$8816.51 & 1.41 & $-$10.80 & 2.82 & $-$4.5279 & 0.0012 \\
2460067.362639 & $-$8800.42 & 1.28 & $-$21.82 & 2.55 & $-$4.4501 & 0.0009 \\
2460068.364604 & $-$8814.20 & 1.27 & $-$5.41 & 2.55 & $-$4.4479 & 0.0009 \\
2460069.366428 & $-$8816.11 & 1.83 & $-$1.46 & 3.67 & $-$4.4271 & 0.0015 \\
2460070.370593 & $-$8819.70 & 1.23 & $-$4.14 & 2.45 & $-$4.4645 & 0.0009 \\
2460073.373663 & $-$8823.96 & 4.50 & 3.43 & 9.01 & $-$4.5419 & 0.0077 \\
2460074.375367 & $-$8819.33 & 2.62 & $-$7.08 & 5.23 & $-$4.5535 & 0.0037 \\
2460075.372929 & $-$8817.50 & 2.64 & $-$13.55 & 5.28 & $-$4.5500 & 0.0037 \\
2460076.363464 & $-$8801.36 & 1.29 & $-$32.99 & 2.59 & $-$4.5023 & 0.0010 \\
2460228.738010 & $-$8810.37 & 0.89 & $-$16.23 & 1.77 & $-$4.4483 & 0.0005 \\
2460235.748317 & $-$8823.56 & 1.33 & 14.11 & 2.66 & $-$4.4269 & 0.0009 \\
2460238.758893 & $-$8811.55 & 1.01 & $-$1.35 & 2.01 & $-$4.4486 & 0.0006 \\
2460243.736822 & $-$8779.95 & 1.43 & $-$29.03 & 2.85 & $-$4.4350 & 0.0009 \\
2460244.690282 & $-$8771.92 & 1.10 & $-$33.92 & 2.19 & $-$4.4043 & 0.0006 \\
2460246.718293 & $-$8828.28 & 1.17 & 24.07 & 2.34 & $-$4.4024 & 0.0007 \\
2460274.653577 & $-$8806.60 & 2.28 & $-$14.01 & 4.57 & $-$4.4776 & 0.0021 \\
2460276.621590 & $-$8797.05 & 1.09 & $-$16.25 & 2.18 & $-$4.4208 & 0.0007 \\
2460288.615907 & $-$8783.29 & 1.17 & $-$28.2 & 2.34 & $-$4.4399 & 0.0007 \\
2460293.563462 & $-$8808.35 & 1.00 & $-$1.64 & 2.00 & $-$4.4350 & 0.0006 \\
2460294.500395 & $-$8805.60 & 0.89 & $-$10.29 & 1.79 & $-$4.4298 & 0.0005 \\
2460296.748428 & $-$8804.38 & 1.99 & $-$1.78 & 3.98 & $-$4.4630 & 0.0019 \\
\hline
\end{tabular}
\end{minipage}\hfill
\begin{minipage}[t]{0.5\textwidth}
\vspace{0pt}
\begin{tabular}{lrrrrrr}
\hline\hline
Epoch & RV & $\sigma_{RV}$  & BIS span  & $\sigma_{BIS}$  & $\log R^\prime_{\rm HK}$ & $\sigma_{\log R^\prime_{\rm HK}}$ \\
\hline
(BJD) & (m/s) & (m/s) & (m/s) & (m/s) & (dex) & (dex) \\
\hline

2460309.642414 & $-$8801.49 & 1.22 & $-$1.14 & 2.44 & $-$4.4549 & 0.0008 \\
2460310.589911 & $-$8793.65 & 1.30 & $-$15.17 & 2.60 & $-$4.4479 & 0.0009 \\
2460313.513376 & $-$8821.25 & 1.09 & 10.62 & 2.17 & $-$4.4273 & 0.0006 \\
2460314.514068 & $-$8814.77 & 1.03 & 6.77 & 2.06 & $-$4.4382 & 0.0006 \\
2460315.615895 & $-$8810.41 & 1.24 & $-$3.51 & 2.48 & $-$4.4519 & 0.0008 \\
2460316.609080 & $-$8820.05 & 1.45 & $-$1.80 & 2.90 & $-$4.4450 & 0.0011 \\
2460317.584404 & $-$8816.46 & 1.05 & $-$6.69 & 2.10 & $-$4.4506 & 0.0006 \\
2460323.714860 & $-$8808.43 & 3.01 & $-$6.04 & 6.02 & $-$4.4555 & 0.0033 \\
2460326.689945 & $-$8813.78 & 2.29 & $-$5.43 & 4.57 & $-$4.4672 & 0.0025 \\
2460331.544730 & $-$8797.01 & 2.34 & $-$4.58 & 4.67 & $-$4.4391 & 0.0021 \\
2460332.545389 & $-$8790.69 & 1.46 & $-$15.82 & 2.92 & $-$4.4331 & 0.0011 \\
2460333.474548 & $-$8784.58 & 3.02 & $-$30.0 & 6.04 & $-$4.4641 & 0.0033 \\
2460335.376083 & $-$8796.24 & 5.86 & $-$5.37 & 11.72 & $-$4.4927 & 0.0090 \\
2460335.388900 & $-$8817.70 & 5.32 & 4.90 & 10.64 & $-$4.4921 & 0.0077 \\
2460338.608660 & $-$8804.50 & 1.06 & $-$2.78 & 2.11 & $-$4.4574 & 0.0007 \\
2460341.557423 & $-$8798.03 & 1.03 & $-$17.78 & 2.06 & $-$4.4416 & 0.0006 \\
2460343.437783 & $-$8789.28 & 1.16 & $-$9.08 & 2.32 & $-$4.4468 & 0.0007 \\
2460349.476362 & $-$8798.75 & 4.11 & $-$31.09 & 8.21 & $-$4.4933 & 0.0052 \\
2460351.535586 & $-$8805.68 & 1.40 & $-$12.5 & 2.80 & $-$4.4354 & 0.0011 \\
2460352.475485 & $-$8799.81 & 1.15 & $-$16.8 & 2.30 & $-$4.4410 & 0.0007 \\
2460356.509249 & $-$8812.56 & 1.31 & $-$4.69 & 2.63 & $-$4.4390 & 0.0009 \\
2460356.535491 & $-$8811.93 & 1.45 & $-$4.53 & 2.90 & $-$4.4380 & 0.0011 \\
2460363.464579 & $-$8792.40 & 1.12 & $-$17.55 & 2.24 & $-$4.4573 & 0.0006 \\
2460364.465036 & $-$8795.80 & 1.14 & $-$10.08 & 2.28 & $-$4.4549 & 0.0007 \\
2460365.475208 & $-$8786.88 & 1.70 & $-$16.92 & 3.41 & $-$4.4488 & 0.0013 \\
2460366.548367 & $-$8788.86 & 1.04 & $-$14.47 & 2.08 & $-$4.4340 & 0.0006 \\
2460367.464551 & $-$8799.14 & 1.01 & $-$6.30 & 2.02 & $-$4.4301 & 0.0006 \\
2460368.393237 & $-$8817.63 & 2.77 & 0.62 & 5.54 & $-$4.4393 & 0.0029 \\
2460369.418187 & $-$8821.60 & 3.12 & 20.16 & 6.25 & $-$4.4686 & 0.0035 \\
2460395.361610 & $-$8817.65 & 1.14 & $-$9.73 & 2.29 & $-$4.4678 & 0.0007 \\
2460397.427184 & $-$8796.67 & 1.44 & $-$29.72 & 2.88 & $-$4.4637 & 0.0011 \\
2460401.351928 & $-$8819.75 & 1.11 & 14.39 & 2.22 & $-$4.4305 & 0.0007 \\
2460404.375480 & $-$8815.28 & 1.50 & $-$10.07 & 3.00 & $-$4.4631 & 0.0012 \\
2460405.384182 & $-$8809.07 & 1.08 & $-$5.68 & 2.15 & $-$4.4564 & 0.0007 \\
2460407.375590 & $-$8814.87 & 1.53 & $-$6.43 & 3.05 & $-$4.4597 & 0.0013 \\
2460408.417507 & $-$8810.51 & 1.16 & $-$17.24 & 2.33 & $-$4.4560 & 0.0008 \\
2460409.377214 & $-$8791.85 & 1.08 & $-$32.43 & 2.15 & $-$4.4572 & 0.0007 \\
2460410.388270 & $-$8774.82 & 2.08 & $-$34.67 & 4.16 & $-$4.4566 & 0.0019 \\
2460410.401916 & $-$8775.25 & 2.15 & $-$39.82 & 4.29 & $-$4.4622 & 0.0021 \\
2460419.384873 & $-$8812.50 & 1.06 & $-$8.01 & 2.12 & $-$4.4588 & 0.0007 \\
2460420.379038 & $-$8802.60 & 1.34 & $-$20.03 & 2.68 & $-$4.4579 & 0.0010 \\
2460422.398055 & $-$8777.42 & 5.02 & $-$21.02 & 10.04 & $-$4.5583 & 0.0080 \\
2460422.413563 & $-$8778.79 & 3.70 & $-$34.46 & 7.40 & $-$4.4535 & 0.0040 \\
2460426.385676 & $-$8817.99 & 1.04 & 8.54 & 2.07 & $-$4.4180 & 0.0006 \\
2460427.433739 & $-$8813.33 & 0.94 & $-$4.19 & 1.89 & $-$4.4384 & 0.0005 \\
2460430.365149 & $-$8819.51 & 1.13 & 8.53 & 2.27 & $-$4.4397 & 0.0007 \\
2460447.392576 & $-$8802.45 & 1.95 & $-$6.80 & 3.91 & $-$4.4274 & 0.0018 \\
2460448.369545 & $-$8803.85 & 1.25 & $-$7.28 & 2.51 & $-$4.4386 & 0.0009 \\
\\
\hline
\end{tabular}
\end{minipage}
\end{table*}

\setlength{\tabcolsep}{10pt}
\begin{table*}[ht]
\tiny
\caption{Nested sampling priors and posteriors (median and the 16$^{th}$ and 84$^{th}$ percentiles) of the GP hyperparameters and the other system parameters used in the analysis of TOI-5734b.}
\label{tab:prior_and_param_other}
\begin{center}
\begin{tabular}{lcrr}
\hline\hline
\noalign{\vskip 1.mm}
Parameter & Unit & Adopted Priors & Posterior \\ \noalign{\vskip 1.mm} 
\hline \noalign{\vskip 1.mm} 
\multicolumn{4}{l}{\textit{Other system parameters}} \\
\noalign{\vskip 1.mm} 
RV offset & m/s & $\mathcal{U}$(-8850, $-$8750) & $-8809.1_{-0.7}^{+0.7}$ \\ \noalign{\vskip 1.mm}
RV jitter & m/s & $\mathcal{J}$(0.30, 10) & $2.5_{-1.1}^{+0.9}$ \\ \noalign{\vskip 1.mm}
RV GP$_{\rm dSHO}$ $\sigma$ & m/s & $\mathcal{U}$(5, 25) & $12.6_{-1.8}^{+2.6}$ \\ \noalign{\vskip 1.mm}
RV GP$_{\rm dSHO}$ $Q_0$ &  & $\mathcal{J}$(2, 100) & $28_{-15}^{+25}$ \\ \noalign{\vskip 1.mm}
RV GP$_{\rm dSHO}$ $dQ$ &  & $\mathcal{J}$(0.1, 50) & $2.8_{-2.5}^{+16.0}$ \\ \noalign{\vskip 1.mm}
RV GP$_{\rm dSHO}$ $f$ &   & $\mathcal{U}$(0.00, 1.00) & $0.36_{-0.17}^{+0.27}$ \\ \noalign{\vskip 1.mm}
Photometry off.$_{\rm TESS-S20}$ & [$\times10^6$] ppm & $\mathcal{U}$(-0.005, 0.005) & $-0.0013_{-0.0011}^{+0.0011}$ \\ \noalign{\vskip 1.mm}
Photometry off.$_{\rm TESS-S47}$ & [$\times10^6$] ppm & $\mathcal{U}$(-0.005, 0.005) & $0.0013_{-0.0011}^{+0.0011}$ \\ \noalign{\vskip 1.mm}
Photometry off.$_{\rm TESS-S60}$ & [$\times10^6$] ppm & $\mathcal{U}$(-0.005, 0.005) & $-0.0009_{-0.0011}^{+0.0011}$ \\ \noalign{\vskip 1.mm}
Photometry jit.$_{\rm TESS-S20}$ & [$\times10^6$] ppm & $\mathcal{U}$(0.000, 0.001) & $0.0001884_{-0.0000058}^{+0.0000060}$ \\ \noalign{\vskip 1.mm}
Photometry jit.$_{\rm TESS-S47}$ & [$\times10^6$] ppm & $\mathcal{U}$(0.000, 0.001) & $0.0002142_{-0.0000046}^{+0.0000045}$ \\ \noalign{\vskip 1.mm}
Photometry jit.$_{\rm TESS-S60}$ & [$\times10^6$] ppm & $\mathcal{U}$(0.000, 0.001) & $0.0003850_{-0.0000049}^{+0.0000050}$ \\ \noalign{\vskip 1.mm}
Photometry GP$_{\rm dSHO}$ $\sigma$ & [$\times10^6$] ppm & $\mathcal{U}$(0.00001, 0.01500) & $0.0053_{-0.0008}^{+0.0011}$ \\ \noalign{\vskip 1.mm}
Photometry $\&$ RV GP$_{\rm dSHO}$ $P_{\rm rot}$ & days & $\mathcal{N}$(11.12, 1.5) & $11.09_{-0.08}^{+0.07}$ \\ \noalign{\vskip 1.mm}
Photometry GP$_{\rm dSHO}$ $Q_0$ &  & $\mathcal{J}$(0.002, 10) & $0.0021_{-0.0001}^{+0.0002}$ \\ \noalign{\vskip 1.mm}
Photometry GP$_{\rm dSHO}$ $dQ$ &  & $\mathcal{J}$(0.002, 200) & $2.6_{-0.9}^{+1.5}$ \\ \noalign{\vskip 1.mm}
Photometry GP$_{\rm dSHO}$ $f$ &  & $\mathcal{U}$(0.00, 1.00) & $0.00077_{-0.00025}^{+0.00033}$ \\ \noalign{\vskip 1.mm}
Limb darkening $u_1$ (TESS) &  & $\mathcal{N}$(0.48, 0.10) & $0.579_{-0.074}^{+0.075}$ \\ \noalign{\vskip 1.mm}
Limb darkening $u_2$ (TESS) &  & $\mathcal{N}$(0.17, 0.10) & $0.210_{-0.080}^{+0.090}$ \\ \noalign{\vskip 1.mm} 
\noalign{\vskip 1.mm} \hline
\end{tabular}
\end{center}
\begin{tablenotes}
    \tiny {
        \item{Notes: photometric and RV data share only the GP$_{\rm dSHO}$ $P_{\rm rot}$ hyperparameter, which is estimated from the light curves. Orbital parameters and the main derived planetary parameters are reported in Table\,\ref{tab:prior_and_param}.
 }
 }
\end{tablenotes}
\end{table*}

\section{Spectral energy distribution analysis}\label{sed_analysis}

\begin{figure}[h]
\begin{center}
    \includegraphics[clip, trim={0 0 0 1.3cm}, width=0.7\linewidth, angle=90]{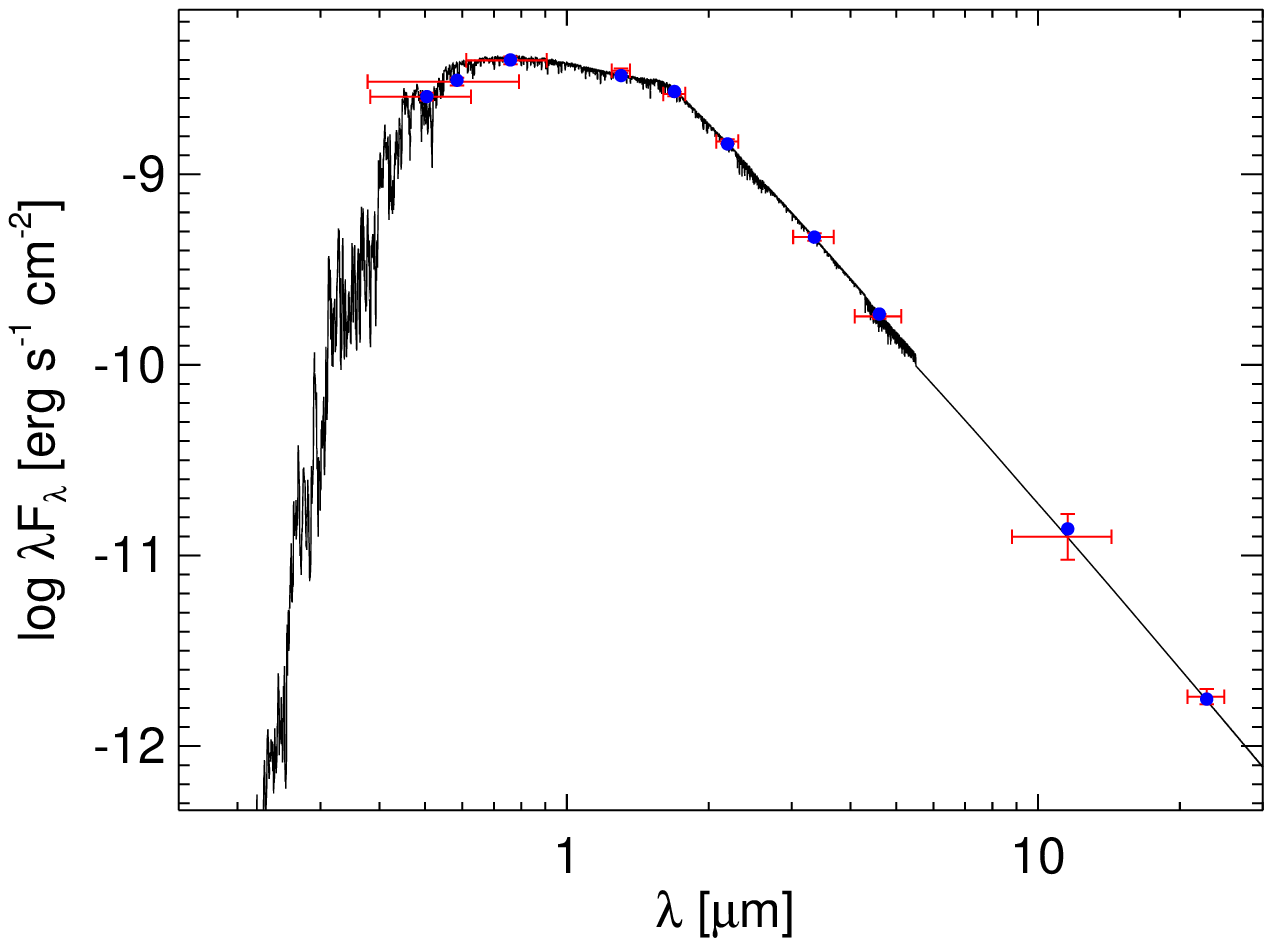}
    \caption{Spectral energy distribution of TOI-5734. Red symbols represent the observed photometric measurements, where the horizontal bars represent the effective width of the passband. Blue symbols are the model fluxes from the best-fit PHOENIX atmosphere model (black).}
    \label{fig:sed}
\end{center}
\end{figure}

As an independent determination of the basic stellar parameters, we performed an analysis of the broadband SED of the star together with the {\it Gaia\/} DR3 parallax \citep[with no systematic offset applied; see, e.g.][]{stassuntorres:2021}, to determine an empirical measurement of the stellar radius, following the procedures described in \citet{Stassun_2016, Stassun_2017, stassun2018_parallaxes}. We pulled the $JHK_S$ magnitudes from {\it 2MASS}, the $G_{\rm BP}-G_{\rm RP}$ magnitudes from {\it Gaia}, and the W1--W4 magnitudes from {\it WISE}. Together, the available photometry spans the full stellar SED over the wavelength range 0.2--20~$\mu$m (see Fig.\,\ref{fig:sed}).  \\
We performed a fit using PHOENIX stellar atmosphere models \citep{husser2013_phoenix_cond}, with the \teff, \logg, and \feh adopted from the spectroscopic analysis. The extinction, $A_{\rm V}$, was limited to the maximum line-of-sight value from the Galactic dust maps of \citet{Schlegel1998}. The resulting fit (see Fig.\,\ref{fig:sed}) has a reduced $\chi^2$ of 1.2, with a best-fit $A_V = 0.13 \pm 0.03$. Integrating the (unreddened) model SED gives the bolometric flux at Earth, $F_{\rm bol} = 6.370 \pm 0.074 \times 10^{-9}$ erg~s$^{-1}$~cm$^{-2}$. Taking the $F_{\rm bol}$ together with the {\it Gaia\/} parallax directly gives the bolometric luminosity, $L_{\rm bol} = 0.2107 \pm 0.0024~L_{\rm \odot}$. The stellar radius follows from the Stefan-Boltzmann relation, giving $R_\star = 0.678 \pm 0.029~R_{\rm \odot}$. In addition, we can estimate the stellar mass from the empirical relations of \citet{torres2010}, giving $M_\star = 0.725 \pm 0.044~M_{\rm \odot}$. The stellar mass and radius estimates obtained through the analysis conducted in this section are similar to the ones found in Sec.\,\ref{subsec:star_mass_radius}.

\section{LCOGT Ground-based Photometry\label{subsec:ground}}

The \textit{TESS} pixel scale is $\sim 21\arcsec$/pixel and photometric apertures typically extend out to roughly 1\,arcminute, generally causing multiple stars to blend in the \textit{TESS} photometric aperture. To attempt to determine the true source of the \textit{TESS} detection, we acquired ground-based time series follow-up photometry of the field around TOI-5734 as part of the \textit{TESS} Follow-up Observing Program \citep[TFOP;][]{collins:2019}\footnote{https://tess.mit.edu/followup}.

\begin{figure}[h]
\centering
\includegraphics[width=\columnwidth]{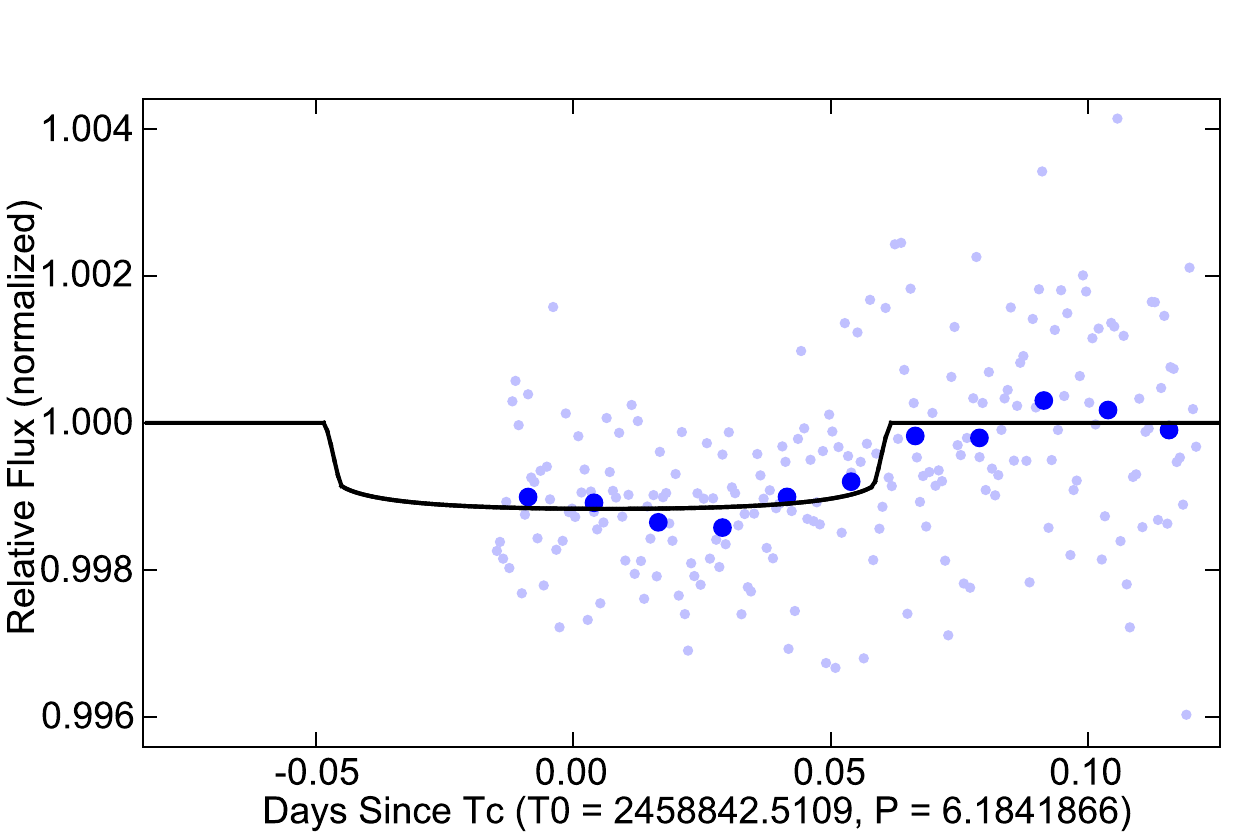}
    \caption{LCOGT 1\,m light curve of TOI-5734b phase folded to the ephemeris reported in this work. The larger blue symbols show the LCOGT data in 18-minute bins. The smaller blue symbols show the unbinned data. An independent transit model fit to the light curve, including ephemeris priors, and fixed $a/R_{\rm *}$ and impact parameters from the global fit in this work, is overplotted using a solid black line. No detrending has been applied since our BIC test preferred an undetrended light curve model. The model has $R_{\rm p}/R_{\rm *}$ = $0.032\pm0.004$, which is $1\sigma$ consistent with the value from the global fit in this work, and $T_{\rm c}$ = $2460289.6098\pm0.005$\,BJD, which is within $1\sigma$ consistent with the value predicted by the global fit ephemeris from this work.
    \label{TOI-5734_ground_lightcurve}}
\end{figure}

We observed a full transit window of TOI-5734b on UTC 2023 December 11 in Pan-STARRS $z_s$ band from the Las Cumbres Observatory Global Telescope (LCOGT) \citep{Brown:2013} 1\,m network node at Teide Observatory on the island of Tenerife (TEID). The 1\,m telescope is equipped with a $4096\times4096$\,px camera having an image scale of $0\farcs389$ per pixel, resulting in a $26\arcmin\times26\arcmin$ field of view. The images were calibrated by the standard LCOGT {\tt BANZAI} pipeline \citep{McCully:2018} and differential photometric data were extracted using {\tt AstroImageJ} \citep{Collins:2017}. We used circular $8\farcs1$ photometric apertures that excluded all flux from all known \textit{Gaia} DR3 catalogue neighbours. The ingress window was marred by passing clouds. With the weather-affected data removed, we ruled out deep events that could be responsible for the \textit{TESS} detection in all neighbouring \textit{Gaia} DR3 stars within $2^\prime.5$ of TOI-5734, and detected an on-time 1\,ppt egress in the target star aperture, which confirms that the TESS-detected event is indeed occurring on-target. The LCOGT light curve is shown in Figure\,\ref{TOI-5734_ground_lightcurve}. 

\section{High Resolution Imaging}\label{app:hri_obs}

As part of our standard process for validating transiting exoplanets to assess the possible contamination of bound or unbound companions on the derived planetary radii \citep{ciardi2015}, we observed TOI-5734 with near-infrared adaptive optics imaging on Palomar and optical speckle imaging at WIYN and SAI. The near-infrared and optical imaging complement each other with differing resolutions and sensitivities.

\subsection{Palomar NIR AO Imaging}

Observations of TOI-5734 were made on November 27th, 2023, with the PHARO instrument \citep{hayward2001} on the Palomar Hale (5\,m) behind the P3K natural guide star AO system \citep{dekany2013}. The pixel scale for PHARO is $0.025$\,\arcsec. The Palomar data were collected in a standard 5-point quincunx dither pattern in the $K_{cont}$ filter. The reduced science frames were combined into a single mosaicked image with final resolutions of $\sim 0.098$\,\arcsec.
        
\begin{figure}[h]
    \centering
    \includegraphics[width=\columnwidth]{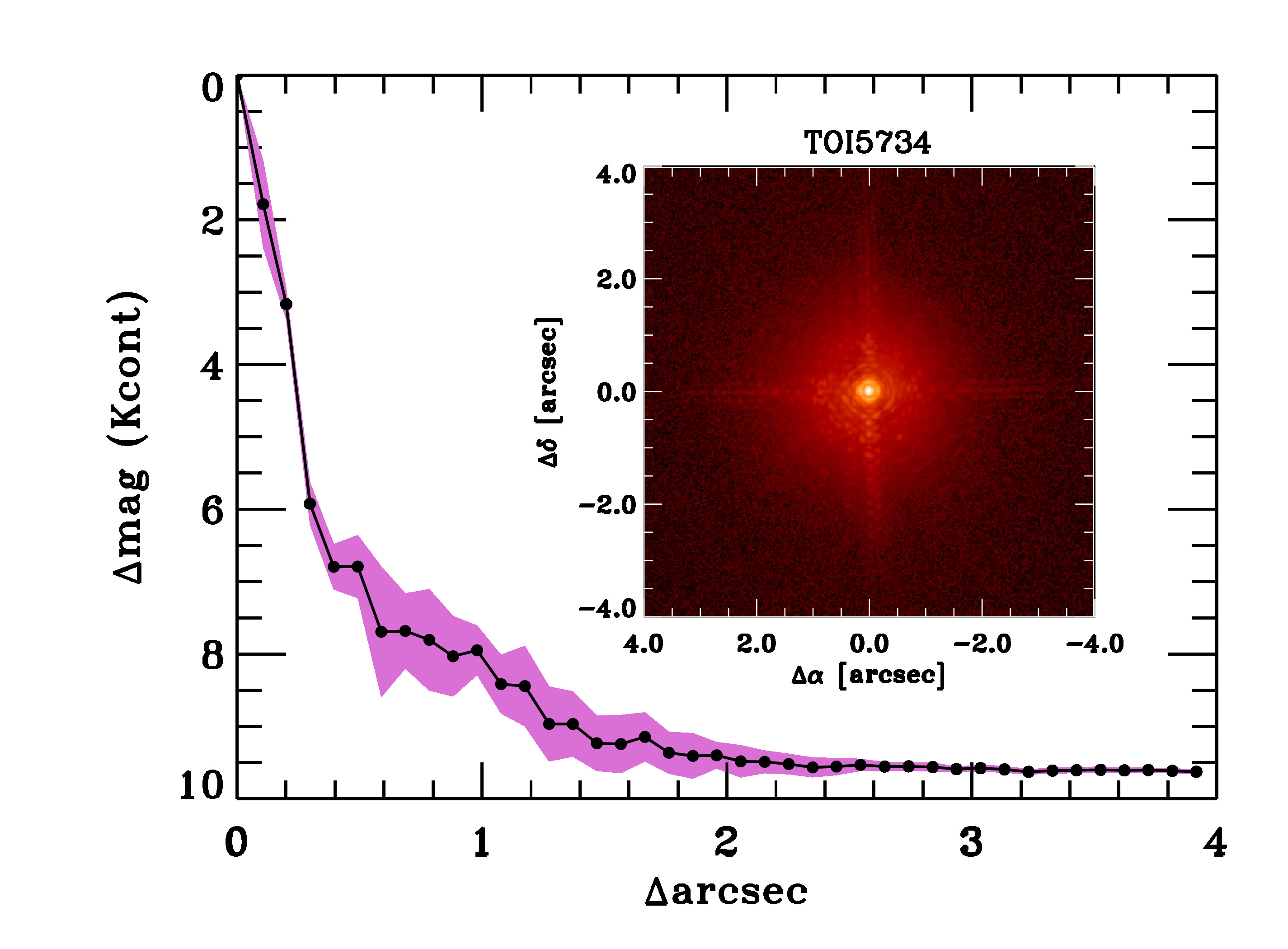}
    \caption{Palomar NIR AO imaging and 5$\sigma$ sensitivity curve in the K$_{cont}$ filter. {\it Inset:} Image of the central portion of the frames.}
    \label{fig:palomarao}
\end{figure}

The sensitivity of the final combined AO images was determined by injecting simulated sources azimuthally around the primary target every $20^\circ$ at separations of integer multiples of the central source's FWHM \citep{furlan2017}. The brightness of each injected source was scaled until standard aperture photometry detected it with 5$\sigma$ significance. The final 5$\sigma$ limit at each separation was determined from the average of all of the determined limits at that separation, and the uncertainty on the limit was set by the rms dispersion of the azimuthal slices at a given radial distance. No stellar companions were detected. The 5$\sigma$ sensitivity as a function of angular separation from TOI-5734 is shown in Figure\,\ref{fig:palomarao}.

\subsection{WIYN Speckle Imaging}

\begin{figure}[h]
    \centering
    \includegraphics[width=0.9\linewidth]{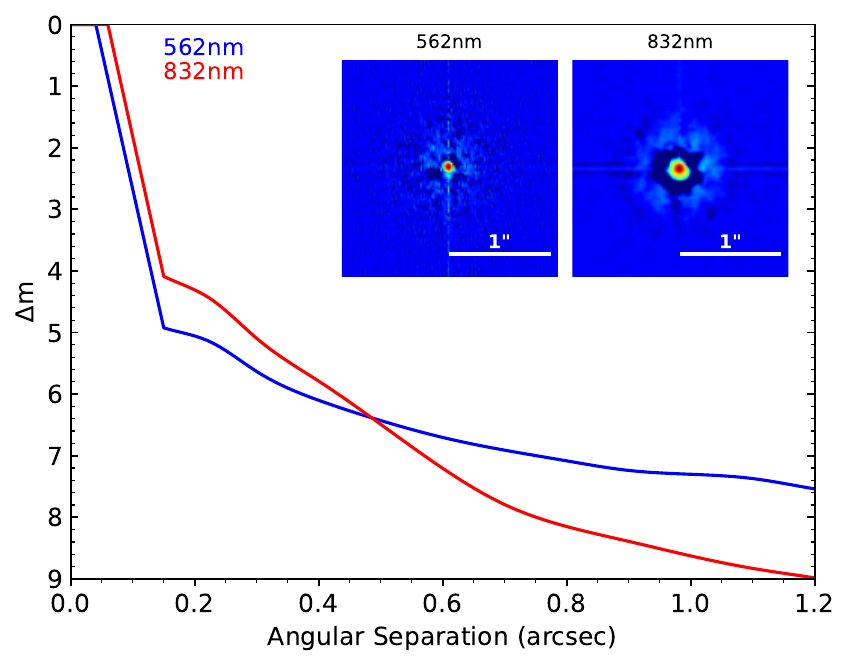}
    \caption{Contrast curves measured from NESSI at the WIYN 3.5\,m telescope for images taken with $\lambda_c=562$ (blue curve) and 832\,nm (red curve) and filters. {\it Inset:} Image of the central portion of the frames in the two filters.}
    \label{fig:speckle_imager_wiyn}
\end{figure}

TOI-5734 was observed on February 18th, 2024, using the NN-EXPLORE Exoplanet Stellar Speckle Imager (NESSI; \citealt{scott_2018_wiyn}) at the WIYN 3.5\,m telescope on Kitt Peak. NESSI data were taken simultaneously in two filters having central wavelengths of $\lambda_c=562$ and 832\,nm and consisted of 9000 40\,ms exposures in a $256\times256$ pixel ($4.6\times4.6$\,arcsec) section of the detectors. A nearby single star was also observed right before the science target to serve as a measure of the point spread function. The NESSI data reduction followed the description given in \cite{howell_2011_wiyn} and resulted in reconstructed images of the field and contrast curves measured from those images (see Figure\,\ref{fig:speckle_imager_wiyn}). No companion sources were detected near TOI-5734.

\subsection{SAI Speckle Imaging}

\begin{figure}[h]
    \centering
    \includegraphics[width=\columnwidth]{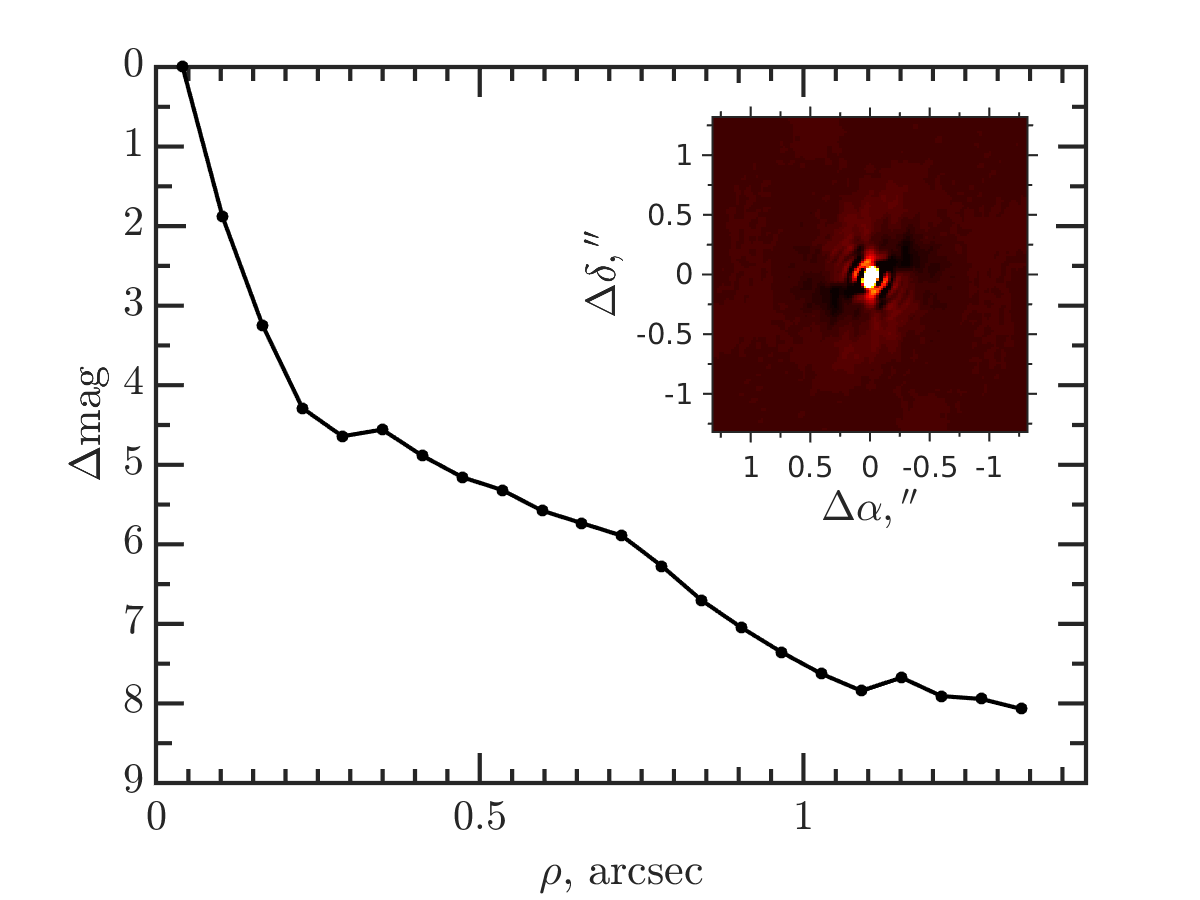}
    \caption{SAI speckle imaging and contrast curve in the I$_c$ band. {\it Inset:} Image of the central portion of the frames.}
    \label{fig:sai_image}
\end{figure}

TOI-5734 was observed on November 8th, 2022, with the speckle polarimeter on the 2.5-m telescope at the Caucasian Observatory of the Sternberg Astronomical Institute (SAI) of Lomonosov Moscow State University. A low-noise CMOS detector, Hamamatsu ORCA-quest (\citealt{Strakhov2023}), was used as a detector. The atmospheric dispersion compensator was active and allowed the use of the $I_\mathrm{c}$ band. The respective angular resolution is $0.083^{\prime\prime}$ and the long-exposure atmospheric seeing was $0.64^{\prime\prime}$. The power spectrum was estimated from 5200 frames with 23\,ms of exposure. The contrast detection limits expressed in units of magnitude difference between the source and the potential companion at angular separations $0.25^{\prime\prime}$ and $1.0^{\prime\prime}$ from the star are $\Delta I_\mathrm{c}=4.4$\,mag and $7.5$\,mag, respectively. From the acquired data, no companions were detected.
The 5$\sigma$ sensitivity as a function of angular separation from TOI-5734 is shown in Figure\,\ref{fig:sai_image}.

\section{RV completeness analysis}\label{app:completeness_rv}

\begin{figure}[h]
    \centering
    \includegraphics[width=0.9\linewidth]{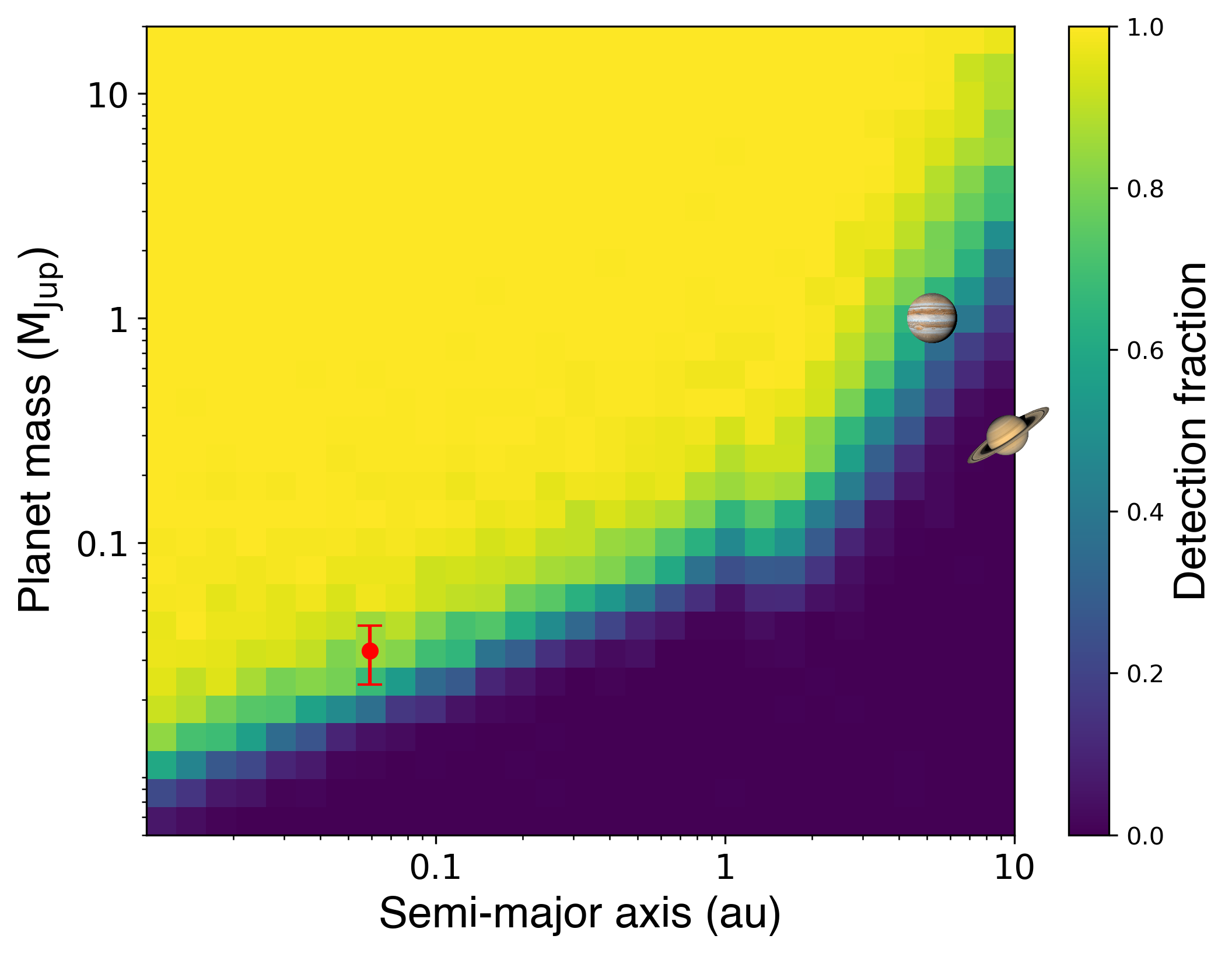}
    \caption{HARPS-N RV detection map for TOI-5734. The colour scale expresses the detection fraction, while the red circle marks the position of TOI-5734b. Jupiter and Saturn have been plotted as a comparison.}
    \label{fig:sensibility_analysis}
\end{figure}

We estimated the completeness of our RV time series by performing injection-recovery simulations, following \citet{Bonomo2023} and \citet{Naponiello2024}. In summary, we injected synthetic RVs with planetary signals at the time of our observations using HARPS-N error bars in addition to the stellar jitter. The signals of additional companions were simulated across a logarithmic $30 \times 30$ grid in planetary mass ($M_{\rm p}$) and semi-major axis ($a$), covering the ranges [0.01,20]\,$M_{\rm Jup}$ and [0.01,10]\,au. For each location of the grid, we generated 200 synthetic planetary signals with a free orbital inclination. Then we fitted the signals by employing either Keplerian orbits or linear and quadratic trends to take into account long-period signals. Finally, we adopted the Bayesian information criterion (BIC) to compare the fitted planetary model with a constant model. We considered the planetary signal significantly detected only when $\Delta \text{BIC} > 10$ in favour of the planet-induced one. The detection fraction was finally computed as the portion of detected signals for each grid element, as illustrated by Fig.\,\ref{fig:sensibility_analysis}.
From this result, we can tell that, with the current dataset, there are no planets of Jovian mass up to 3\,au; otherwise, they should have been seen through RV analysis. Jupiter and Saturn have been plotted as a comparison, and they would be at the detection limit and off-limits, respectively.

\begin{figure}[h]
    \centering
    \includegraphics[width=\linewidth]{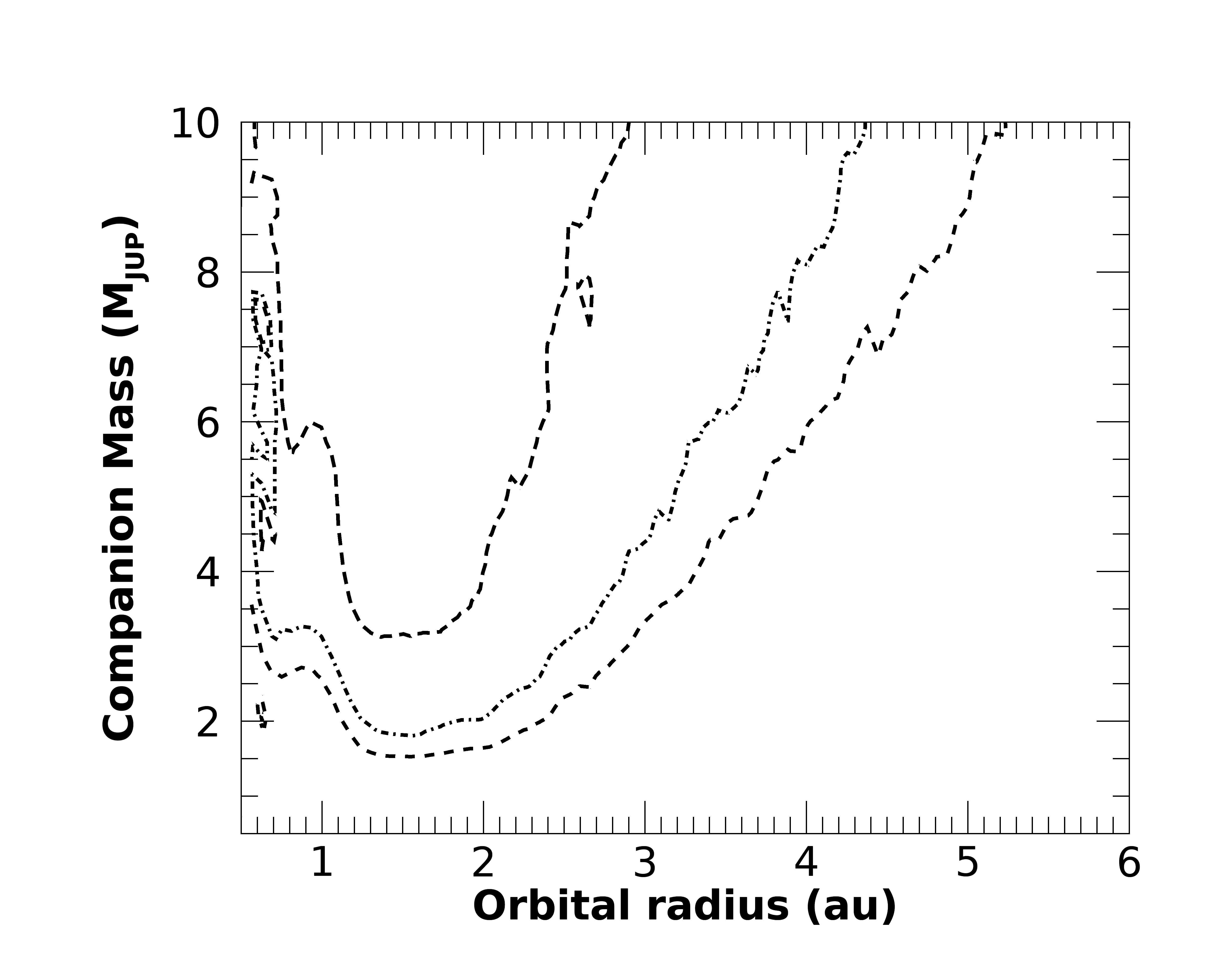}
    \caption{Sensitivity of \textit{Gaia} DR3 astrometry. The plot of companion mass vs. orbital radius is based on simulations of planetary signals that would have produced a RUWE value in excess with respect to the one reported in the \textit{Gaia} DR3 archive. Dashed contours represent the 80$\%$, 90$\%$, and 99$\%$ probability, from the outer to the inner.}
    \label{fig:ruwe_sensitivity}
\end{figure}

Furthermore, data from \textit{Gaia} DR3 for TOI-5734 indicate a RUWE of 0.93. RUWE serves as an indicator of how well a single-star astrometric solution fits the data, and values below 1.4 are generally interpreted as consistent with a good single-star model (see, e.g. \citealt{Lindegren2018, Lindegren2021}). By performing a Monte Carlo analysis to evaluate which combinations of companion mass and orbital separation would lead to RUWE values exceeding those observed in DR3 (following the methodology of \citealt{Sozzetti2023}), we determined that the \textit{Gaia} DR3 astrometry rules out, at a 99$\%$ confidence level, the presence of companions with masses around 6 times that of Jupiter at separations between approximately 1 and 2.5\,au (see Fig.\,\ref{fig:ruwe_sensitivity}). While RV measurements impose more stringent limits, it's worth noting that \textit{Gaia} DR3 astrometry is nearly immune to inclination-related uncertainties.

\end{appendix}

\end{document}